\documentclass{amsart}
\usepackage[foot]{amsaddr}
\makeatletter
\renewcommand{\email}[2][]{%
  \ifx\emails\@empty\relax\else{\g@addto@macro\emails{,\space}}\fi%
  \@ifnotempty{#1}{\g@addto@macro\emails{\textrm{(#1)}\space}}%
  \g@addto@macro\emails{#2}%
}
\makeatother

\usepackage{graphicx,float}
\usepackage{tabls}
\usepackage{multirow}

\usepackage{tikz,amsfonts,bbm}

\usepackage{graphicx,lastpage,subcaption}%
\usepackage{amsmath,color,url}

\begin{document}

\newcommand{\EX}{{\Bbb{E}}}
\newcommand{\PX}{{\Bbb{P}}}

\newcommand{\lp}{\left(}
\newcommand{\rp}{\right)}
\newcommand{\lb}{\left[}
\newcommand{\rb}{\right]}
\newcommand{\lbr}{\left\{}
\newcommand{\rbr}{\right\}}
\newcommand{\lnorm}{\left\|}
\newcommand{\rnorm}{\right\|}

\newtheorem{remark}{Remark}[section]
\newtheorem{lemma}{Lemma}[section]
\newtheorem{theorem}{Theorem}[section]
\newtheorem{corollary}{Corollary}[section]
\newtheorem{proposition}{Proposition}[section]
\newtheorem{definition}{Definition}[section]
\newtheorem{assumption}{Assumption}[section]

\def\PP{{{\rm l}\kern - .15em {\rm P} }}
\def\PN2{{\PP_{N}-\PP_{N-2}}}

\newcommand{\erf}[1]{\mbox{erf}\left(#1\right)}

\newcommand{\D}{\mathbbm{D}}
\newcommand{\I}{\mathbbm{I}}
\newcommand{\N}{\mathbbm{N}}
\newcommand{\R}{\mathbbm{R}}
\newcommand{\Z}{\mathbbm{Z}}

\newcommand{\cD}{\mathcal{D}}
\newcommand{\cE}{\mathcal{E}}
\newcommand{\cF}{\mathcal{F}}
\newcommand{\cH}{\mathcal{H}}
\newcommand{\cO}{\mathcal{O}}
\newcommand{\cP}{\mathcal{P}}

\newcommand{\bfeta}{\boldsymbol{\eta}}
\newcommand{\bLambdar}{\boldsymbol{\Lambda}_r}
\newcommand{\bLambdarL}{\boldsymbol{\Lambda}_r^{L^2}}
\newcommand{\bLambdarH}{\boldsymbol{\Lambda}_r^{H^1}}
\newcommand{\bmu}{\boldsymbol{\mu}}
\newcommand{\bPhi}{\boldsymbol{\Phi}}
\newcommand{\bPhir}{\boldsymbol{\Phi}_r}
\newcommand{\bphi}{\boldsymbol{\varphi}}
\newcommand{\bphir}{\boldsymbol{\varphi}_r}
\newcommand{\bPsi}{\boldsymbol{\Psi}}
\newcommand{\btau}{\boldsymbol{\tau}}

\newcommand{\ba}{\boldsymbol{a}}
\newcommand{\bas}{{\boldsymbol a}^{snap}}
\newcommand{\bA}{\boldsymbol{A}}
\newcommand{\bb}{\boldsymbol{b}}
\newcommand{\bB}{\boldsymbol{B}}
\newcommand{\bc}{\boldsymbol{c}}
\newcommand{\bd}{\boldsymbol{d}}
\newcommand{\be}{\boldsymbol{e}}
\newcommand{\bff}{\boldsymbol{f}}
\newcommand{\bFF}{{\boldsymbol F}}
\newcommand{\bG}{{\boldsymbol G}}
\newcommand{\bGs}{{\boldsymbol G}^{snap}}
\newcommand{\bh}{\boldsymbol{h}}
\newcommand{\bH}{\boldsymbol{H}}
\newcommand{\bk}{\boldsymbol{k}}
\newcommand{\bL}{\boldsymbol{L}}
\newcommand{\bM}{\boldsymbol{M}}
\newcommand{\bq}{{\boldsymbol q}}
\newcommand{\bqs}{{\boldsymbol q}^{snap}}
\newcommand{\br}{\boldsymbol{r}}
\newcommand{\bS}{\boldsymbol{S}}
\newcommand{\bu}{\boldsymbol{u}}
\newcommand{\bU}{\boldsymbol{U}}
\newcommand{\bur}{{\boldsymbol{u}}_r}
\newcommand{\bUr}{{\boldsymbol{U}}_r}
\newcommand{\bv}{\boldsymbol{v}}
\newcommand{\bV}{\boldsymbol{V}}
\newcommand{\bvr}{{\boldsymbol{v}}_r}
\newcommand{\bw}{\boldsymbol{w}}
\newcommand{\bW}{\boldsymbol{W}}
\newcommand{\bwr}{{\boldsymbol{w}}_r}
\newcommand{\bWr}{{\boldsymbol{W}}_r}
\newcommand{\bx}{\boldsymbol{x}}
\newcommand{\bX}{\boldsymbol{X}}
\newcommand{\bXh}{{\bf X}^h}
\newcommand{\bXr}{{\bf X}^r}
\newcommand{\bY}{\boldsymbol{Y}}
\newcommand{\bz}{\boldsymbol{z}}

\newcommand{\tA}{\tilde{A}}
\newcommand{\tB}{\widetilde{B}}
\newcommand{\tC}{\widetilde{C}}

\newcommand{\oa}{\overline{a}}
\newcommand{\oA}{\overline{A}}
\newcommand{\oc}{\overline{c}}
\newcommand{\obc}{\overline{\boldsymbol c}}
\newcommand{\op}{\overline{p}}
\newcommand{\oU}{\overline{U}}
\newcommand{\obu}{\overline{\boldsymbol u}}
\newcommand{\obU}{\overline{\boldsymbol U}}
\newcommand{\obur}{\overline{\boldsymbol{u}_r}}
\newcommand{\obUr}{\overline{\boldsymbol{U}_r}}
\newcommand{\obv}{\overline{\boldsymbol v}}
\newcommand{\obx}{\overline{\boldsymbol x}}
\newcommand{\ob}[1]{\overline{\boldsymbol{#1}}}
\newcommand{\orr}[1]{\overline{#1}^r}
\newcommand{\obr}[1]{\overline{\boldsymbol{#1}}^r}
\newcommand{\oh}[1]{\overline{#1}^h}
\newcommand{\obh}[1]{\overline{\boldsymbol{#1}}^h}

\newcommand{\as}{a^{snap}}
\newcommand{\CinvNabla}{C_{inv}^{\nabla}(r)}
\newcommand{\CinvDelta}{C_{inv}^{\Delta}(r)}
\newcommand{\Deltar}{\Delta_r}
\newcommand{\Gs}{G^{snap}}
\newcommand{\ct}{\bu_h^{avg}}

\newcommand{\bus}{{\bf u}^*}
\newcommand{\By}{\mathcal B(\by)}
\newcommand{\eci}[1]{\mathcal E_{#1}}
\newcommand{\dpyi}[1]{\delta_{#1}^+(\by)}
\newcommand{\dmyi}[1]{\delta_{#1}^-(\by)}
\newcommand{\cA}{{\mathcal A(\by)}}
\newcommand{\dyi}[1]{\delta_{#1}(\by)}
\newcommand{\cG}{{\mathcal G(\bx,\by)}}
\newcommand{\cGi}[1]{{\mathcal G_{#1}(\bx,\by)}}
\newcommand{\pti}{\partial_i}
\newcommand{\ptii}[1]{\partial_{#1}}
\newcommand{\vertiii}[1]{{|\!|\!| #1 |\!|\!|}}

\newcommand{\half}{\frac{1}{2}}


\newcommand{\red}[1]{{\color{red}#1}}
\newcommand{\blue}[1]{{\color{blue}#1}}
\definecolor{vargreen}{rgb}{0.0, 0.5, 0.0}
\newcommand{\green}[1]{{\color{vargreen} #1}}

\newcommand{\todo}[1]{{\color{red}#1}}
\newcommand{\inserted}[1]{\blue{#1}}
\newcommand{\deleted}[1]{{}}

\newcommand{\TI}[1]{{\color{blue}TI: #1}}
\newcommand{\DRW}[1]{{\color{vargreen}DRW: #1}}
\newcommand{\ZW}[1]{{\color{red}ZW: #1}}
\newcommand{\XX}[1]{{\color{brown}XX: #1}}
\newcommand{\CM}[1]{{\color{purple}CM: #1}}

\title{
A NUMERICAL INVESTIGATION OF THE LENGTHSCALE IN THE MIXING-LENGTH REDUCED ORDER MODEL OF THE \\ TURBULENT CHANNEL FLOW
  }

\author{Changhong Mou}
\address[CM]{
  Department of Mathematics \\
  University of Wisconsin-Madison \\
  Madison, WI 53706, USA 
  }
\email[CM]{  cmou@vt.edu
}

\author{Elia Merzari}
\address[EM]{
  Department of Nuclear Engineering \\
  The Pennsylvania State University \\
  University Park, PA 16802, USA
    }
\email[EM]{ 
  ebm5351@psu.edu
}

\author{Omer San}
\address[OS]{
  School of Mechanical and Aerospace Engineering \\
  Oklahoma State University \\
  Stillwater, OK 74078, USA 
    }
\email[OS]{ 
  osan@okstate.edu
}

\author{Traian Iliescu}
\address[TI]{
  Department of Mathematics \\
  Virginia Tech \\
  Blacksburg, VA 24061, USA 
    }
\email[TI]{ 
  iliescu@vt.edu}

\maketitle

\begin{abstract}
In this paper, we propose a novel reduced order model (ROM) lengthscale definition that is based on energy distribution arguments.
This novel ROM lengthscale is fundamentally different from the current ROM lengthscales, which are generally based on dimensional arguments.
As a first step in the assessment of the new, energy based ROM lengthscale, we compare it with a standard, dimensional based ROM lengthscale in the mixing-length ROM (ML-ROM) simulation of the turbulent channel flow at $Re_{\tau} = 395$.
The numerical investigation shows that the energy based ROM lengthscale yields a significantly more stable ML-ROM than the dimensional based ROM lengthscale.
The new energy based lengthscale definition could allow the development of scale-aware reduced order modeling strategies that are better suited for flow-specific applications.

\smallskip
\noindent
\textbf{Keywords.}
Reduced order model,
lengthscale,
mixing-length,
turbulent channel flow
\end{abstract}

\section{Introduction}

Reduced order models (ROMs) are models whose dimension is dramatically lower than the dimension of full order models (FOMs), i.e., computational models obtained by using classical numerical discretizations (e.g., finite element or finite volume methods).
Because of their relatively low dimensionality, ROMs can be used as efficient alternatives to FOMs in computationally intensive applications, e.g., flow control, shape optimization, and uncertainty quantification.
The Galerkin ROM (G-ROM) framework has been often used in the numerical simulation of fluid flows~\cite{HLB96,ahmed2021closures}.
The G-ROM is constructed as follows:
First, in an offline phase, the FOM is used to produce snapshots, which are then utilized to construct a low-dimensional (i.e., $r \ll N$) ROM basis $\{ \varphi_{1}, \ldots, \varphi_{r}\}$, where $r$ is the ROM dimension and $N$ is the FOM dimension. 
Next, the ROM basis is used together with a Galerkin projection to build the G-ROM, which has the following form:
\begin{eqnarray}
	\frac{d\ba}{dt} 
	= {\bf F}(\ba), 	
	\label{eqn:g-rom}
\end{eqnarray}
where $\ba$ is the vector of coefficients in the ROM approximation $\sum_{i=1}^{r} a_{i}(t) \bphi_{i}(\bx)$ of the variable of interest and the vector $\bff$ comprises the ROM operators that are preassembled in the offline phase. 
In the online phase, the G-ROM~\eqref{eqn:g-rom} is employed  for parameters values and/or time intervals that are different from those used in the training stage. 

The G-ROM~\eqref{eqn:g-rom} is computationally efficient and relatively accurate in the numerical simulation of laminar flows.
However, the G-ROM generally yields inaccurate results in the numerical simulation of turbulent flows.
The main reason for the G-ROM's inaccuracy is that it is used in the under-resolved regime, i.e., when the number of ROM basis functions, $r$, is not large enough to accurately represent the complex dynamics of the turbulent flow.
Thus, for turbulent flows, the standard G-ROM is replaced with 
\begin{eqnarray}
	\frac{d\ba}{dt} 
	= {\bf F}(\ba)
	+ \btau, 	
	\label{eqn:les-rom}
\end{eqnarray}
where $\btau(\ba)$ is the ROM closure model, which represents the effect of the discarded modes on the G-ROM dynamics.
There are different types of ROM closures, which are surveyed in~\cite{ahmed2021closures}.
In this paper, we consider the mixing-length ROM closure model, which increases the physical viscosity by the following constant:
\begin{eqnarray}
	\nu_{ML}
	= \alpha \, U_{ML} \, L_{ML},
	\label{eqn:ml}
\end{eqnarray}
where $L_{ML}$ is a characteristic lengthscale, $U_{ML}$ is a characteristic velocity scale, and $\alpha$ is a constant.
The ML model~\eqref{eqn:ml} is a functional closure model, which aims at increasing the ROM viscosity in order to dissipate energy and mimic the effect of the discarded modes~\cite{CSB03}.  
The ML-ROM~\eqref{eqn:les-rom}--\eqref{eqn:ml} was first used in~\cite{AHLS88,HLB96} and was further investigated in~\cite{wang2012proper}.

The main goal of this paper is to investigate the role of the lengthscale, $L_{ML}$, used in the ML-ROM~\eqref{eqn:les-rom}--\eqref{eqn:ml}.
Specifically, we first propose a new lengthscale definition, which is based on energy distribution arguments.
Then, we compare this new lengthscale with the classical lengthscale definition used in~\cite{AHLS88,HLB96,wang2012proper}, which is based on dimensional arguments.
To compare these two lengthscales, we utilize them to build the ML-ROM~\eqref{eqn:les-rom}--\eqref{eqn:ml}, which we then test in the numerical simulation of the turbulent channel flow at $Re_{\tau}=395$.

The rest of the paper is organized as follows:
In Section~\ref{sec:g-rom}, we outline the standard G-ROM and the ML-ROM.
In Section~\ref{sec:rom-lengthscale}, we define the new, energy based lengthscale and 
the standard dimensional based lengthscale.
In Section~\ref{sec:numerical-results}, we present results for our investigation of the ML-ROM
equipped with the two lengthscales in numerical simulation of the turbulent channel flow at $Re_{\tau}=395$.
Finally, in Section~\ref{sec:conclusions}, we draw conclusions and outline directions of future research.

\section{Galerkin ROM and Mixing-Length ROM}
    \label{sec:g-rom}

In this section, we outline the construction of the standard Galerkin ROM (G-ROM) and mixing-length ROM (ML-ROM).    
As a mathematical model, we consider the incompressible Navier-Stokes equation (NSE): 
\begin{eqnarray}
\frac{\partial \bu}{\partial t} - Re^{-1} \Delta\bu +\bigl(\bu\cdot\nabla\bigr)\bu+\nabla p \,&=\, \bff,\label{eqn:nse-1} \\
\nabla\cdot\bu\,&=\,0, \label{eqn:nse-2}
\end{eqnarray}
where  $\bu = [u_1,u_2,u_3]^\top$ is the velocity vector field, $p$ the pressure field, $Re$ the Reynolds number, and $\bff$ the forcing term.
The NSE are equipped with appropriate boundary and initial conditions.

\subsection{G-ROM}

To build the G-ROM, we consider the centering trajectory of the flow,
\begin{eqnarray}
\bU(\bx) = \frac{1}{T}\int_t^{t+T} \bu(\bx,t)dt,
\end{eqnarray}
and we assume that the ROM velocity approximation can be written as follows:
\begin{eqnarray}
\bu_r(\bx,t) \, 
=\,\bU(\bx)+\sum_{j=1}^ra_j(t)\bphi(\bx)\, ,\label{eqn:rom-soln}
\end{eqnarray}
where $\{\bphi_j\}_{j=1}^r$ are the ROM basis functions and $\ba = [a_1,\cdots,a_r]^\top$ are the sought ROM coefficients.
In our numerical experiments, we use the proper orthogonal decomposition (POD)~\cite{HLB96} to construct the ROM basis, but other ROM bases could be used~\cite{brunton2019data,hesthaven2015certified,quarteroni2015reduced}.
The next step in the G-ROM construction is to replace $\bu$ with $\bu_r$ in
~\eqref{eqn:nse-1} and project the resulting equations onto the space spanned by the ROM basis, $\{\bphi_j\}_{j=1}^r$.
This yields the G-ROM:
\begin{eqnarray}
\left(\frac{\partial \bu_r}{\partial t},\bphi_i \right)
+\bigl((\bu_r\cdot\nabla)\bu_r,\bphi_i\bigr)
+ Re^{-1} \bigl(\nabla\bu_r,\nabla\bphi_i\bigr) = \bigl(\bff,\bphi_i\bigr)\,,\qquad i =1,\cdots,r\, .\label{eqn:g-rom-1}
\end{eqnarray}
The G-ROM can be 
written as the following dynamical system for the vector of time coefficients, $\ba(t)$:
\begin{eqnarray}
	\overset{\bullet}{\ba} 
	= \bb+A  \ba 
	+ \ba^\top B  \ba \, , 
	\label{eqn:g-rom-U}
\end{eqnarray}
where
\begin{align}
&\bb_i \,=\, \bigl(\bphi_i,\bff\bigr)-\bigl(\bphi_i,\bU\cdot\nabla\bU\bigr)
- Re^{-1} \bigl(\nabla\bphi_i,\nabla\bU\bigr)\, ,\\
&\bA_{im} \,=\, -\bigl(\bphi_i,\bU\cdot\nabla\bphi\bigr) - \bigl(\bphi_i,\bphi_m\cdot\nabla\bU\bigr)
- Re^{-1} \bigl(\nabla\bphi_i,\nabla\bphi_m\bigr)\, ,\\
&\bB_{imn} \, =\, -\bigl(\bphi_i,\bphi_m\cdot\nabla\bphi_n\bigr)\, .
\end{align}


\subsection{ML-ROM}

As mentioned in the introduction, the G-ROM~\eqref{eqn:g-rom-U} generally yields inaccurate results in the numerical simulation of turbulent flows.
Thus, in those cases, the G-ROM is generally equipped with a ROM closure model, which models the effect of the discarded ROM modes $\{ \bphi_{r+1}, \ldots \}$ on the G-ROM dynamics.
In general, the G-ROM with a closure model can be written as 
\begin{eqnarray}
	\overset{\bullet}{\ba} 
	= \bb
	+ A  \ba 
	+ \ba^\top B  \ba 
	+ \btau \, , 
	\label{eqn:g-rom-closed}
\end{eqnarray}
where $\btau$ is the ROM closure model.
The current ROM closure models are surveyed in~\cite{ahmed2021closures}.
Some of these ROM closure models are inspired from classical large eddy simulation (LES) closure modeling~\cite{sagaut2006large}.
These LES-ROM closure models generally involve a lengthscale.
There are only a few ROM lengthscales in current use.
In the next section, we define a novel ROM lengthscale.
To assess this new ROM lengthscale, we consider one of the simplest ROM closure models, the ML-ROM~\cite{HLB96,wang2012proper}, in which the ROM closure term $\btau$ in~\eqref{eqn:g-rom-closed} is written as
\begin{eqnarray}
	\btau
	= - \bigl( \alpha \, U_{ML} \, \delta \bigr) \, S \, \ba \, ,
	\label{eqn:ml-v2}
\end{eqnarray}
where $\delta$ is one of the two ROM lengthscales defined in Section~\ref{sec:rom-lengthscale}, $U_{ML}$ is a characteristic velocity scale, $\alpha$ is a constant, and $S$ is the ROM stiffness matrix with entries $S_{ij} = \bigl(\nabla\bphi_i,\nabla\bphi_j\bigr)$.

\section{ROM Lengthscale}
    \label{sec:rom-lengthscale}

In this section, we present two different ROM lengthscales: 
In Section~\ref{sec:rom-lengthscale-dimensional}, we present the first ROM lengthscale, denoted $\delta_1$, which is based on dimensional analysis arguments.
In Section~\ref{sec:rom-lengthscale-energy}, we propose a new ROM lengthscale, denoted $\delta_2$, which is based on energy balance arguments.

Both definitions aim at expressing the ROM lengthscale as a function of the following two types of input variables: 
(i) the ROM variables (e.g., the ROM dimension, $r$, the total number of ROM basis functions, $R$, the eigenvalues, $\lambda_i$, and the ROM basis functions, $\bphi_i$).
(ii) the FOM variables (e.g., the fine FOM mesh size, $h$, the FOM solution, $\bu^{FOM}$, and the computational domain characteristic lengthscale, $L$).
Given these input variables, we then try to answer the following natural question:
{\it For a given ROM dimension, $r$, what is the corresponding ROM lengthscale, $\delta$?}

\subsection{ROM Lengthscale $\delta_1$: \ Dimensional Analysis}
    \label{sec:rom-lengthscale-dimensional}

In this section, we use dimensional analysis to construct the first ROM lengthscale, $\delta_1$.
To this end, we follow the approach used in Section 3.2 in \cite{wang2012proper}, which, in turn, is based on the pioneering ML-ROM proposed in~\cite{AHLS88} for a turbulent pipe flow.

To construct the ROM lengthscale $\delta_1$, we first define the componentwise FOM velocity fluctuations:
\begin{eqnarray}
    {u^{'}_{i}}^{FOM} 
    = \sum_{j=r+1}^{R} a_j^{FOM} \, \varphi_j^i ,
    \qquad
    i = 1, 2, 3,
    \label{eqn:u-prime}
\end{eqnarray}
where $R$ is the total number of ROM modes and $\varphi_j^i$ are the componentwise ROM basis functions.
Using the componentwise FOM velocity fluctuations ${u^{'}_{1}}^{FOM}, {u^{'}_{2}}^{FOM}$, and ${u^{'}_{3}}^{FOM}$ in the $x, y$, and $z$ directions, respectively, we construct the FOM velocity fluctuation vector field ${\bu'}^{FOM} =  [ {u^{'}_{1}}^{FOM}, {u^{'}_{2}}^{FOM}, {u^{'}_{3}}^{FOM} ]$.
Since ${\bu'}^{FOM}$ varies with time, we calculate the time averaged value of ${\bu'}^{FOM}$, i.e.,
\begin{eqnarray}
    \langle {\bu'}^{FOM} \rangle_t (\bx) 
    = \frac{1}{M} \sum_{k=1}^{M} {\bu'}^{FOM}(\bx, t_k) 
    = \frac{1}{M} \sum_{k=1}^{M} \sum_{l= r+1}^{R} \biggl(\bu^{FOM}(\cdot,t_k),\bphi_l(\cdot)\biggr)\bphi_l(\bx),
\end{eqnarray}
where $M$ is the number of snapshots. 

To construct the ROM lengthscale $\delta_1$, we adapt equation (22) in~\cite{wang2012proper} to our computational setting (i.e., the turbulent channel flow in Section~\ref{sec:numerical-results}):
\begin{eqnarray}
\delta_{1}
:= \left( \frac{\int_{0}^{L_1} \, \int_{0}^{L_2} \, \int_{0}^{L_3} 
\sum_{i=1}^{3} {u^{'}_{i}}^{FOM} {u^{'}_{i}}^{FOM}  
\, dx_1 \, dx_2 \, dx_3}
{\int_{0}^{L_1} \, \int_{0}^{L_2} \, \int_{0}^{L_3} 
\sum_{i=1}^{3} \sum_{j=1}^{3}  \frac{\partial {u^{'}_{i}}^{FOM}}{\partial x_j\hfill} \, \frac{\partial {u^{'}_{i}}^{FOM}}{\partial x_j\hfill}  
\, dx_1 \, dx_2 \, dx_3} \right)^{1/2} ,
\label{eqn:delta-1}
\end{eqnarray}
where 
$L_1, L_2$, and $L_3$ are the streamwise, wall-normal, and spanwise dimensions of the 
computational domain of the turbulent channel flow test problem, respectively.

Note that a quick dimensional analysis shows that the quantity defined in \eqref{eqn:delta-1}
has the units of a lengthscale:
\begin{eqnarray}
[ \delta_{1} ] 
= \left( 
\frac{\frac{m}{s} \, \frac{m}{s} \, m^3}
{\frac{1}{s} \, \frac{1}{s} \, m^3 }
\right)^{1/2}
= m \, .
\label{eqn:delta-1-dimensional-analysis}
\end{eqnarray}

We also note that an alternative lengthscale was defined in equation (23) in~\cite{wang2012proper}.
Since this alternative lengthscale was not used in the numerical investigation in~\cite{wang2012proper}, we do not consider it in this study.

The ROM lengthscale, $\delta_1$, defined in~\eqref{eqn:delta-1}, depends on the FOM velocity fluctuation vector field, ${\bu'}^{FOM}$.

\subsection{ROM Lengthscale $\delta_2$: \ Energy Balancing}
    \label{sec:rom-lengthscale-energy}

In this section, we use energy balancing arguments and propose a new ROM lengthscale, $\delta_2$.
Noticing that the ROM truncation level, $r$, has the role of dividing the kinetic energy of the system, we can require that $\delta_2$ do the same.
Specifically, we require that the ratio of kinetic energy contained in the first $r$ ROM modes,  $\sum_{i=1}^{r} \lambda_i$, to the kinetic energy contained in the total number of ROM modes, 
$\sum_{i=1}^{R} \lambda_i$, is equal to the ratio of the kinetic energy that can be represented on 
the mesh of size $\delta_2$, $KE(\delta_2)$, to the kinetic energy that can be represented on the 
FOM mesh, $KE(h)$:
\begin{eqnarray}
\frac{\sum_{i=1}^{r} \lambda_i}{\sum_{i=1}^{R} \lambda_i} 
= \frac{KE(\delta_2)}{KE(h)} \, .
\label{energy_balancing}
\end{eqnarray}
To compute the ratio $\frac{KE(\delta_2)}{KE(h)}$ in \eqref{energy_balancing}, we transfer the problem to the usual Fourier space.
To this end, we first notice that $\delta_2$ defines a cutoff wavenumber:
\begin{eqnarray}
k_{\delta_2} := \frac{2 \, \pi}{\delta_2} \, .
\label{def_k_c}
\end{eqnarray}
We then notice that the kinetic energy in the system can be written in terms of the energy 
spectrum, $E(\cdot)$:
\begin{eqnarray}
KE(k) 
= \int_{k_0}^{k} E(k') \, dk' \, ,
\label{def_E}
\end{eqnarray}
where $\displaystyle k_0 = \frac{2 \, \pi}{L}$ is the Fourier wavenumber that corresponds to the computational domain characteristic lengthscale, $L$.
In the case of isotropic, homogeneous turbulence, we have the usual energy spectrum given by
Kolmogorov's theory~\cite{sagaut2006large, Pop00}
\begin{eqnarray}
E(k) 
\sim C \, \varepsilon^{2/3} \, k^{-5/3} \, .
\label{energy spectrum}
\end{eqnarray}
Thus, the condition imposed in \eqref{energy_balancing} can be written as
\begin{eqnarray}
\frac{\int_{k_0}^{k_{\delta_2}} E(k) \, dk}{\int_{k_0}^{k_h} E(k) \, dk}
= \frac{\sum_{i=1}^{r} \lambda_i}{\sum_{i=1}^{R} \lambda_i} 
\stackrel{\text{notation}}{=} \Lambda \, ,
\label{delta_energy}
\end{eqnarray}
where $\displaystyle k_h = \frac{2 \, \pi}{h}$ is the highest Fourier wavenumber that can be resolved on the given FOM meshsize, $h$.
The LHS of \eqref{delta_energy} can be evaluated by using \eqref{energy spectrum}:
\begin{eqnarray}
\int_{k_0}^{k_{\delta_2}} E(k) \, dk
= C \, \varepsilon^{2/3} \, \int_{k_0}^{k_{\delta_2}} k^{-5/3}\, dk
= C \, \varepsilon^{2/3} \, \frac{k_{\delta_2}^{-2/3} - k_{0}^{-2/3}}{-2/3} \, ,
\label{delta_energy_1}
\end{eqnarray}
and, similarly, 
\begin{eqnarray}
\int_{k_0}^{k_h} E(k) \, dk
= C \, \varepsilon^{2/3} \, \int_{k_0}^{k_h} k^{-5/3}\, dk
= C \, \varepsilon^{2/3} \, \frac{k_h^{-2/3} - k_{0}^{-2/3}}{-2/3} .
\label{delta_energy_2}
\end{eqnarray}
Plugging \eqref{delta_energy_1} and \eqref{delta_energy_2} back into \eqref{delta_energy}, simplifying, and rearranging,
we obtain
\begin{eqnarray}
k_{\delta_2}^{-2/3}
= \Lambda \, k_{h}^{-2/3}
+ \left( 1 - \Lambda \right) \, k_{0}^{-2/3} \, .
\label{delta_energy_3}
\end{eqnarray}
Since $1 \leq r \leq R$, $\Lambda$ satisfies the inequality $0 < \Lambda \leq 1$.
Thus,~\eqref{delta_energy_3} 
implies that $k_{\delta_2}^{-2/3}$ is a convex combination of $k_{h}^{-2/3}$ and $k_{0}^{-2/3}$.
Furthermore, 
\begin{eqnarray}
\biggl[ (r \rightarrow R) \, \Longrightarrow \, (k_{\delta_2} \longrightarrow k_h) \biggr] 
\qquad \text{and} \qquad 
\biggl[ (r \rightarrow 0) \, \Longrightarrow \, (k_{\delta_2} \longrightarrow k_0) \biggr] \, ,
\label{delta_energy_3b}
\end{eqnarray}
as expected. 
Using~\eqref{delta_energy_3} together with \eqref{def_k_c}, gives us a formula for $\delta_2$:
\begin{eqnarray}
\begin{aligned}
\delta_2
&= \frac{2 \, \pi}{k_{\delta_2}}
= 2 \, \pi \,
\left[ 
\Lambda \, \left( \frac{2 \, \pi}{h} \right)^{-2/3}
+ \left( 1 - \Lambda \right) \, \left( \frac{2 \, \pi}{L} \right)^{-2/3}
\right]^{3/2} 
\\
&= \left[ 
\Lambda \, h^{2/3}
+ \left( 1 - \Lambda \right) \, L^{2/3}
\right]^{3/2} \, .\\
\label{eqn:delta-2}
\end{aligned}
\end{eqnarray}
The new ROM lengthscale, $\delta_2$, defined in~\eqref{eqn:delta-2}, depends on the FOM mesh size, $h$, the ROM dimension, $r$, the total number of ROM basis functions, $R$, the eigenvalues, $\lambda_i$, and the computational domain characterisitic lengthscale, $L$.
We note that, as expected, as $r$ approaches $R$, $\delta_2$ approaches $h$, and as $r$ approaches $1$, $\delta_2$ approaches $L$.


\section{Numerical Results}
    \label{sec:numerical-results}

In this section, we preform a numerical investigation of the two lengthscales discussed in Section~\ref{sec:rom-lengthscale}: the ROM lengthscale defined in~\eqref{eqn:delta-1} and the new ROM lengthscale defined in~\eqref{eqn:delta-2}.
Specifically, we fix the velocity scale, $U_{ML}$,  in the ML-ROM~\eqref{eqn:les-rom}--\eqref{eqn:ml}, and use two different characteristic lengthscales: $L_{ML} = \delta_1$ and $L_{ML} = \delta_2$.
We denote the two resulting ML-ROMs as ML-ROM1 and ML-ROM2, respectively.
To compare the two lengthscales, we compare ML-ROM1 with ML-ROM2 in the numerical simulation of the 3D turbulent channel flow at $Re_{\tau} = 395$.
We emphasize that the goal of this section is not to find the best ML-ROM.
Instead, we aim at investigating whether the two lengthscales are different and, if so, quantify their differences.

\subsection{Numerical Setting}

The computational domain is a rectangular box, $\Omega = (-2\pi,2\pi)\times (0,2)\times (-2\pi/3,2\pi/3)$. 
We enforce no slip boundary conditions 
on the walls $y=0$ and $y=2$, and periodic boundary conditions 
on the remaining walls. 
We also use the forcing term $\bff = [1,0,0]^\top$
and the Reynolds number $Re_\tau =395$ ($Re=13,750$).

To generate the snapshots, we run 
an LES model using the rNS-$\alpha$ scheme~\cite{rebholz2017global,rebholz2017accurate} with the time step size $\Delta t
= 0.002$.

We collect a total of $5000$ snapshots from $t = 60$ to $t = 70$ and use the POD to generate the ROM basis.
For the ROM time discretization, we utilize the commonly used linearized BDF2 temporal discretization with a time step size $\Delta t = 0.002$.
As the ROM initial conditions, we use the ROM projections of the LES approximations at $t = 60$ and $t = 60.002$.
For convenience, in our ROM simulations, $t=0$ corresponds to $t=60$ in the LES model.

To assess the ROMs’ performance, we use two different criteria: 
(i) the time evolution of the kinetic energy, $E(t)$, and (ii) second-order statistics.

We define the ROM kinetic energy as follows:
\begin{eqnarray}
E(t) 
= \frac{1}{2}\int_\Omega\bigl((u_1(\bx,t)^2+u_2(\bx,t)^2+u_3(\bx,t)^2)\bigr) \, d\bx \, ,
\end{eqnarray}
where $u_1, u_2$, and $u_3$ are the components of the ROM velocity field approximation.

Following~\cite{rebholz2017accurate}, we consider the 
following two second-order statistics:
(i) the normalized root mean square (RMS) of the streamwise vlocity component, $U_{RMS}$:
\begin{eqnarray}
U_{RMS} := \frac{\biggl|\widetilde{\mathbb{R}}_{11}-\frac{1}{3} \sum_{j=1}^3\widetilde{\mathbb{R}}_{jj}\biggr|^{1/2}}{u_{1,\tau}},
\end{eqnarray}
and (ii) the normalized streamwise-spanwise Reynolds stress tensor component, $\mathbb{R}_{12}$:
\begin{eqnarray}
\mathbb{R}_{12} := \frac{\widetilde{\mathbb{R}}_{12}}{u_{1,\tau}^2}.
\end{eqnarray}
In these second-order statistics, the Reynolds stress tensor components are calculated as follows: 
\begin{eqnarray}
\widetilde{\mathbb{R}}_{ij} = \left\langle \left\langle u_{i}u_{j}\right\rangle_s\right\rangle_t
-\left\langle \left\langle u_{i}\right\rangle_s\right\rangle_t \left\langle\left\langle u_{j}\right\rangle_s\right\rangle_t \, ,
\end{eqnarray}
where $\langle \cdot \rangle_s$ denotes spatial averaging, $\langle \cdot \rangle_t$ denotes time averaging, and $u_i$ are the components of the given ROM or FOM velocity field approximations.
$u_{1,\tau}$ is the friction velocity, which is calculated by the following formula:
\begin{align}
    u_{1,\tau} = 
\frac{U_{\text{mean}}(y_{min})}{y_{min}},
\end{align}
where $y_{min}$ is the minimum positive $y$-value of the FOM mesh and
$U_{\text{mean}} = \left\langle \left\langle 
u_1 \right\rangle_s\right\rangle_t$ the average ROM or FOM velocity flow profile.

\subsection{Numerical Results}

First, we investigate the relative size of the two ROM lengthscales, $\delta_1$ defined in~\eqref{eqn:delta-1} and $\delta_2$ defined in~\eqref{eqn:delta-2}.
To calculate $\delta_2$ in equation~\eqref{eqn:delta-2}, we define the global mesh size, $h$, as  $h=\max_{K\in\mathcal{K}}h_K$, where the mesh $\mathcal{K}$ is the set $\{K\}$ of tetrahedrons K and $h_K$ is the length of the longest edge of the local tetrahedron, $K$.

In Table~\ref{table:delta}, we list the $\delta_1$ and $\delta_2$ values for $r$ values from $4$ to $50$.
These results show that the two ROM lengthscales have very different behaviors: 
While the magnitudes of both ROM lengthscales do not vary significantly when $r$ varies, the magnitude of $\delta_2$ is almost two orders of magnitude larger than the magnitude of $\delta_1$.  
\begin{table}[H]
    \centering
    \begin{tabular}{c|c c c c c c c c c c c c }
    \hline\hline
        $r$ & 4 & 8 & 16 & 32 &40  & 50 
    \\ \hline
        $\delta_1$ &4.64e-02 &4.65e-02 
        &4.68e-02  &4.68e-02 &4.66e-02   &4.62e-02  
    \\ 
        $\delta_2$
        &2.18e+00  &2.30e+00 &2.49e+00  &2.77e+00 
        &2.87e+00  &2.97e+00
    \\     
    \hline 
    \end{tabular}
    \caption{ROM lengthscale values for different $r$ values.}
    \label{table:delta}
\end{table}

\begin{figure}[H]
\centering
    \includegraphics[width=.45\textwidth]{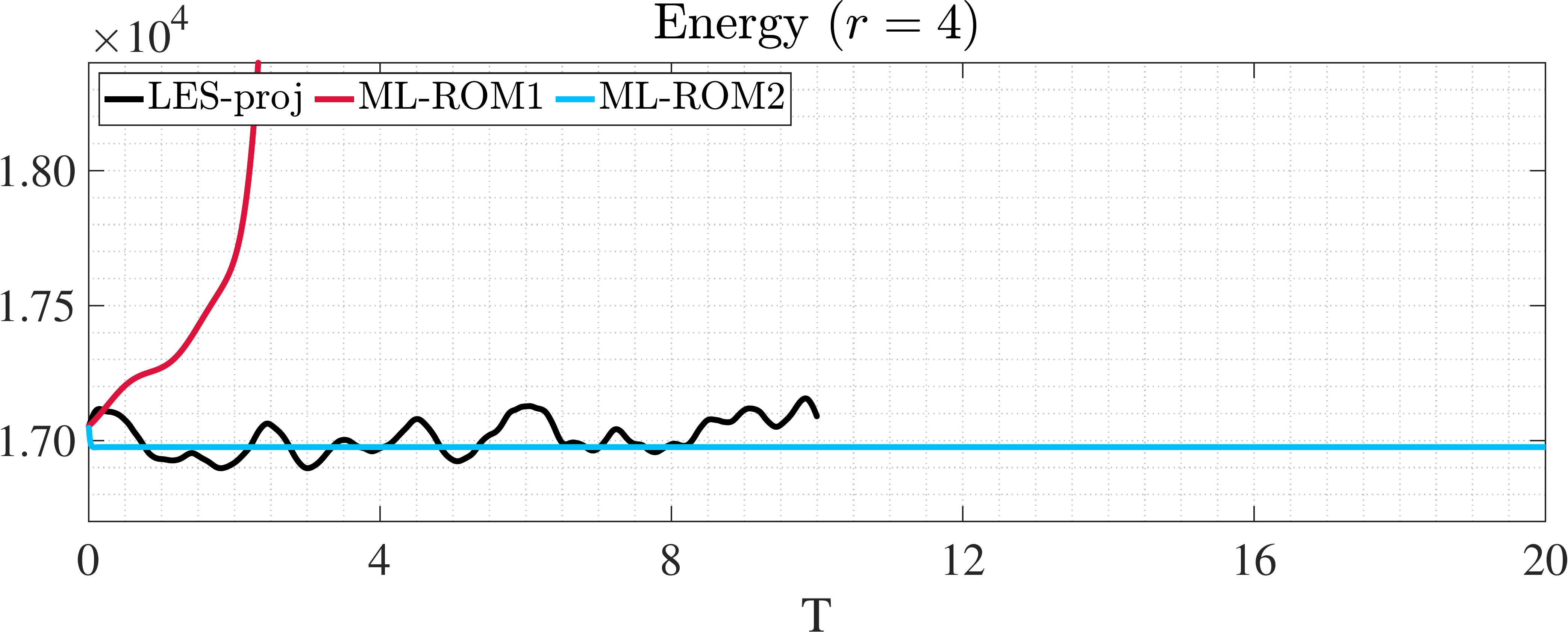}
    \includegraphics[width=.45\textwidth]{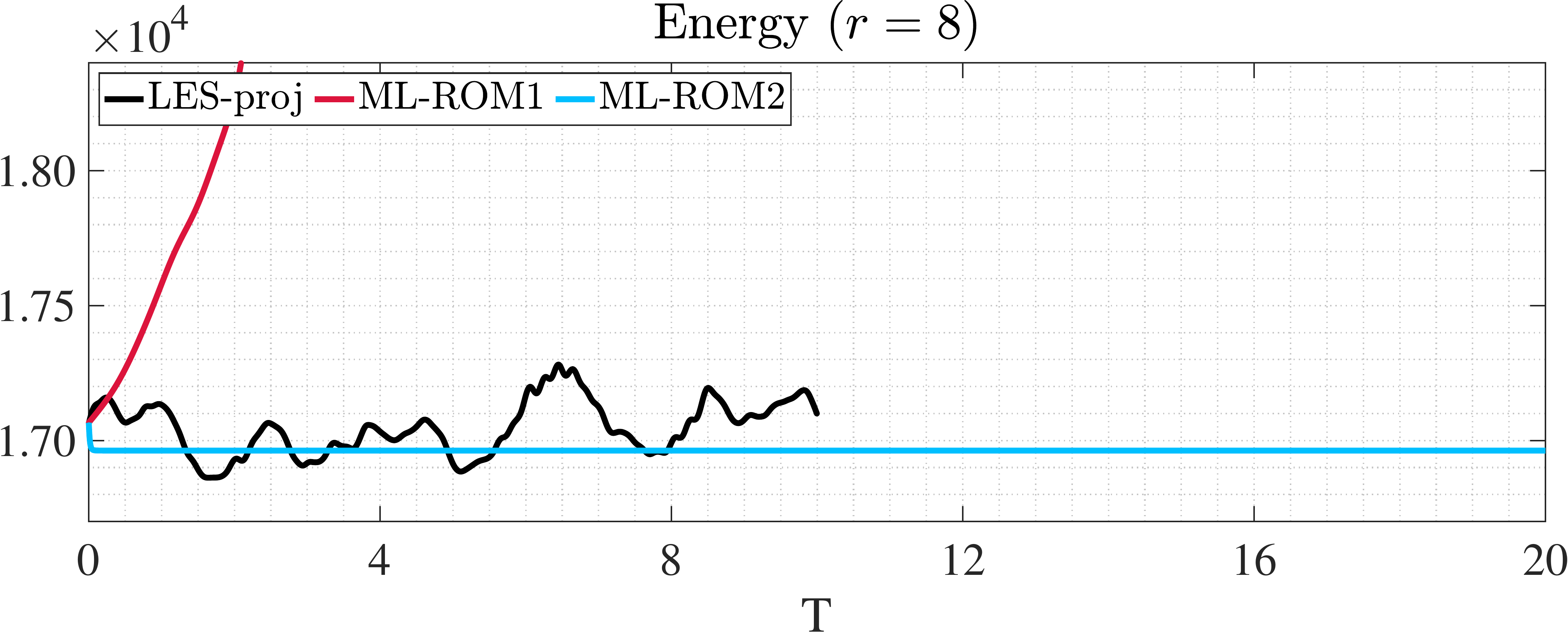}
    \includegraphics[width=.45\textwidth]{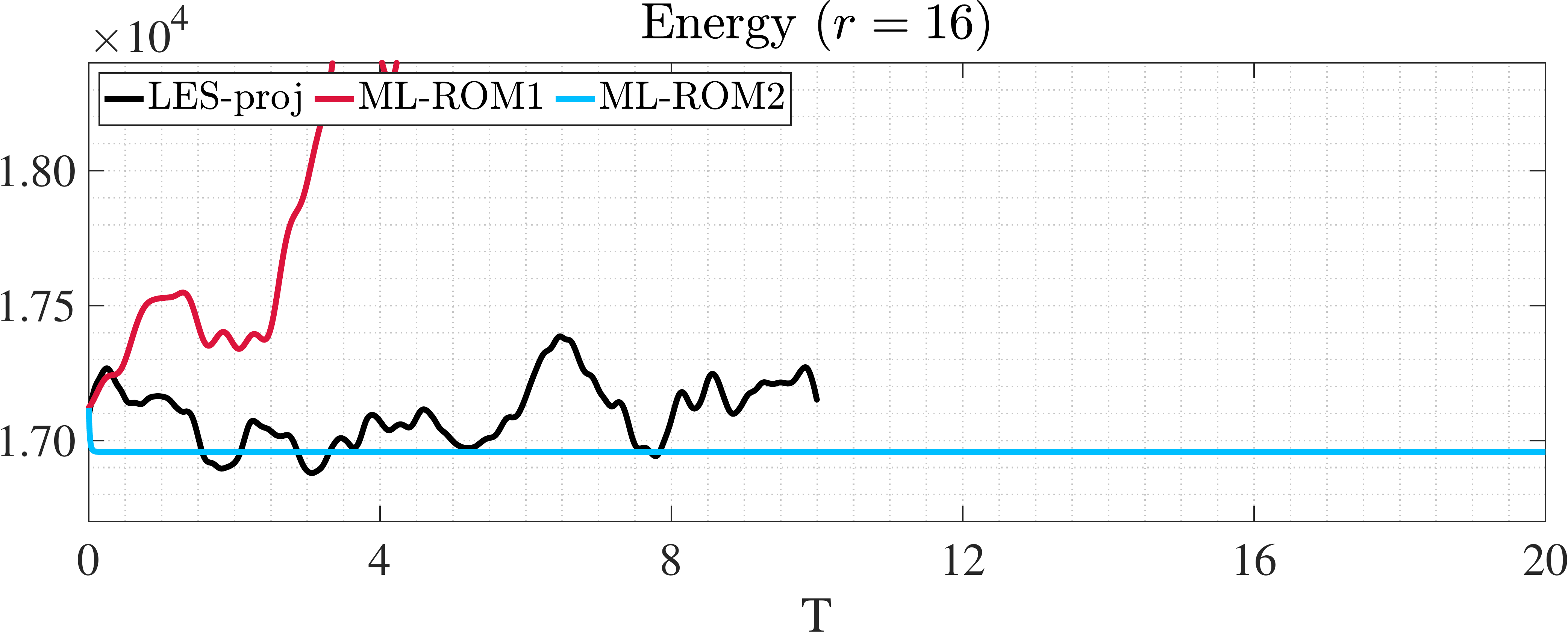}
    \includegraphics[width=.45\textwidth]{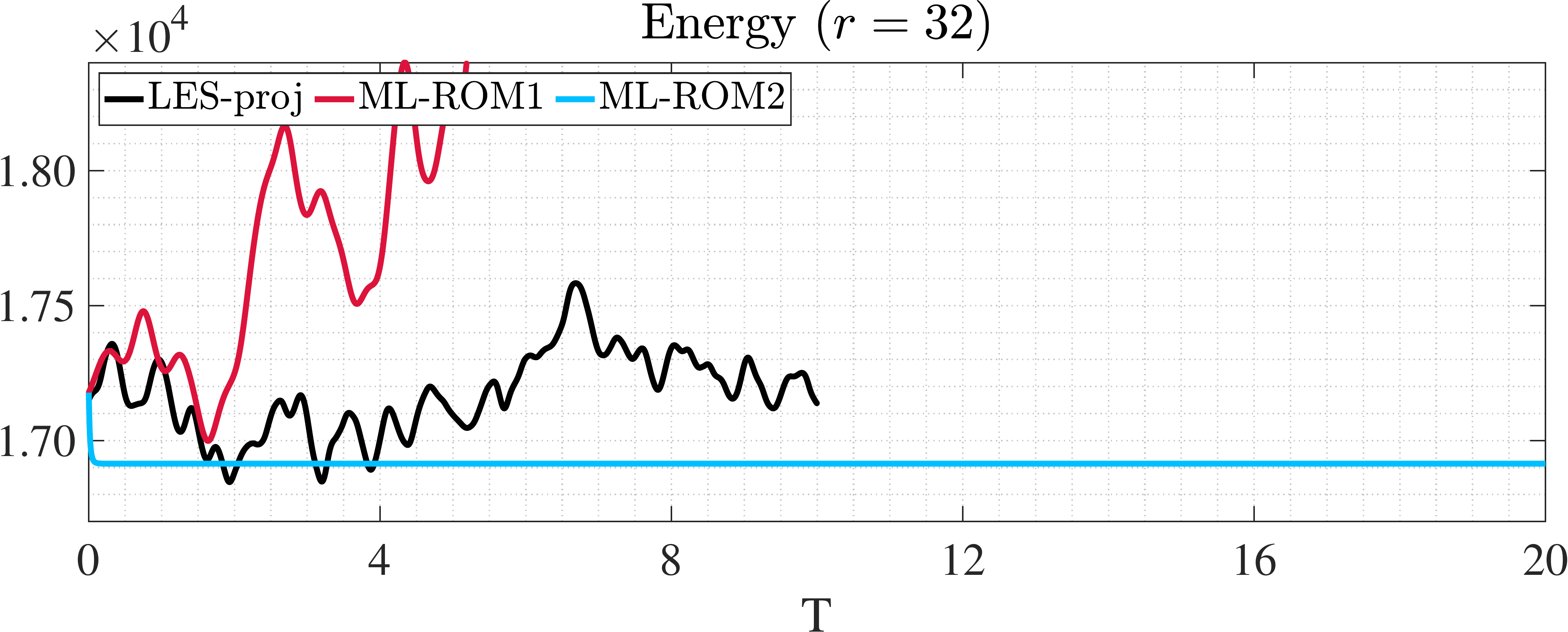}
    \includegraphics[width=.45\textwidth]{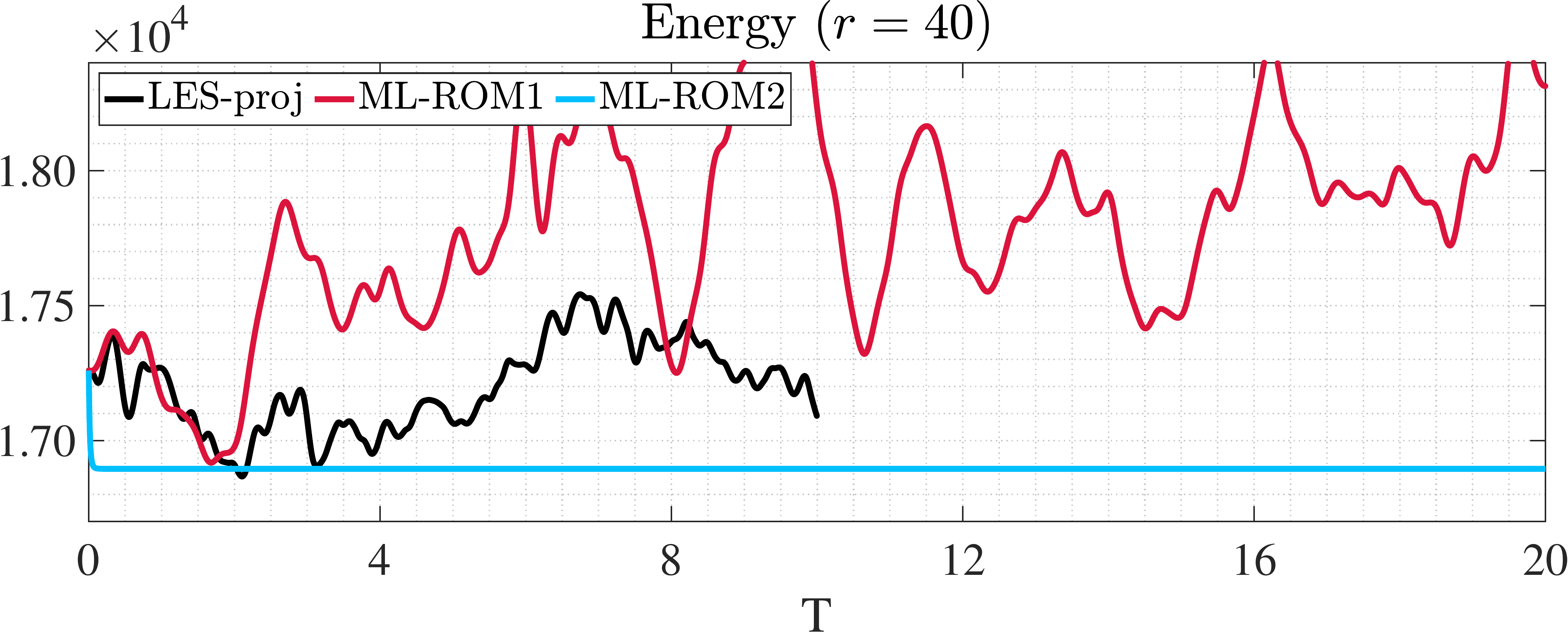}
    \includegraphics[width=.45\textwidth]{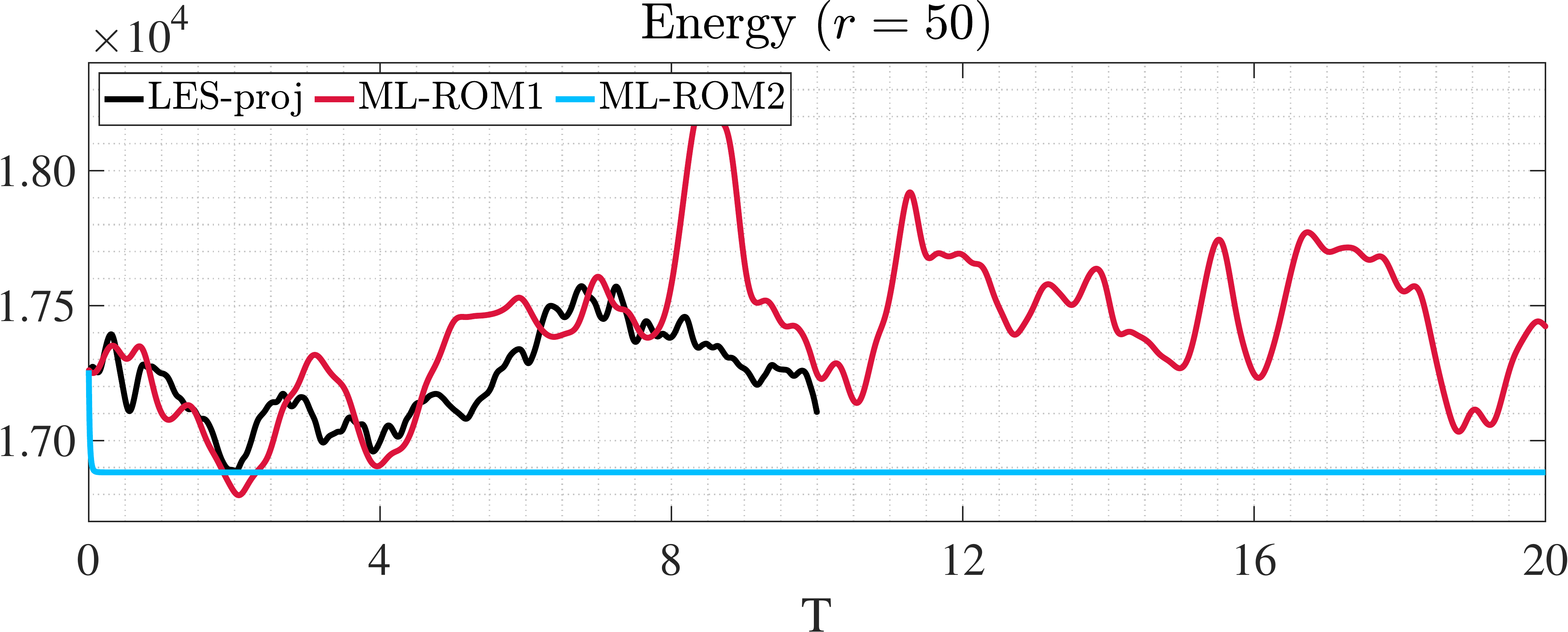}   
    \caption{Time evolution of the kinetic energy for $\alpha=6\times 10^{-3}$
    }    
    \label{fig:ke-alpha-1}
\end{figure}

\begin{figure}[H]
\centering
     \begin{subfigure}[b]{0.48\textwidth}
         \centering
    \includegraphics[width=.45\textwidth]{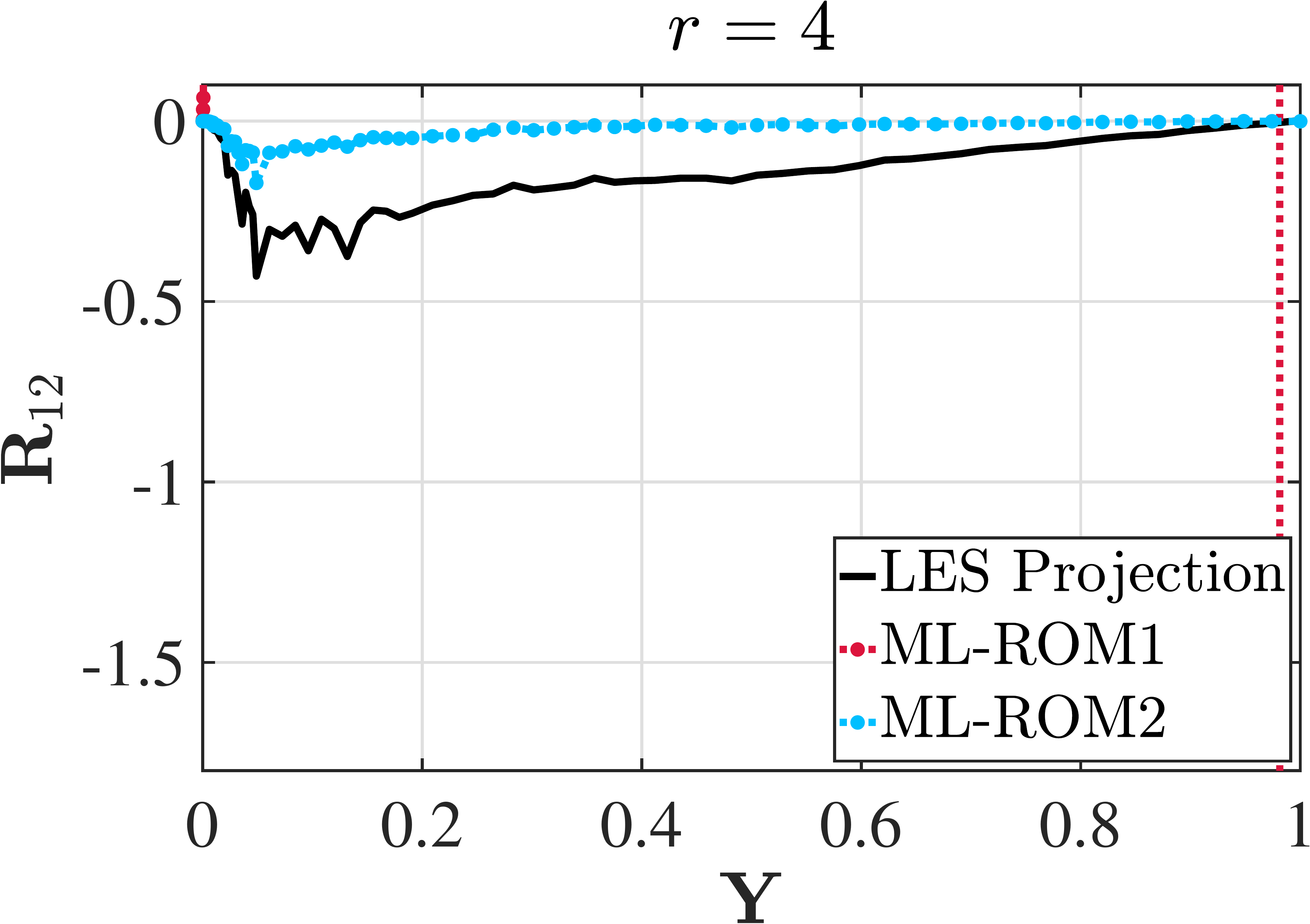}
    \includegraphics[width=.45\textwidth]{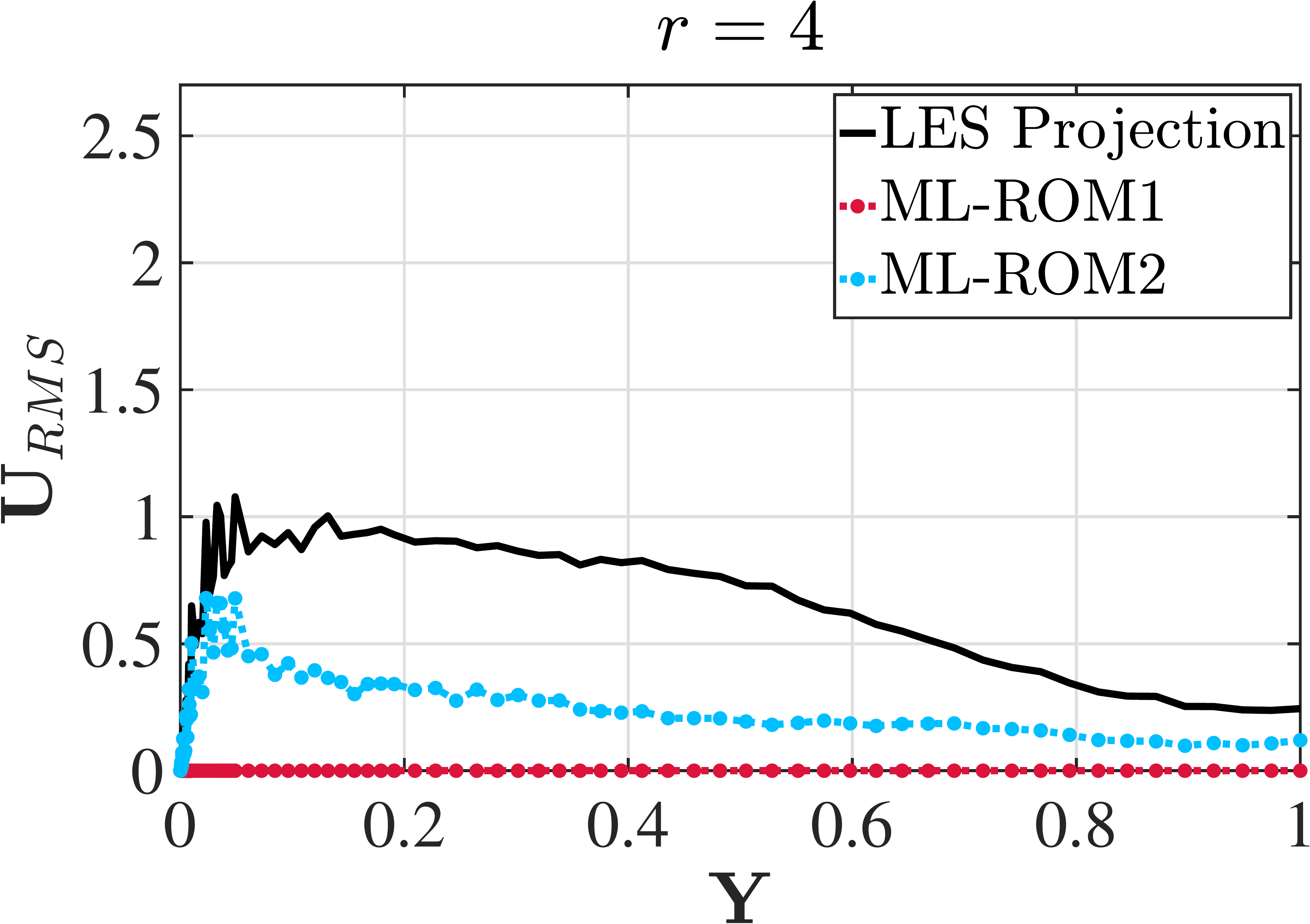}         \caption{$r=4$}
         \label{fig:stat-r-4}
     \end{subfigure}
     \begin{subfigure}[b]{0.48\textwidth}
         \centering
    \includegraphics[width=.45\textwidth]{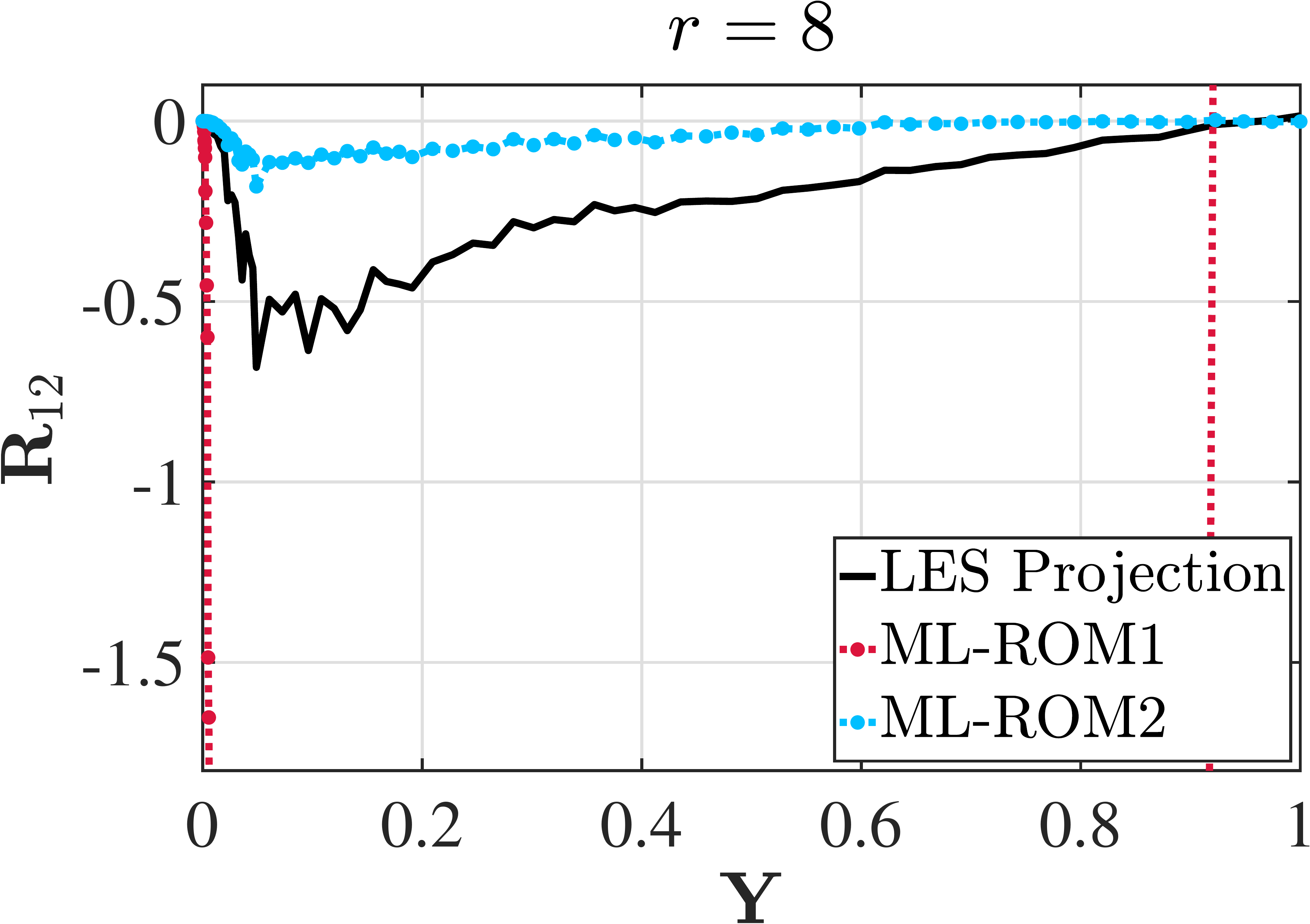}
    \includegraphics[width=.45\textwidth]{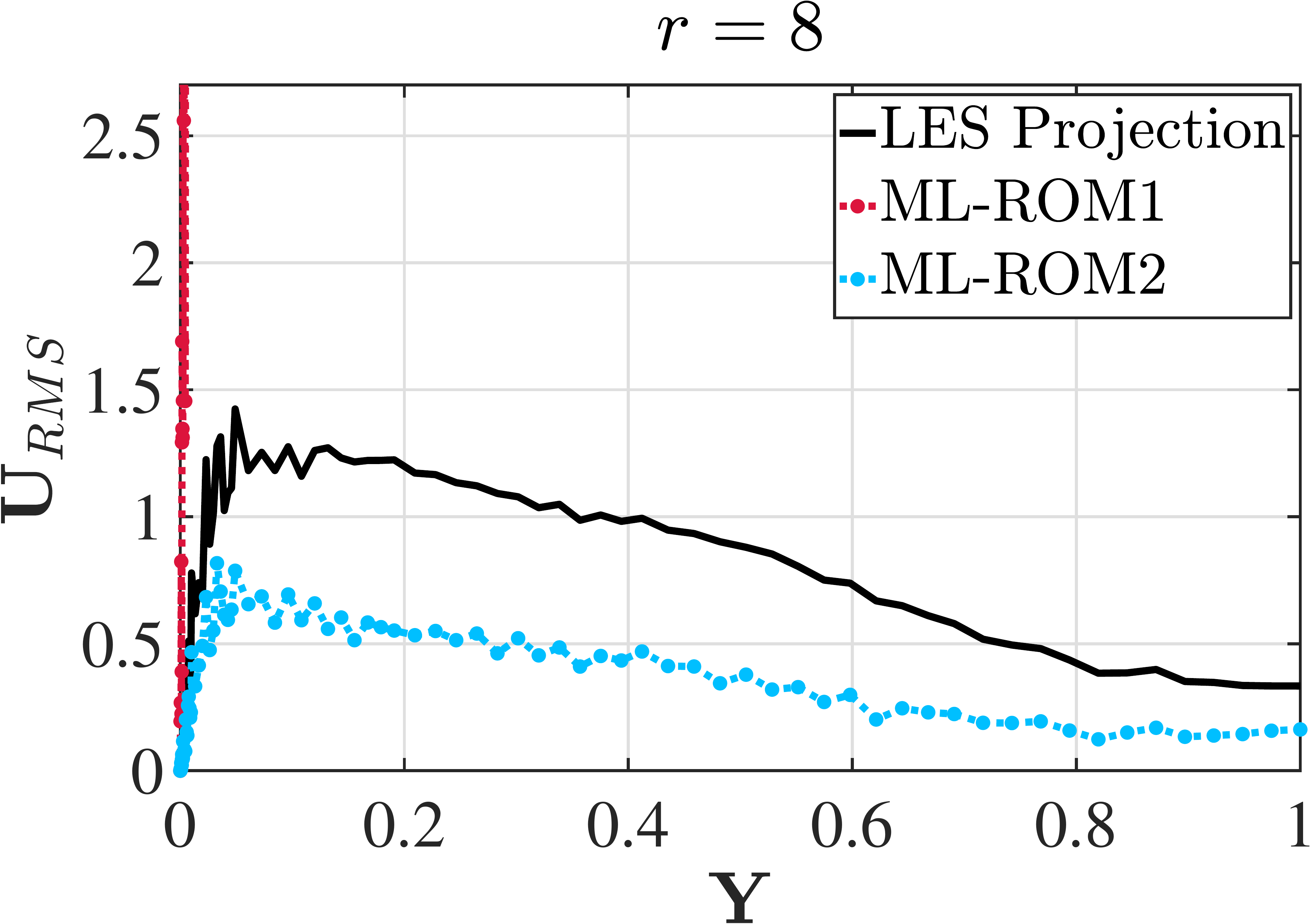}         \caption{$r=8$}
         \label{fig:stat-r-8}
     \end{subfigure}
     \begin{subfigure}[b]{0.48\textwidth}
         \centering
    \includegraphics[width=.45\textwidth]{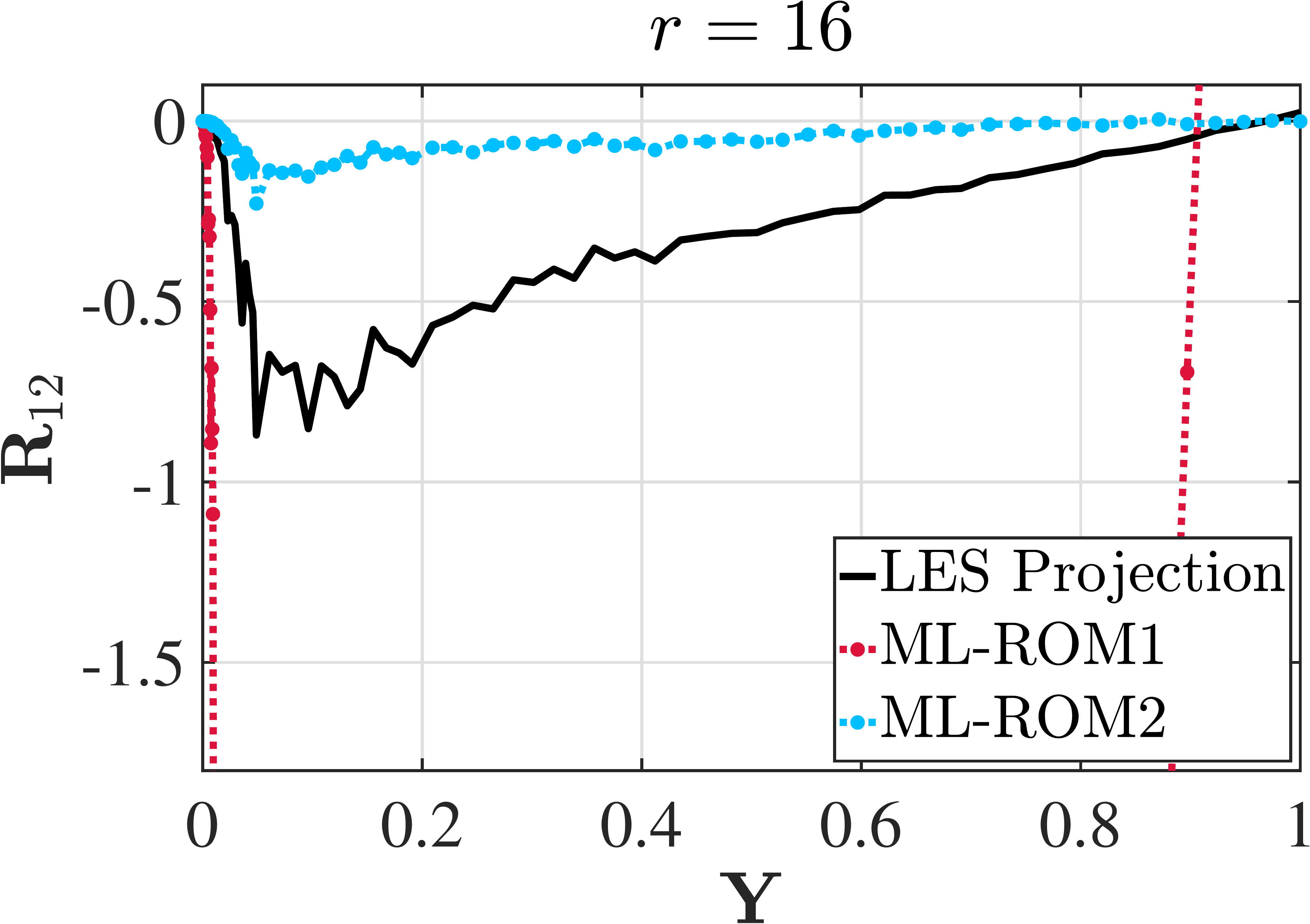}
    \includegraphics[width=.45\textwidth]{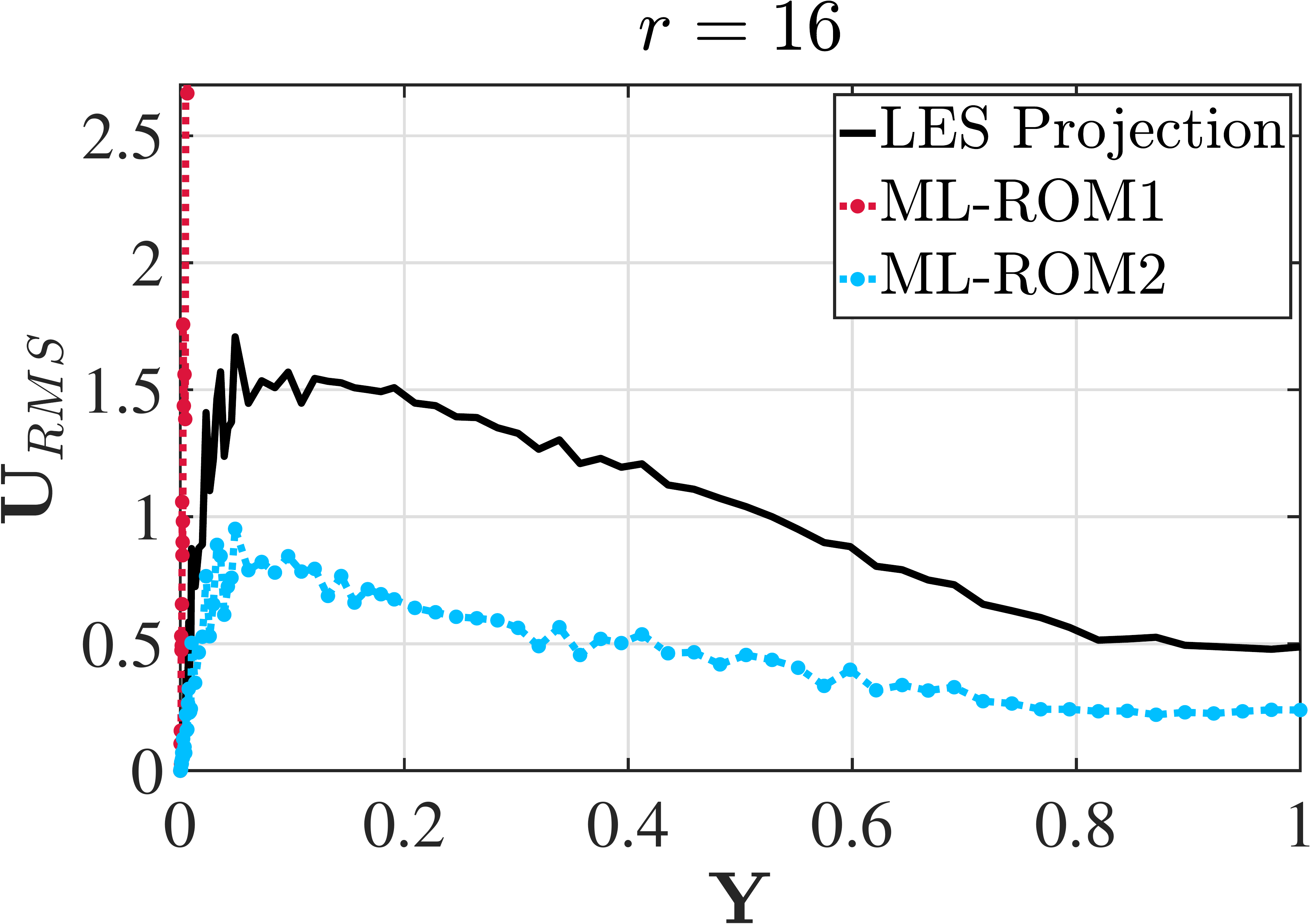}         \caption{$r=16$}
         \label{fig:stat-r-16}
     \end{subfigure}
     \begin{subfigure}[b]{0.48\textwidth}
         \centering
    \includegraphics[width=.45\textwidth]{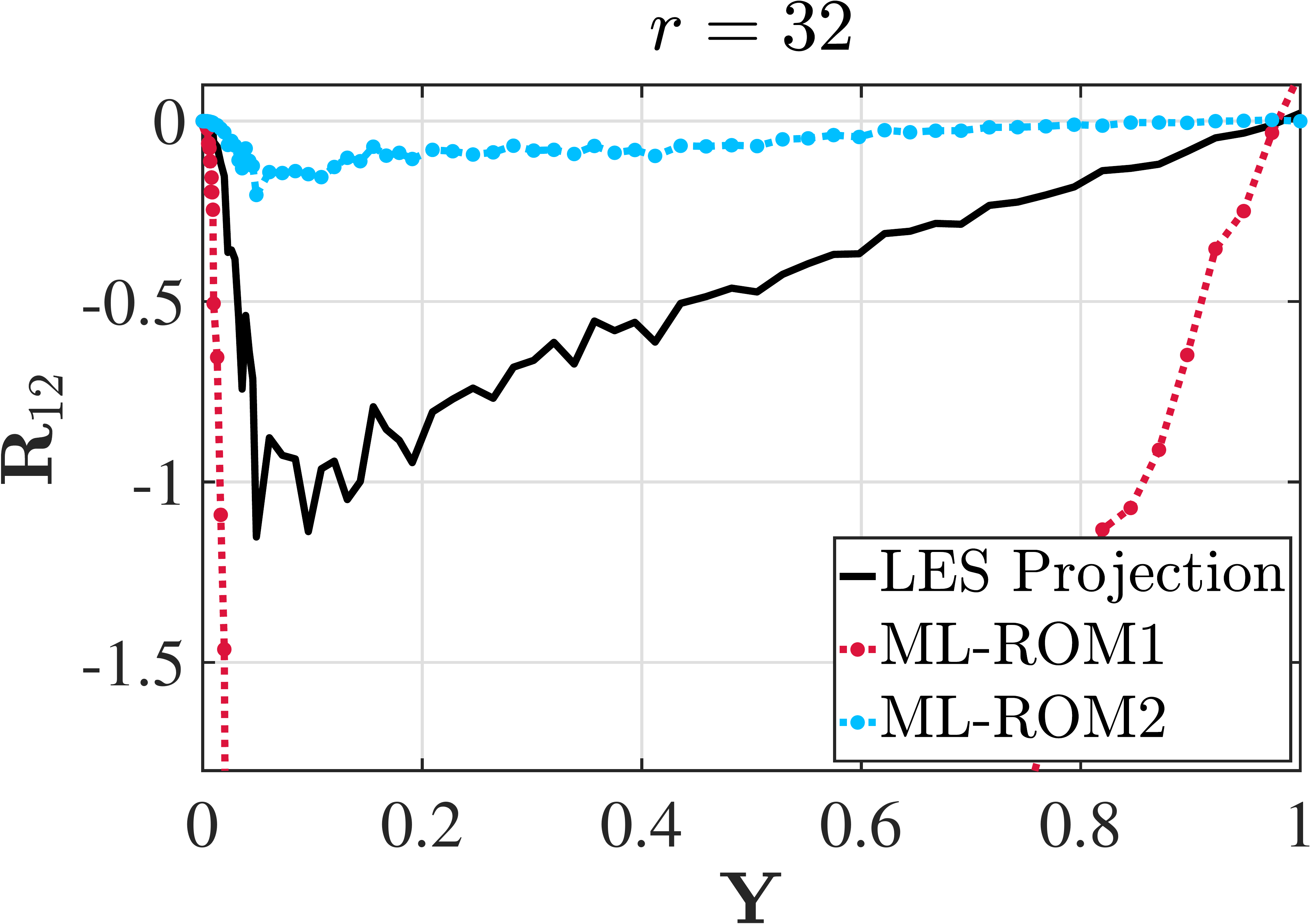}
    \includegraphics[width=.45\textwidth]{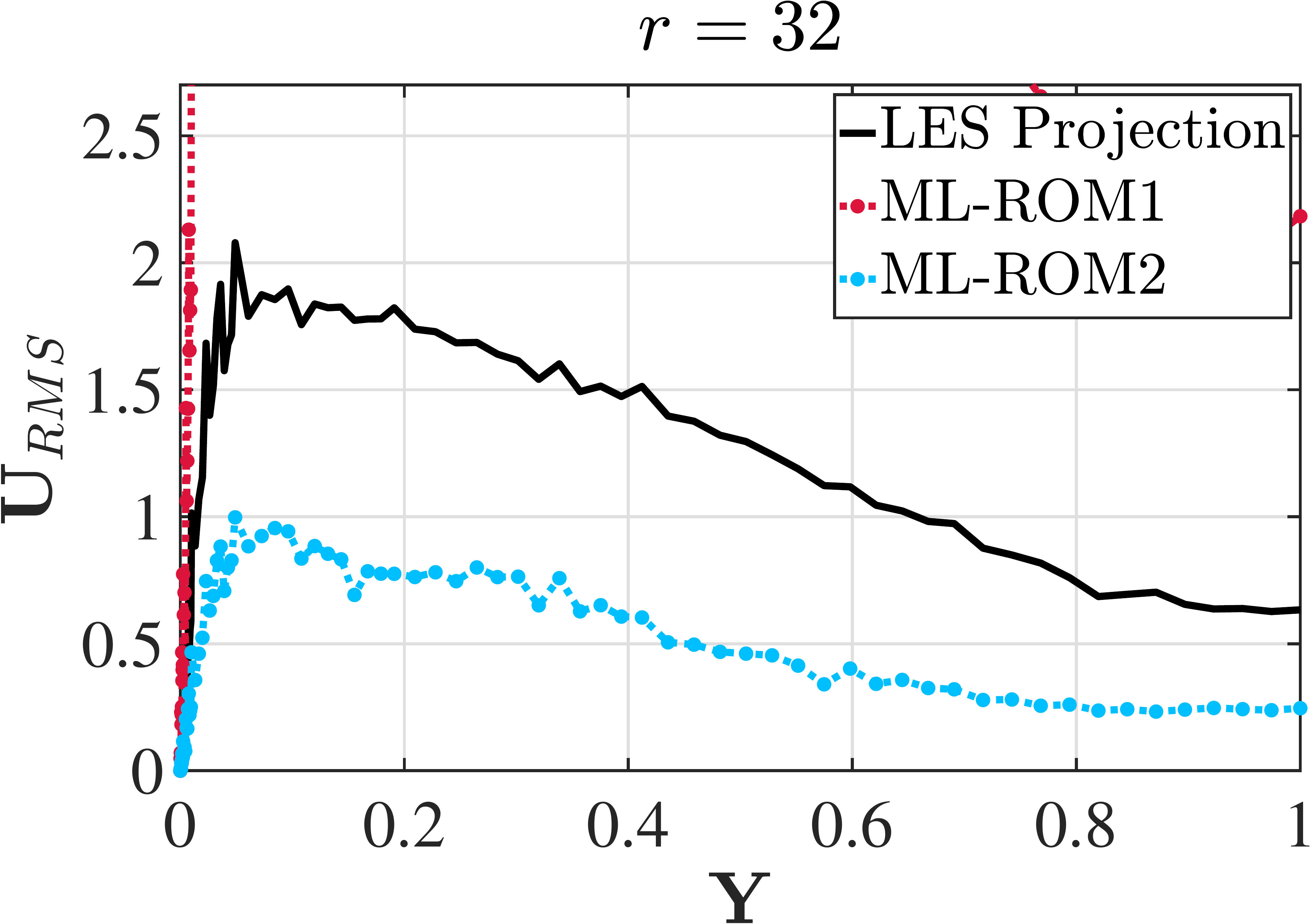}         \caption{$r=32$}
         \label{fig:stat-r-32}
    \end{subfigure}     
     \begin{subfigure}[b]{0.48\textwidth}
         \centering
    \includegraphics[width=.45\textwidth]{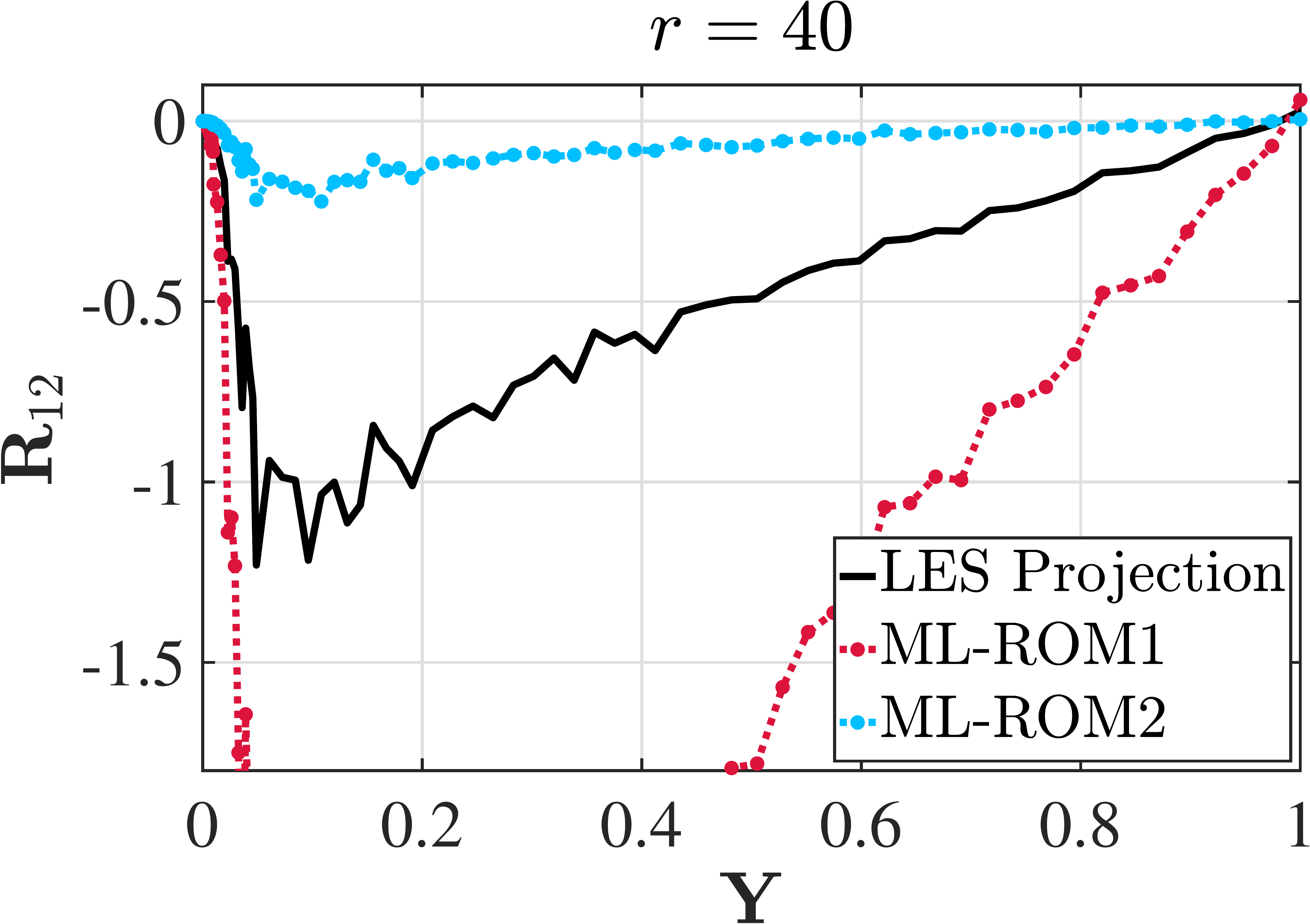}
    \includegraphics[width=.45\textwidth]{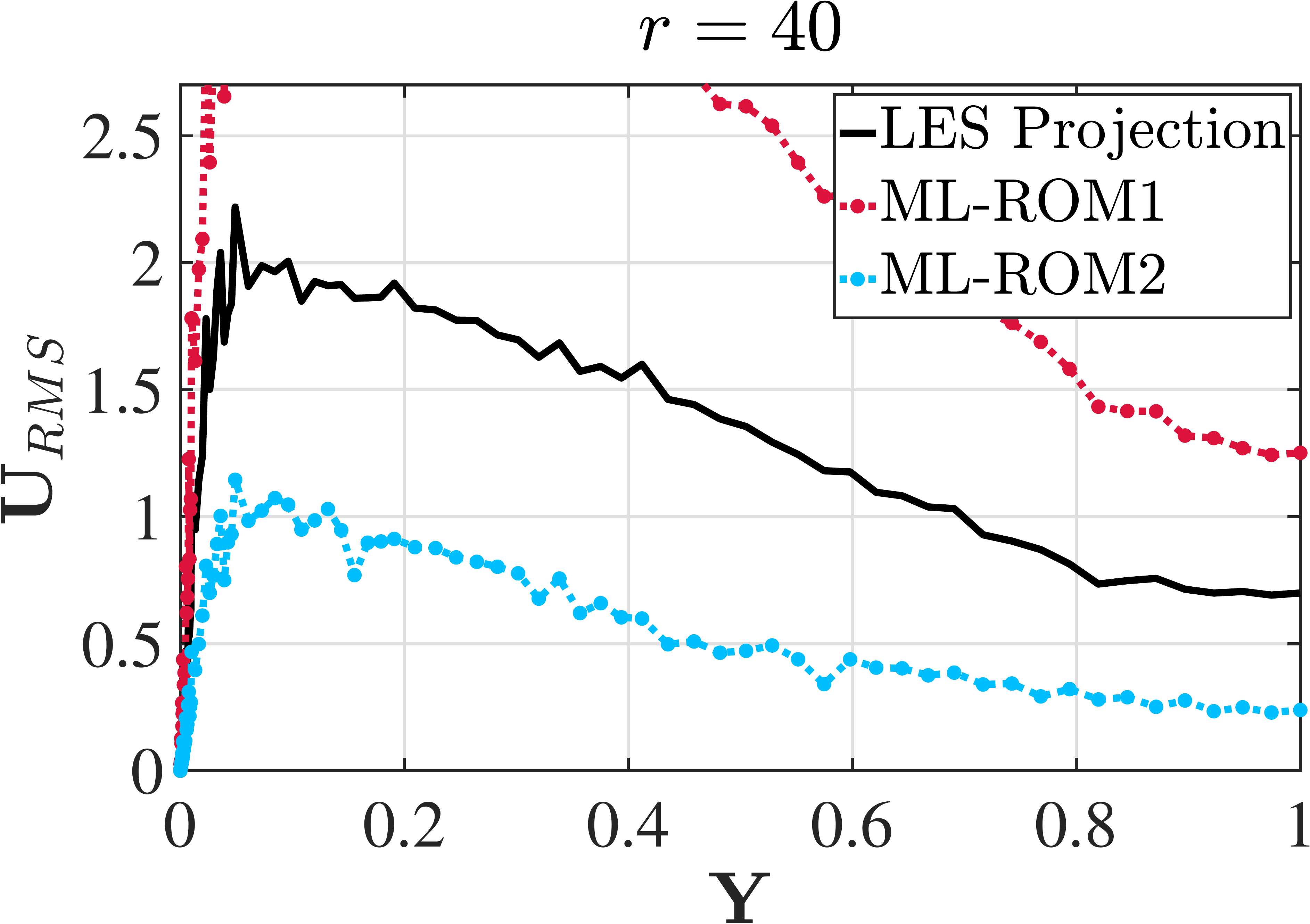}         \caption{$r=40$}
         \label{fig:stat-r-40}
     \end{subfigure} 
     \begin{subfigure}[b]{0.48\textwidth}
         \centering
    \includegraphics[width=.45\textwidth]{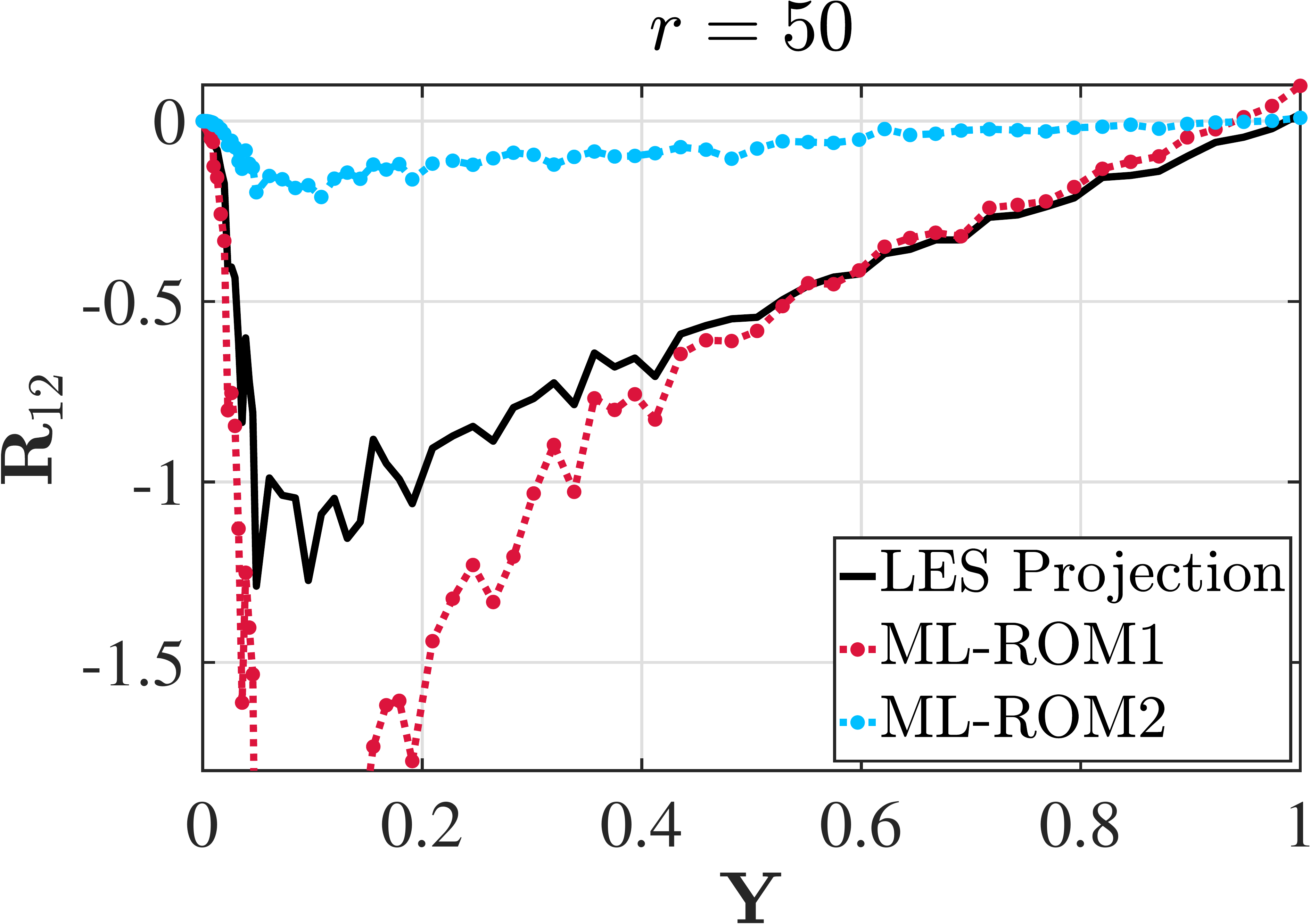}
    \includegraphics[width=.45\textwidth]{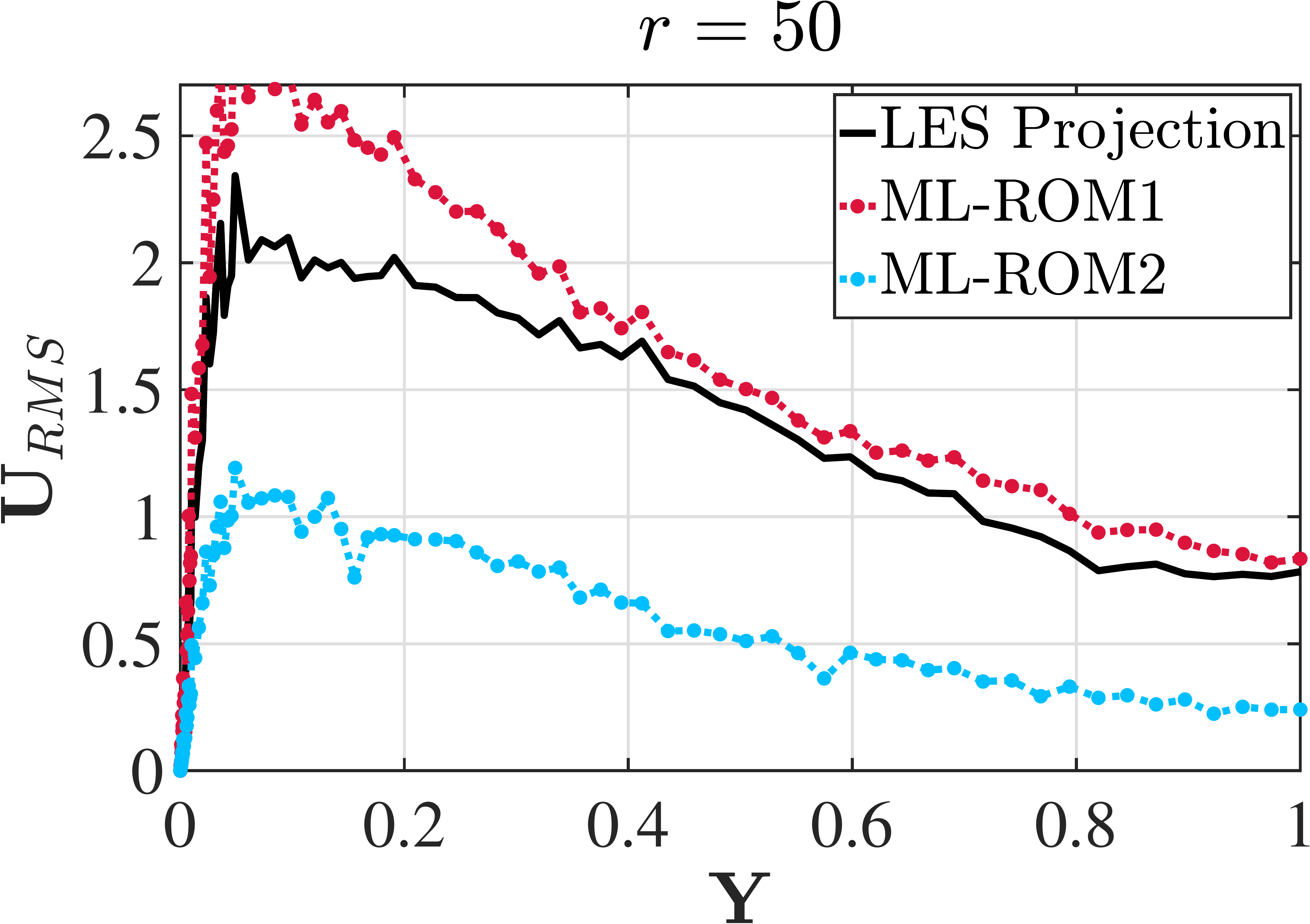}         \caption{$r=50$}
         \label{fig:stat-r-50}
     \end{subfigure} 
     \caption{Second-order statistics for $\alpha=6\times 10^{-3}$
     }   
    \label{fig:stat-alpha-1}
\end{figure}

\begin{figure}[H]
\centering
    \includegraphics[width=.45\textwidth]{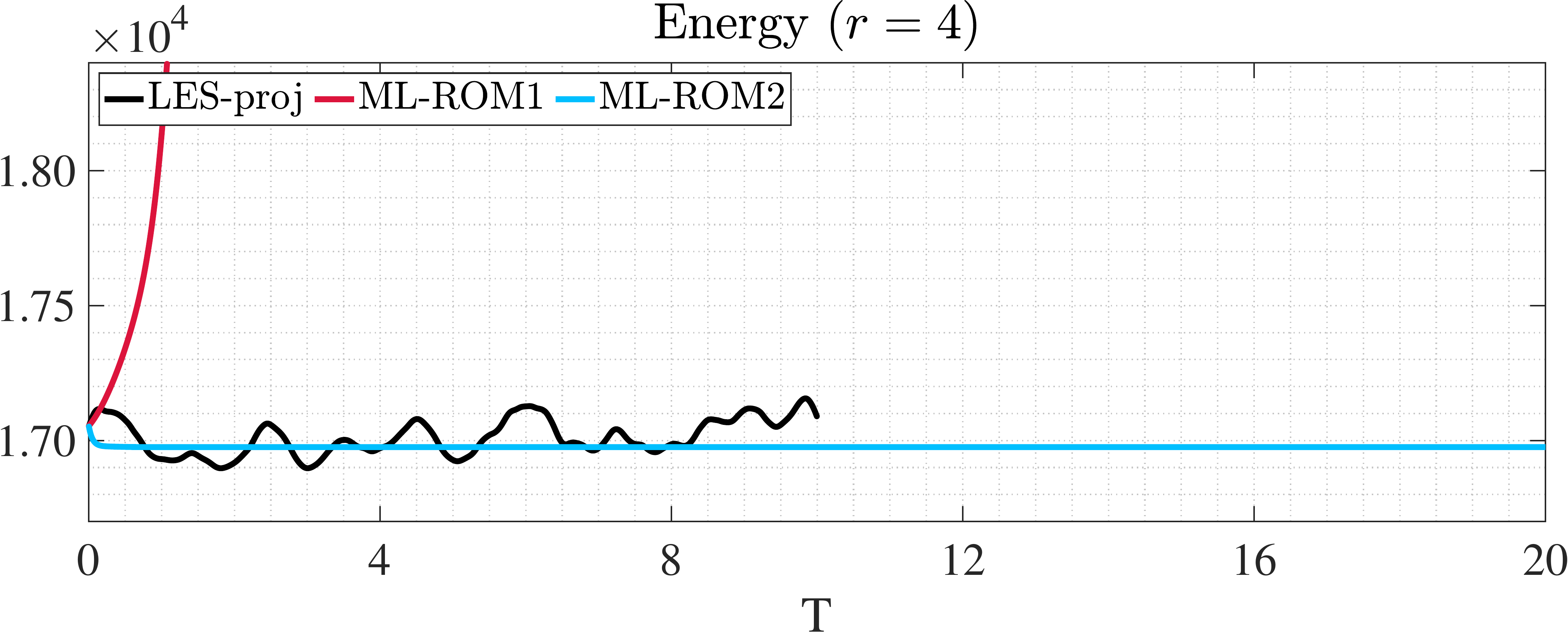}
    \includegraphics[width=.45\textwidth]{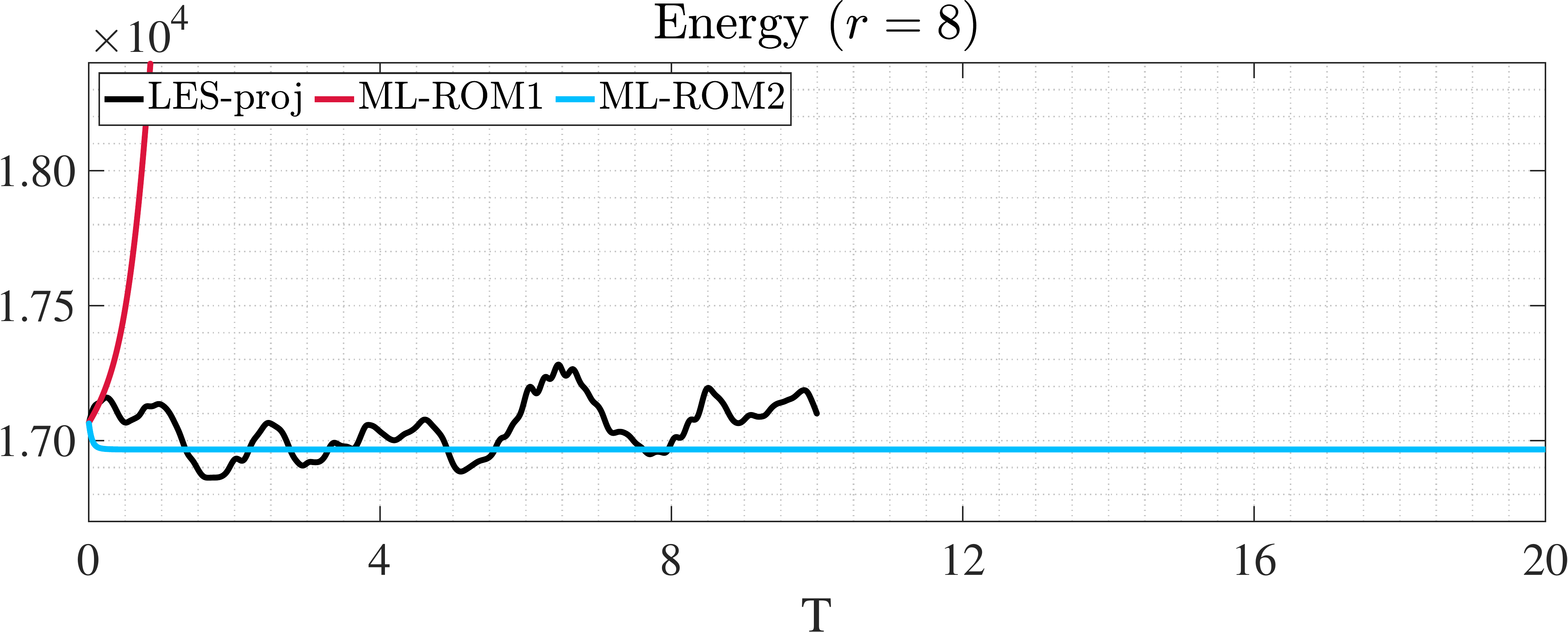}
    \includegraphics[width=.45\textwidth]{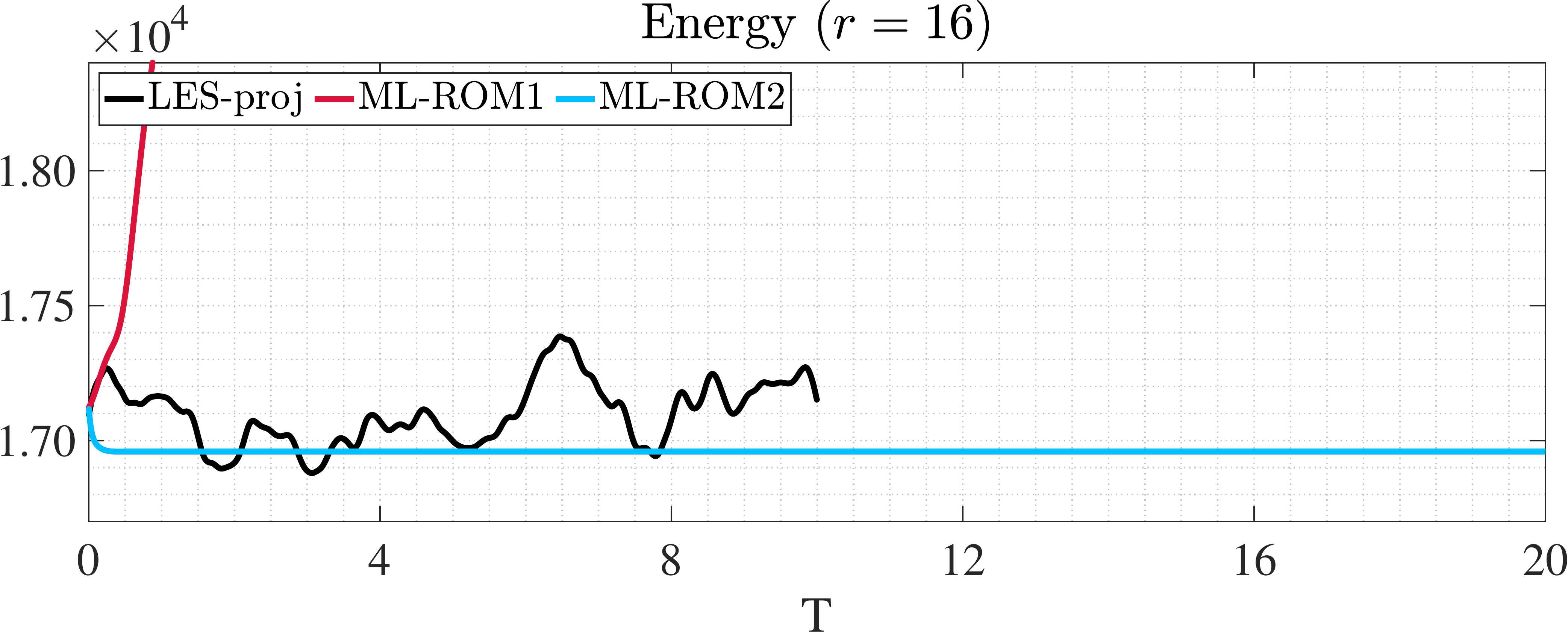}
    \includegraphics[width=.45\textwidth]{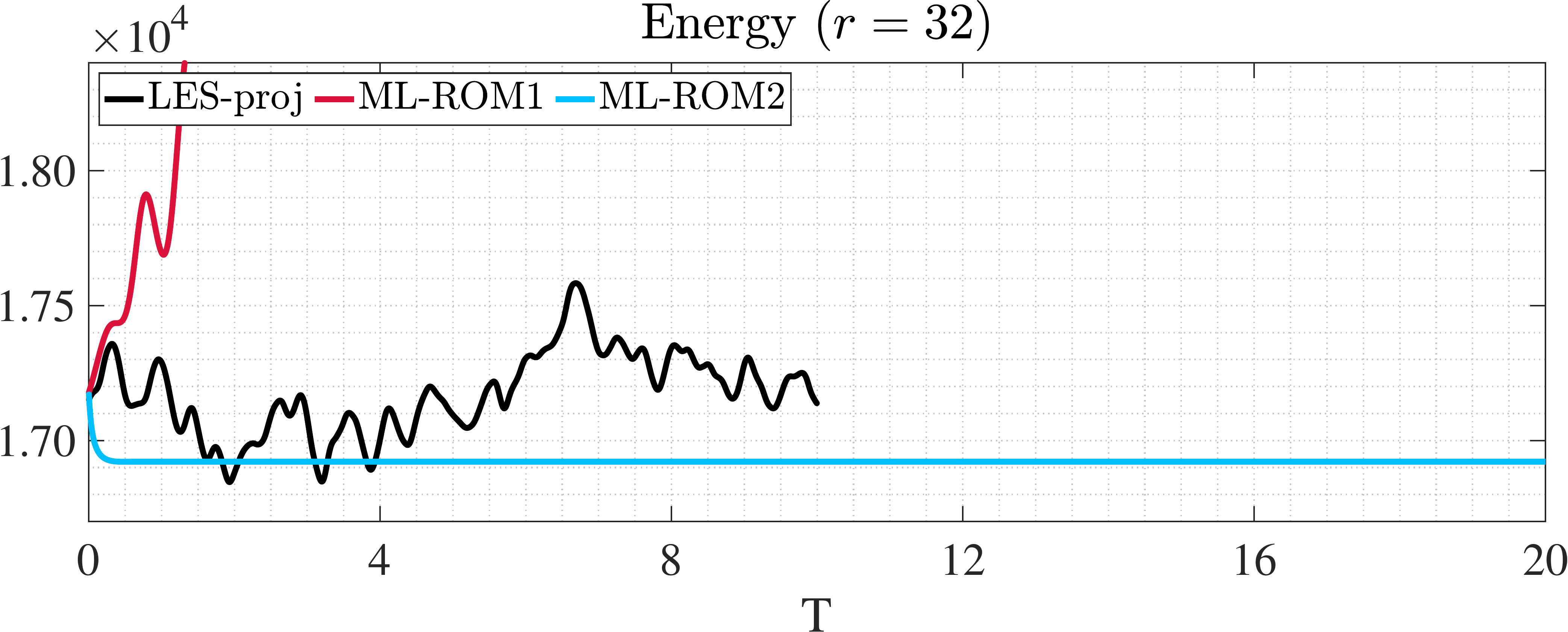}
    \includegraphics[width=.45\textwidth]{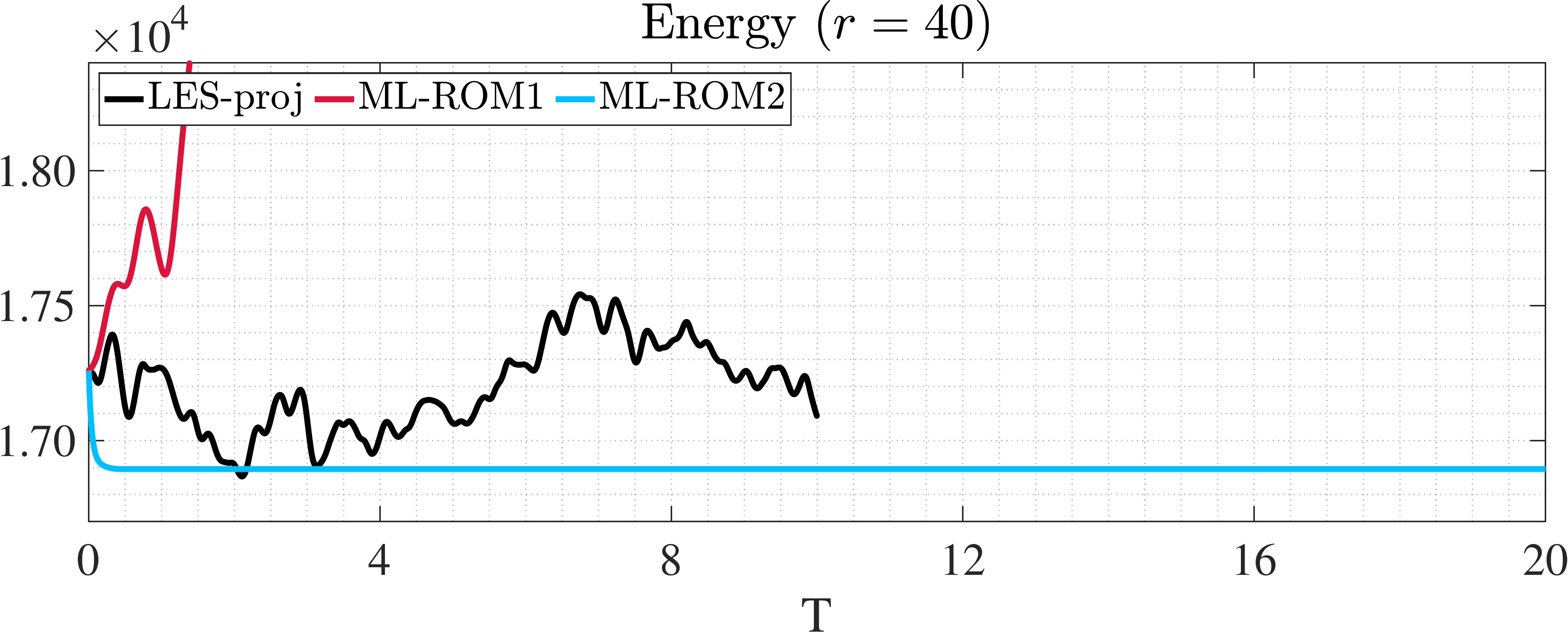}
    \includegraphics[width=.45\textwidth]{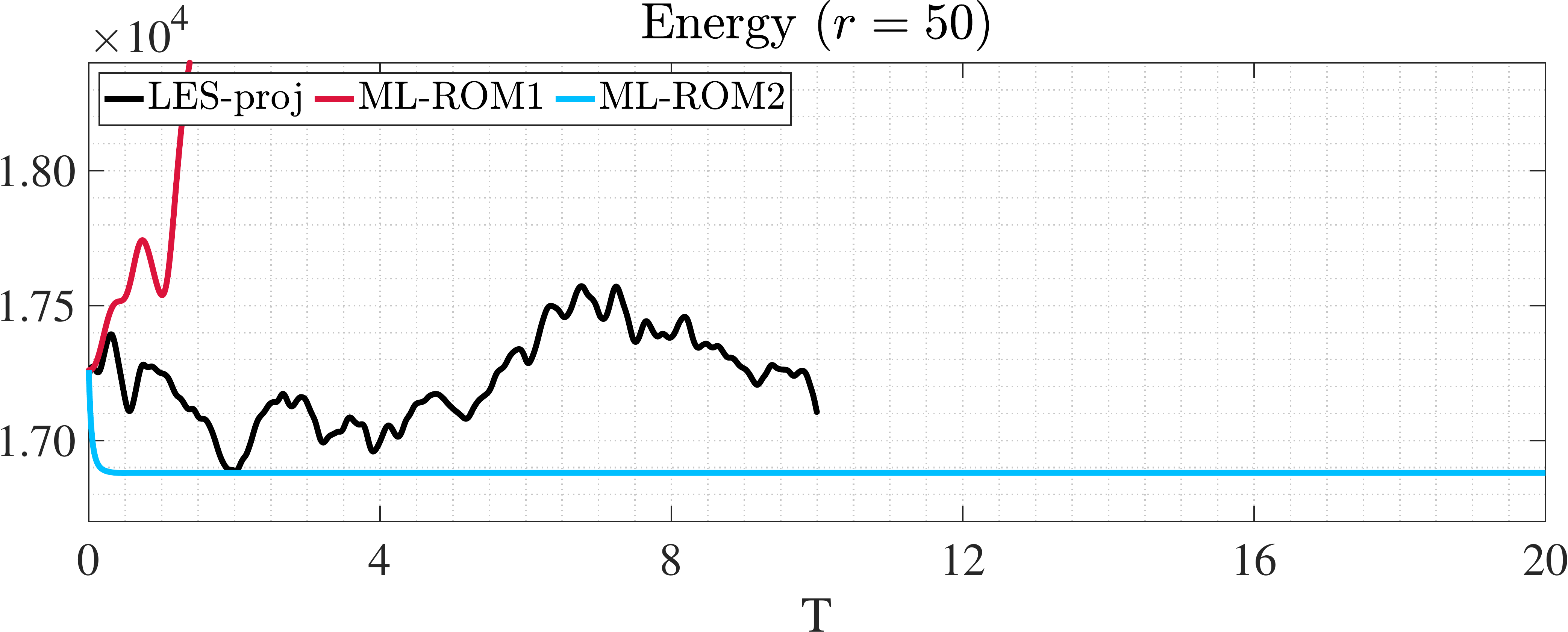}   
    \caption{Time evolution of the kinetic energy for $\alpha=2\times 10^{-3}$
    }    
    \label{fig:ke-alpha-2}
\end{figure}

\begin{figure}[H]
\centering
     \begin{subfigure}[b]{0.48\textwidth}
         \centering
    \includegraphics[width=.45\textwidth]{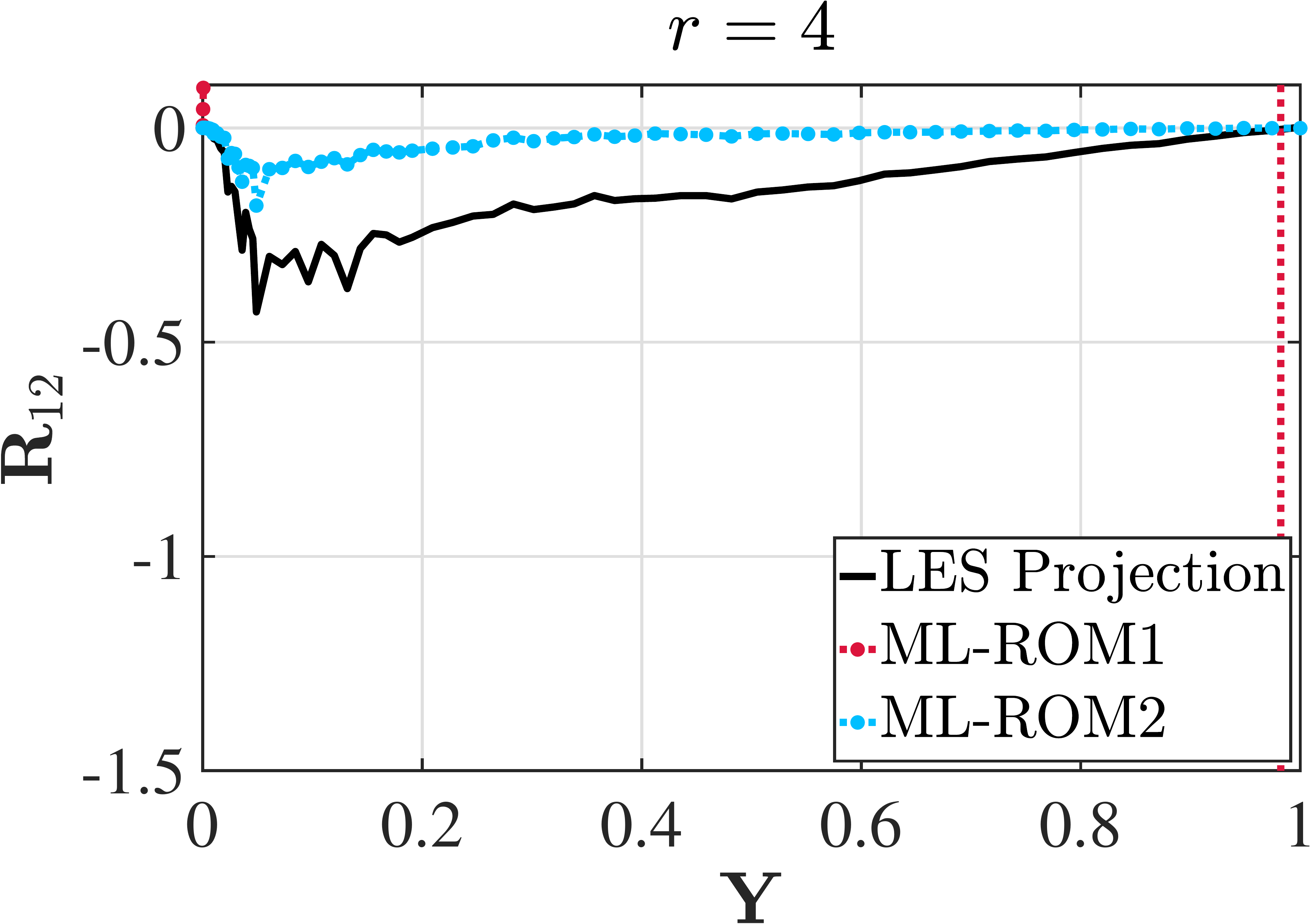}
    \includegraphics[width=.45\textwidth]{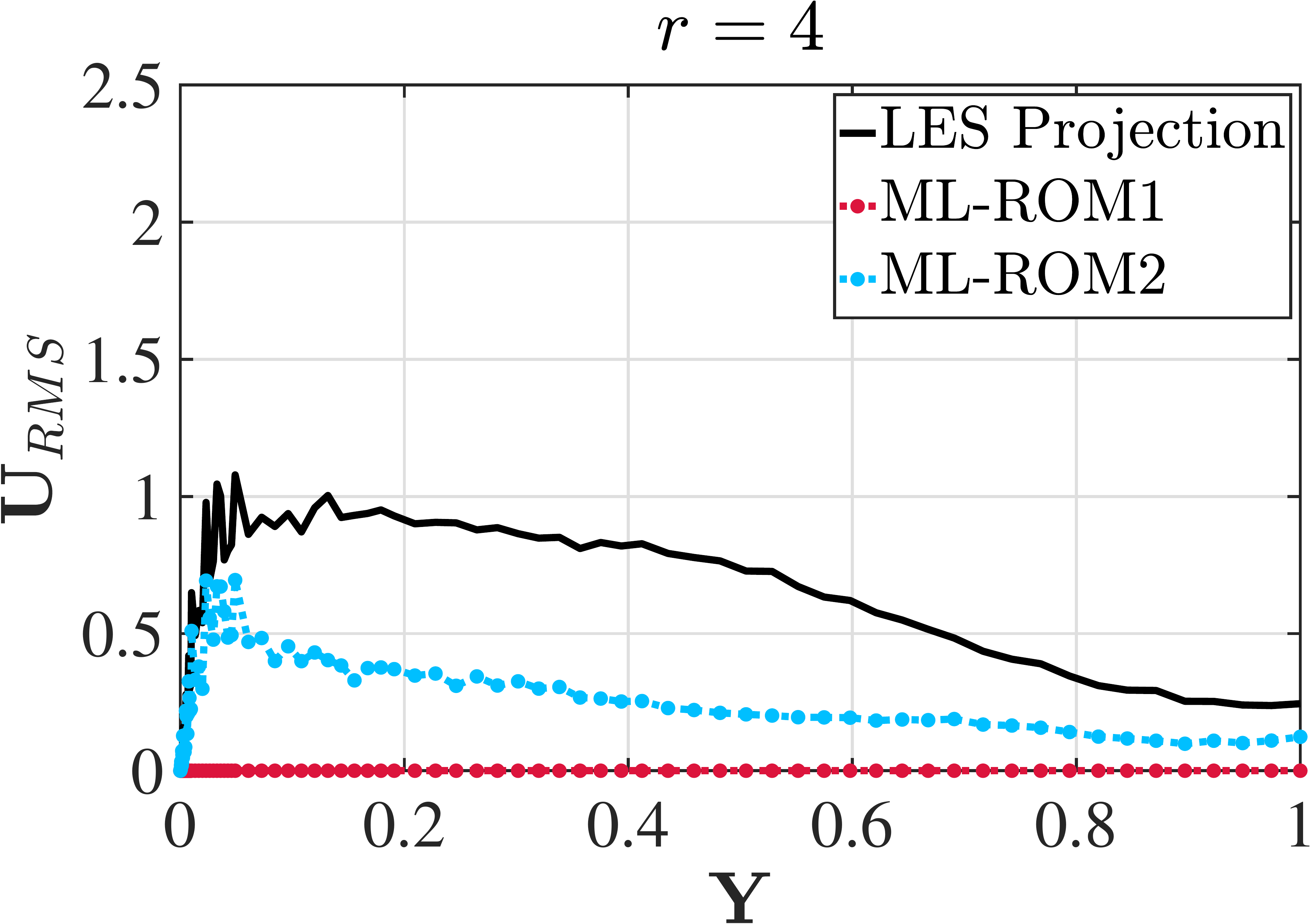}         \caption{$r=4$}
         \label{fig:stat-r-4}
     \end{subfigure}
     \begin{subfigure}[b]{0.48\textwidth}
         \centering
    \includegraphics[width=.45\textwidth]{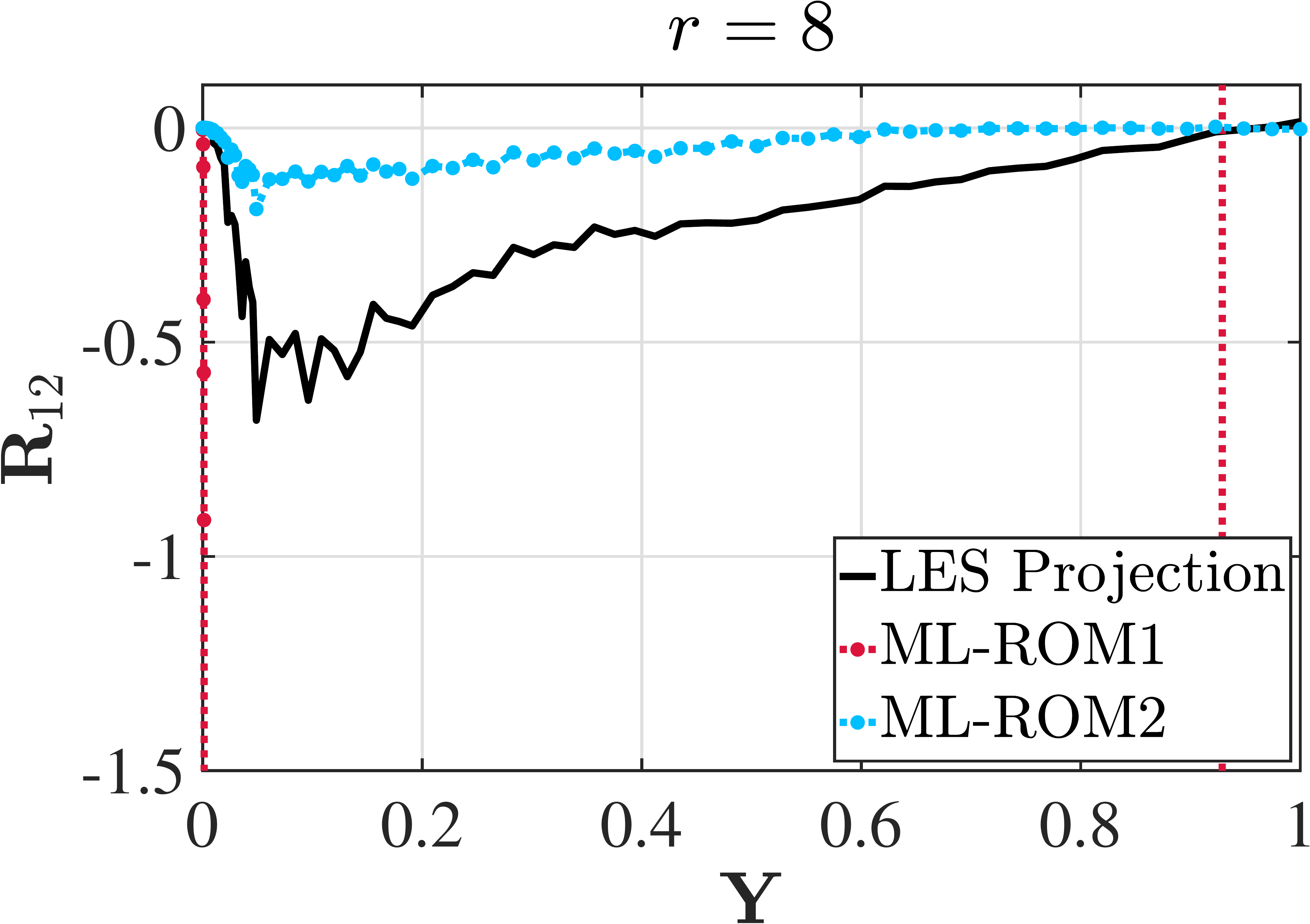}
    \includegraphics[width=.45\textwidth]{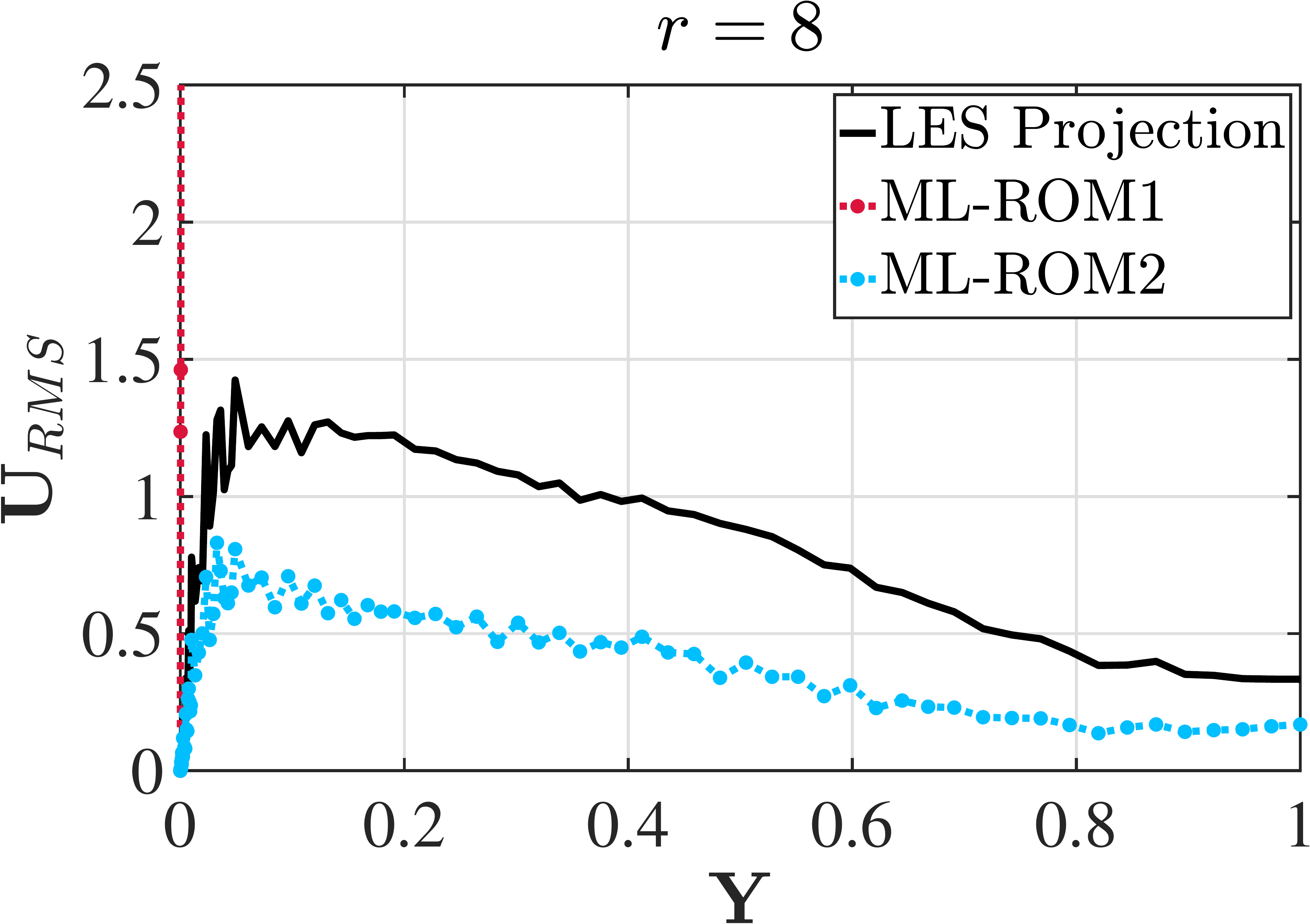}         \caption{$r=8$}
         \label{fig:stat-r-8}
     \end{subfigure}
     \begin{subfigure}[b]{0.48\textwidth}
         \centering
    \includegraphics[width=.45\textwidth]{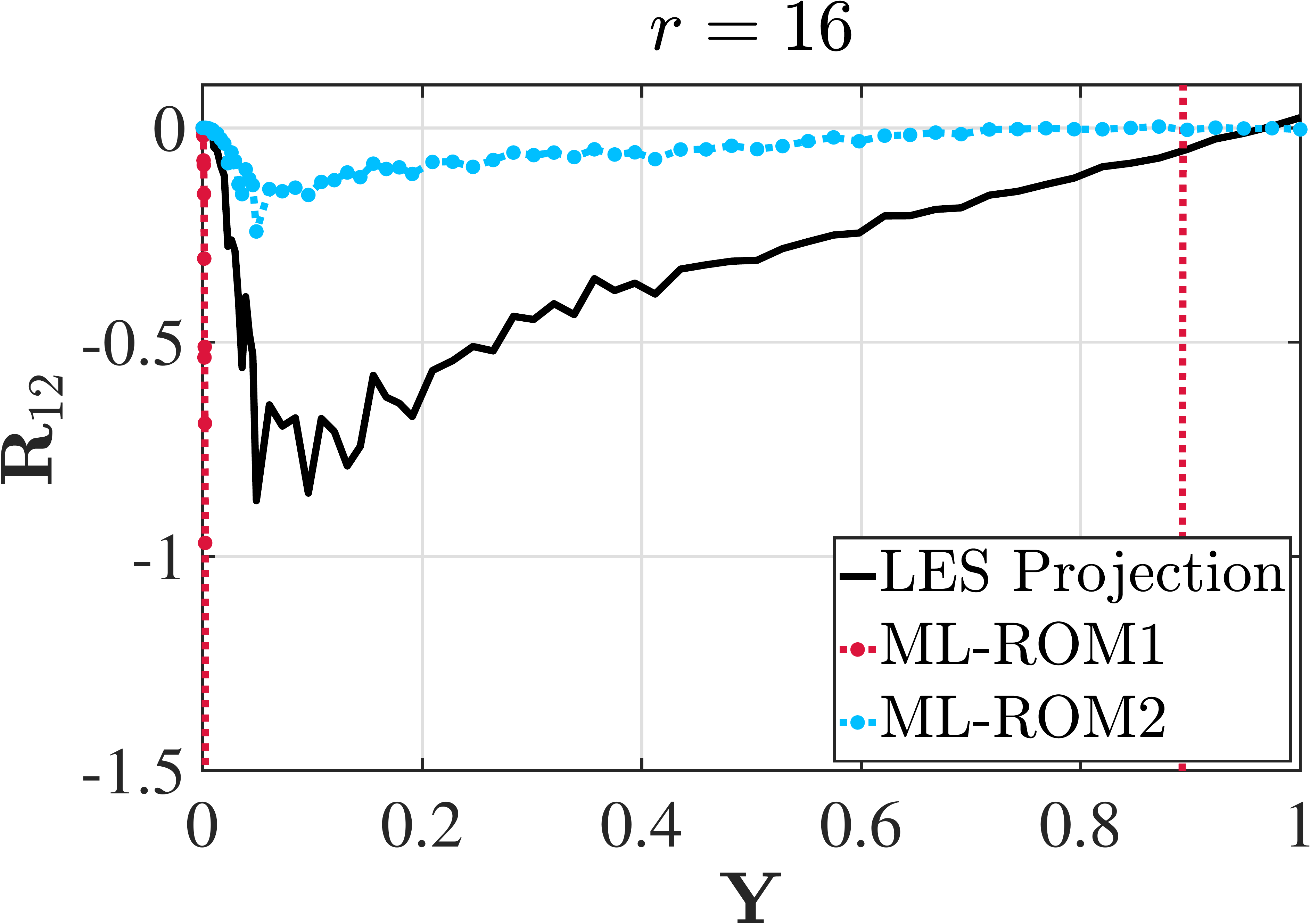}
    \includegraphics[width=.45\textwidth]{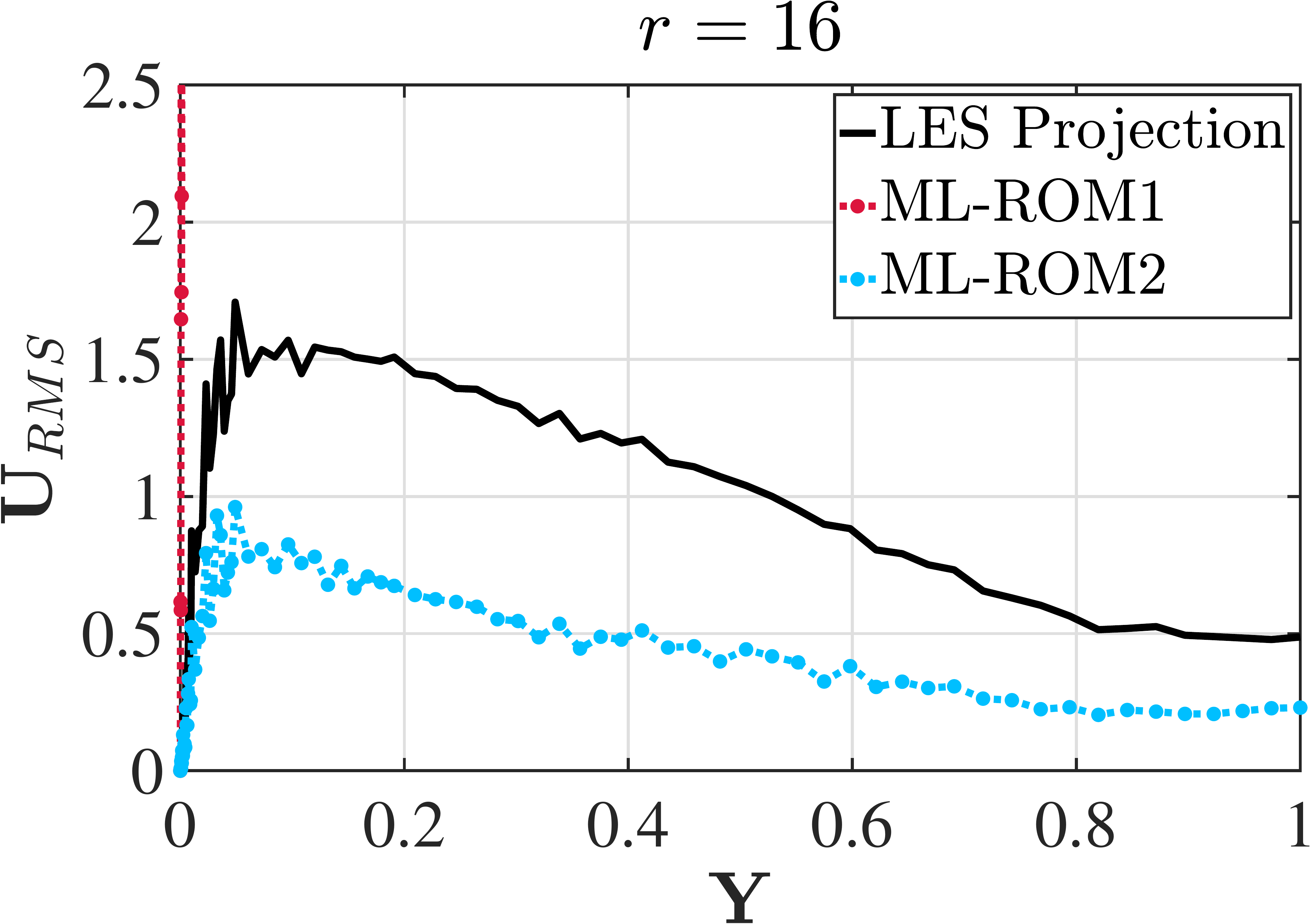}         \caption{$r=16$}
         \label{fig:stat-r-16}
     \end{subfigure}
     \begin{subfigure}[b]{0.48\textwidth}
         \centering
    \includegraphics[width=.45\textwidth]{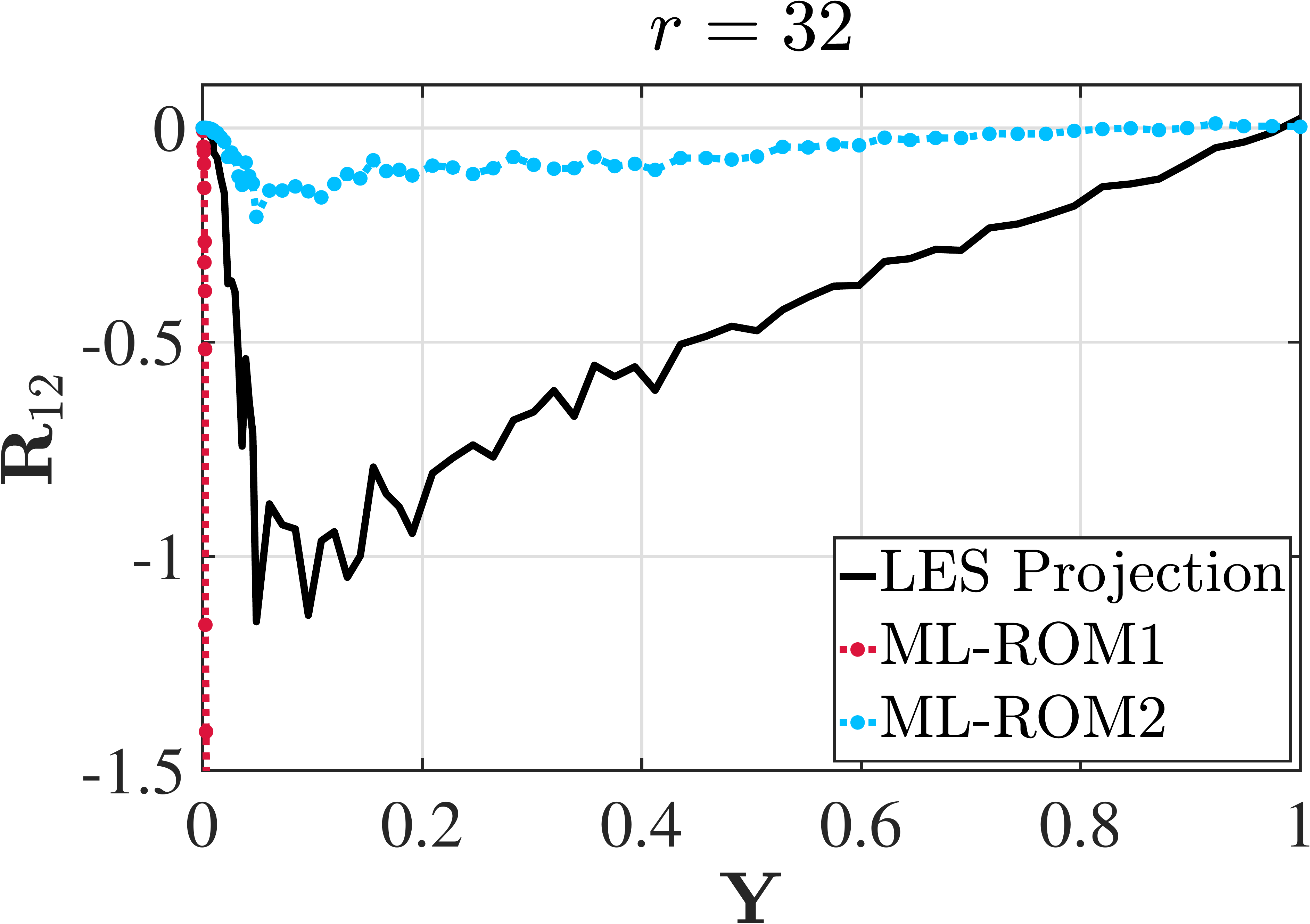}
    \includegraphics[width=.45\textwidth]{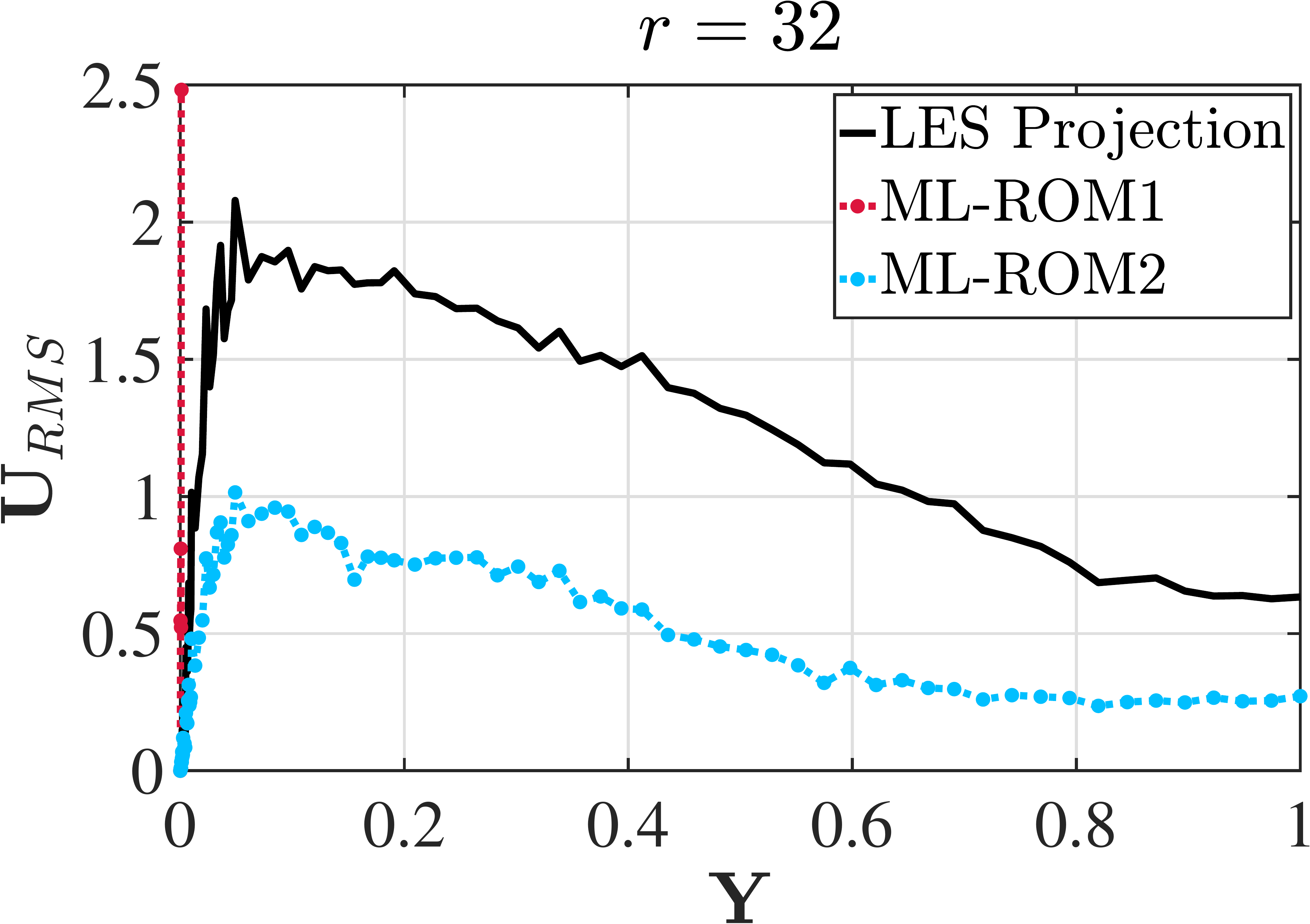}         \caption{$r=32$}
         \label{fig:stat-r-32}
    \end{subfigure}     
     \begin{subfigure}[b]{0.48\textwidth}
         \centering
    \includegraphics[width=.45\textwidth]{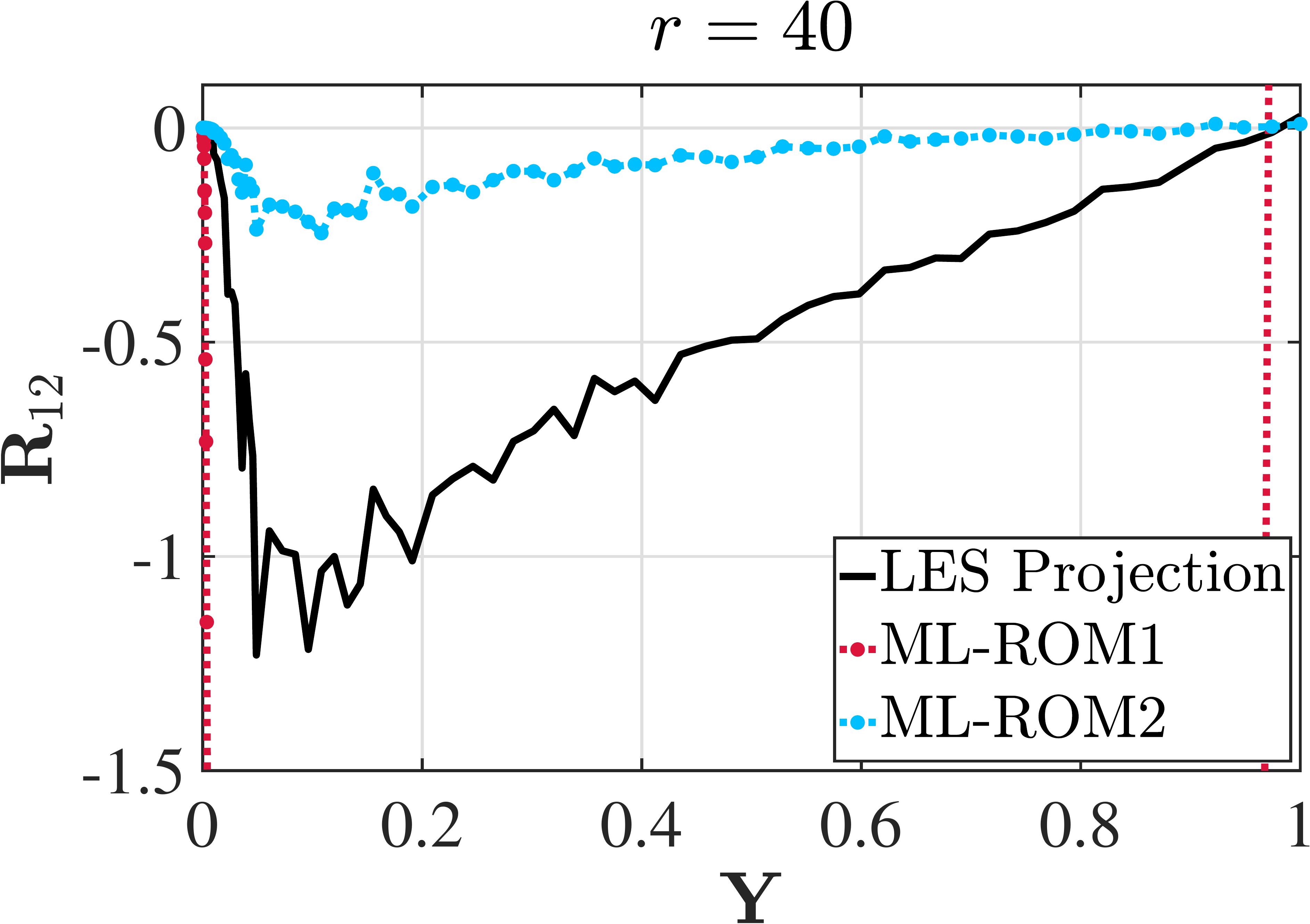}
    \includegraphics[width=.45\textwidth]{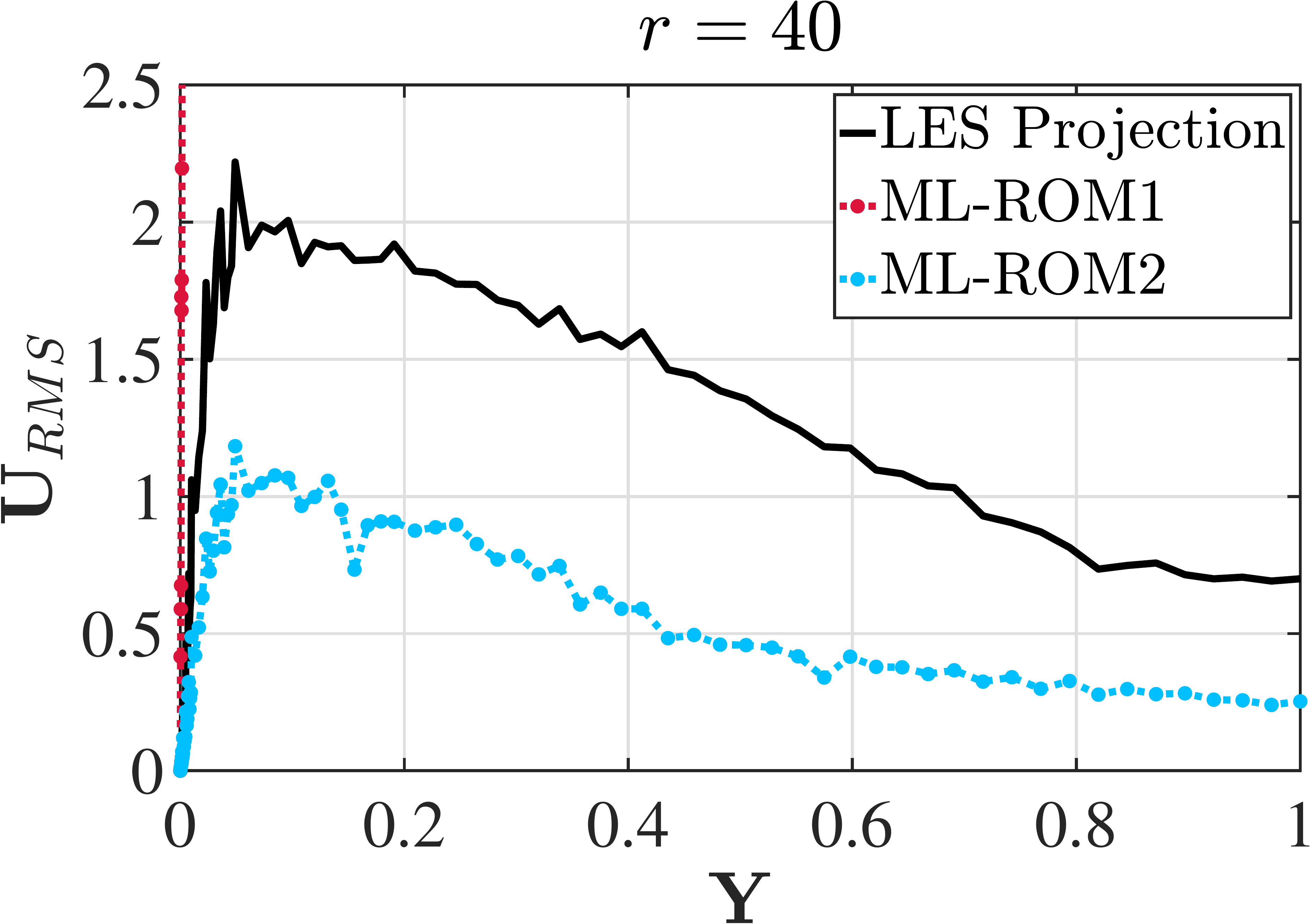}         \caption{$r=40$}
         \label{fig:stat-r-40}
     \end{subfigure} 
     \begin{subfigure}[b]{0.48\textwidth}
         \centering
    \includegraphics[width=.45\textwidth]{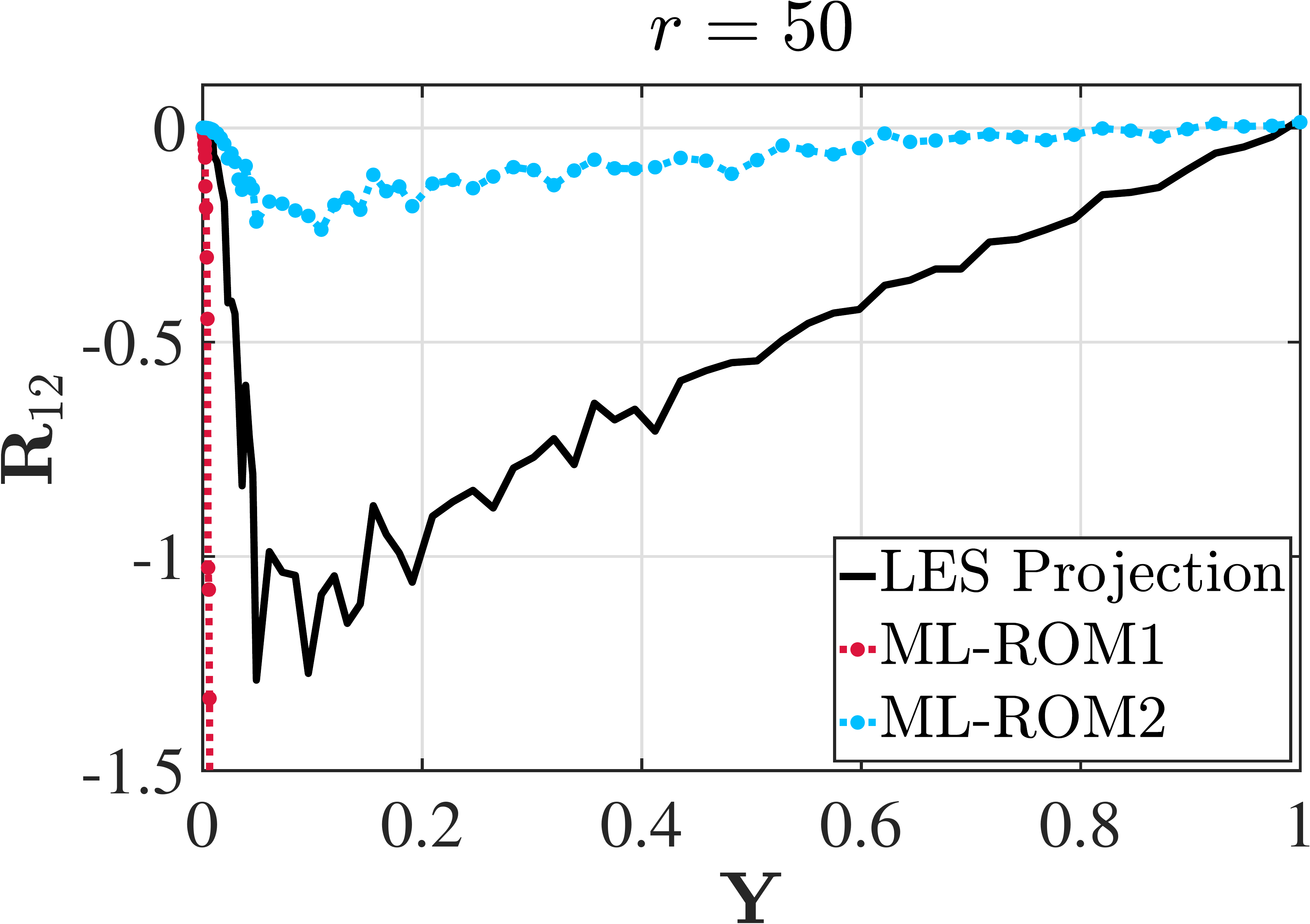}
    \includegraphics[width=.45\textwidth]{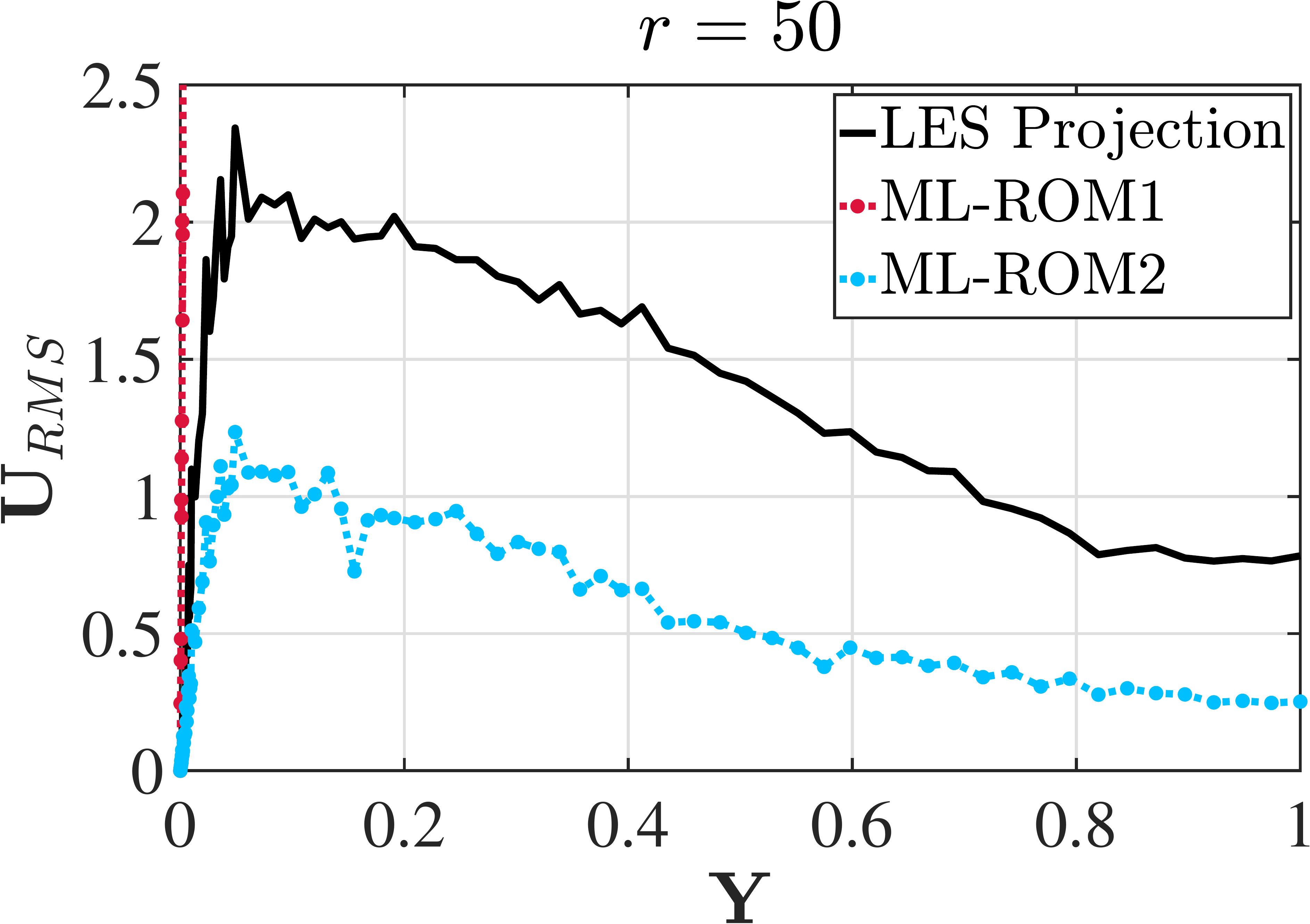}         \caption{$r=50$}
         \label{fig:stat-r-50}
     \end{subfigure} 
     \caption{Second-order statistics for $\alpha=2\times 10^{-3}$
    }    
    \label{fig:stat-alpha-2}
\end{figure}

\begin{figure}[H]
\centering
    \includegraphics[width=.45\textwidth]{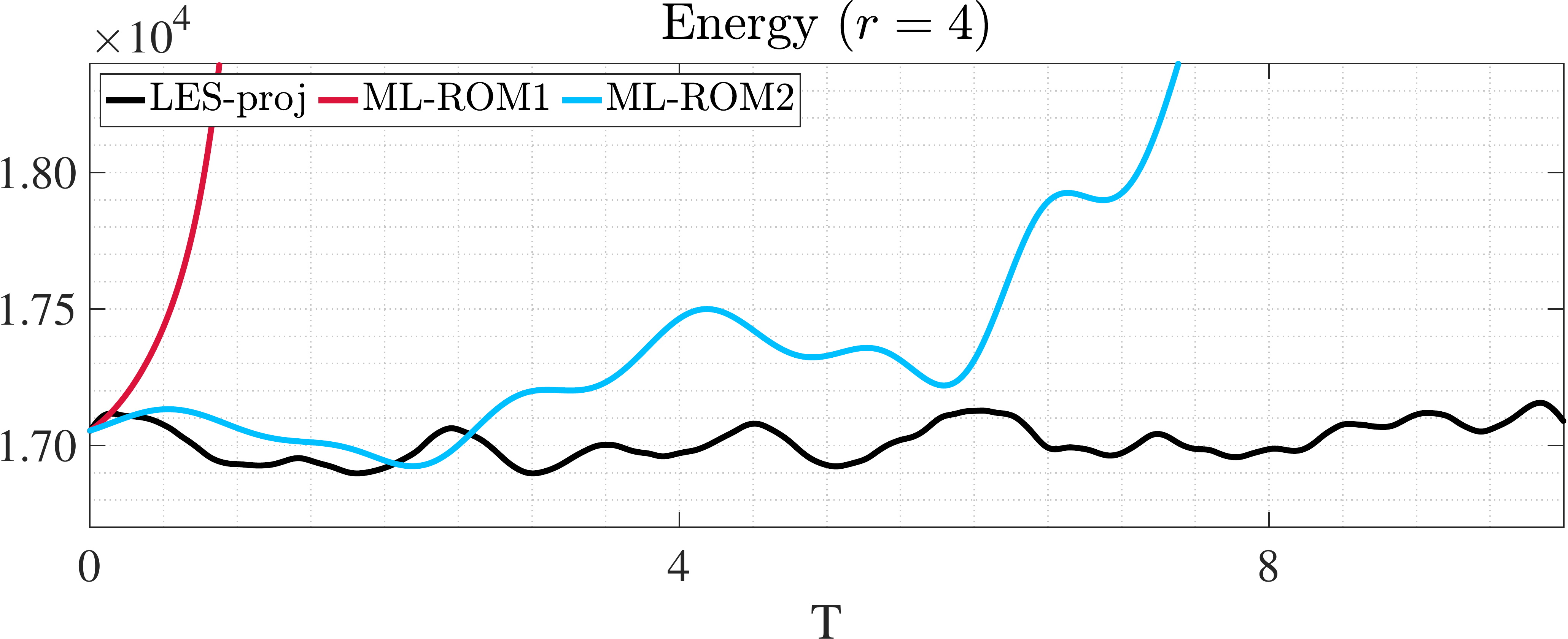}
    \includegraphics[width=.45\textwidth]{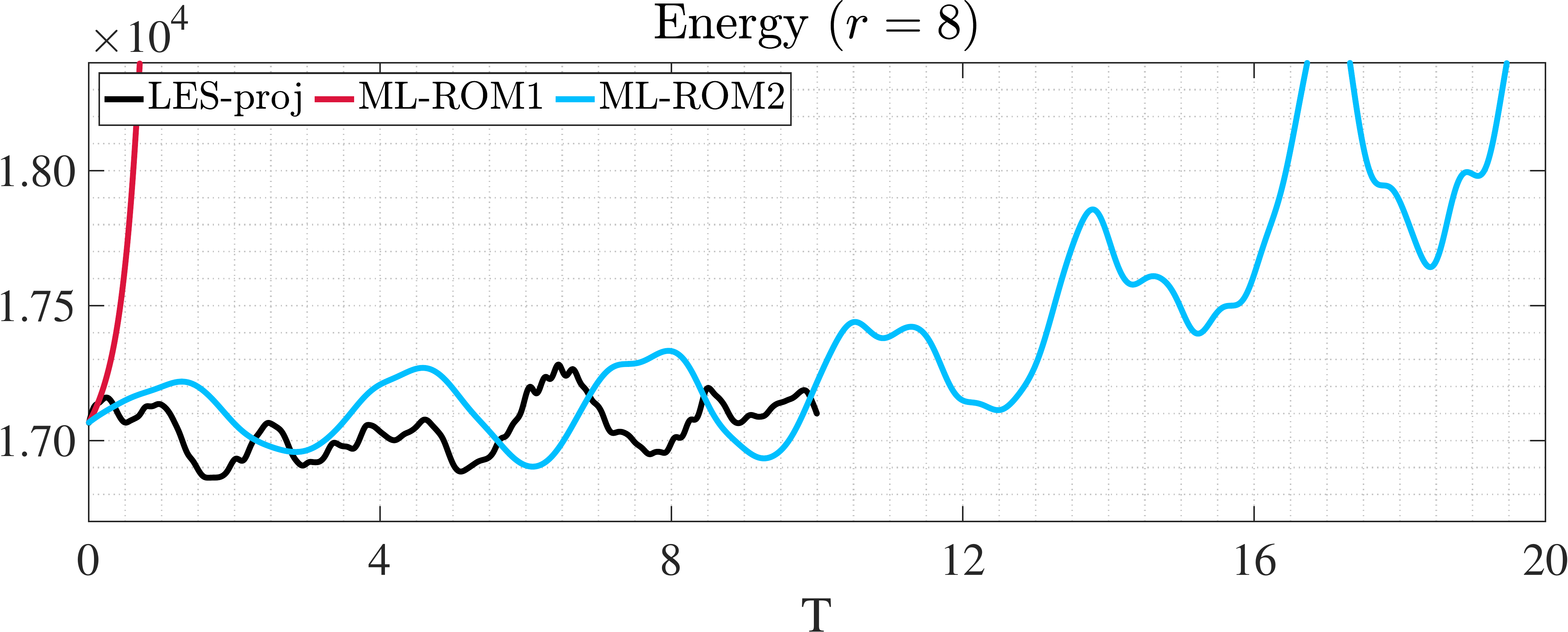}
    \includegraphics[width=.45\textwidth]{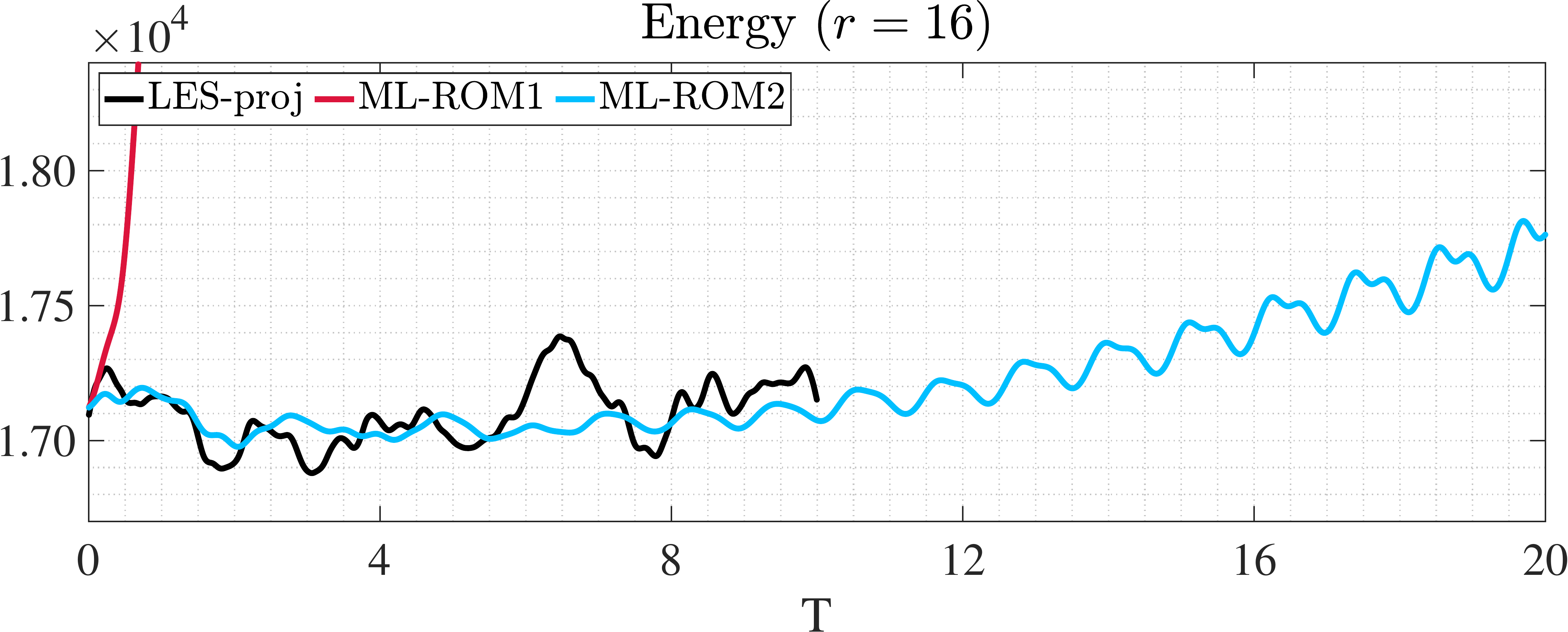}
    \includegraphics[width=.45\textwidth]{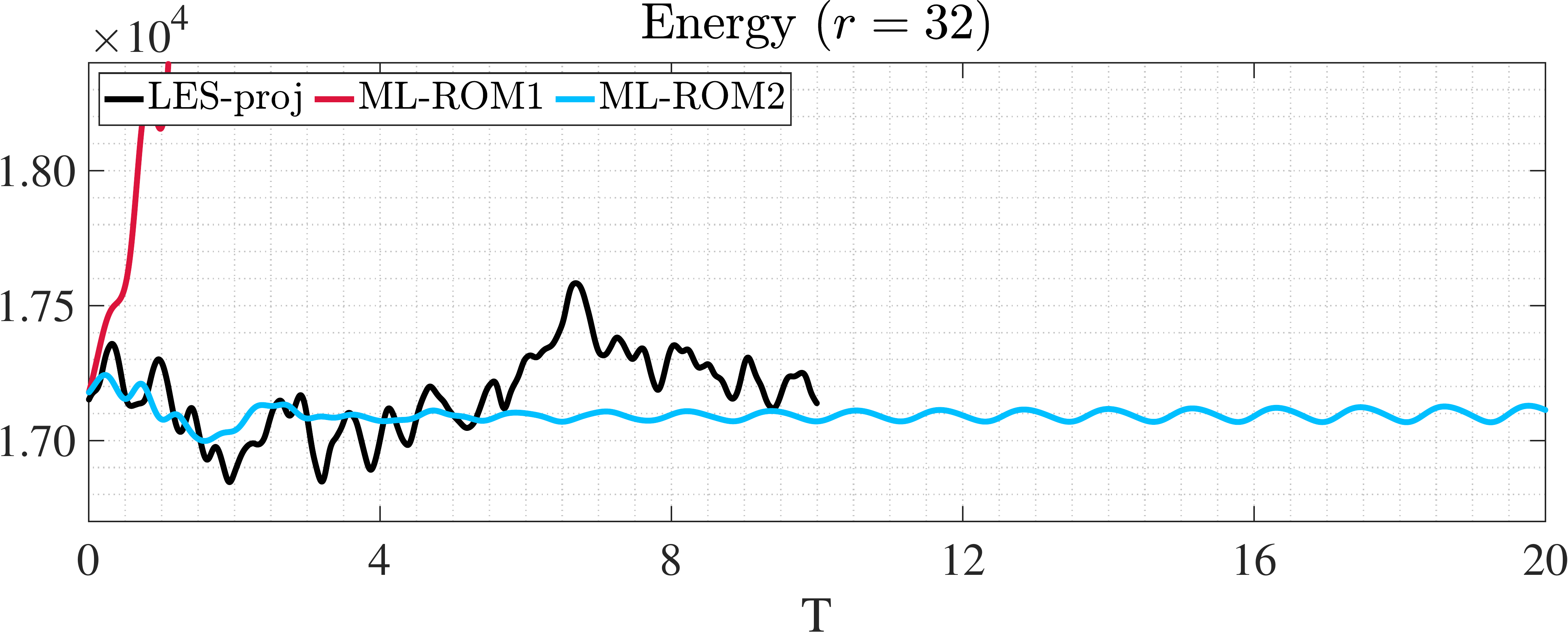}
    \includegraphics[width=.45\textwidth]{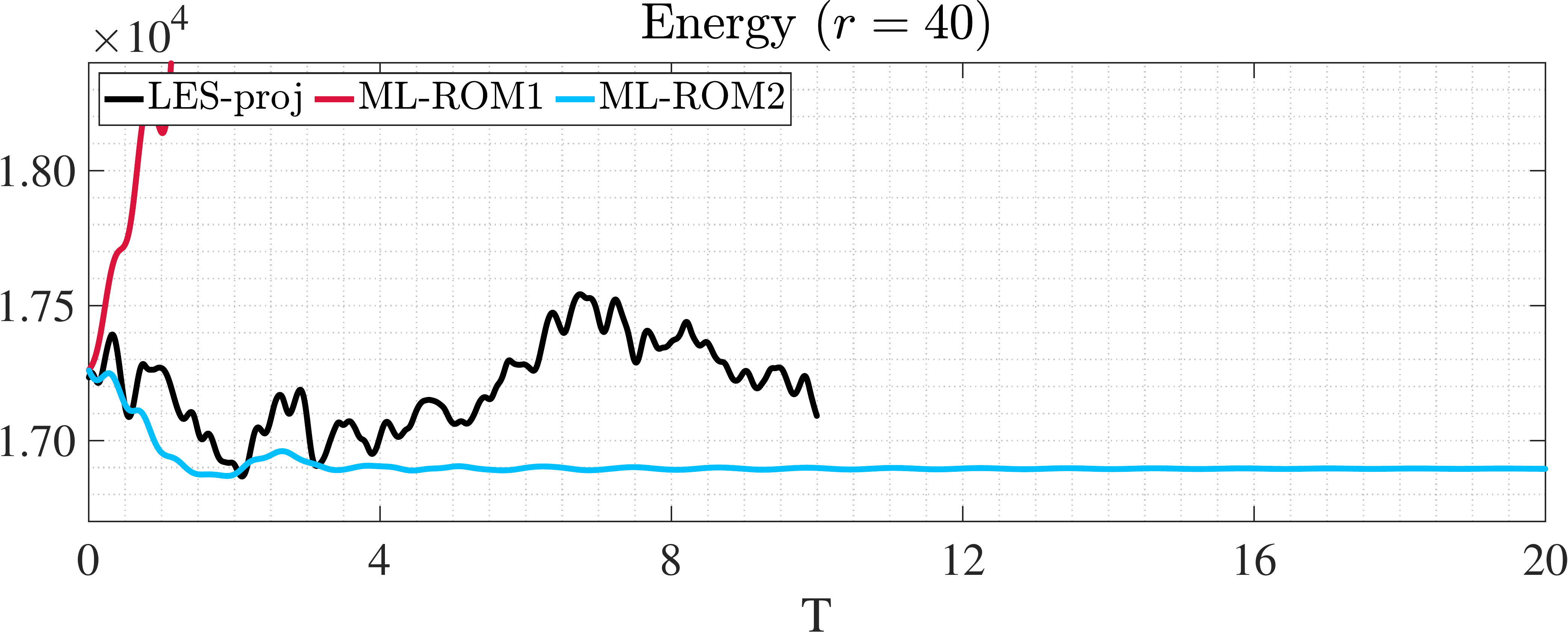}
    \includegraphics[width=.45\textwidth]{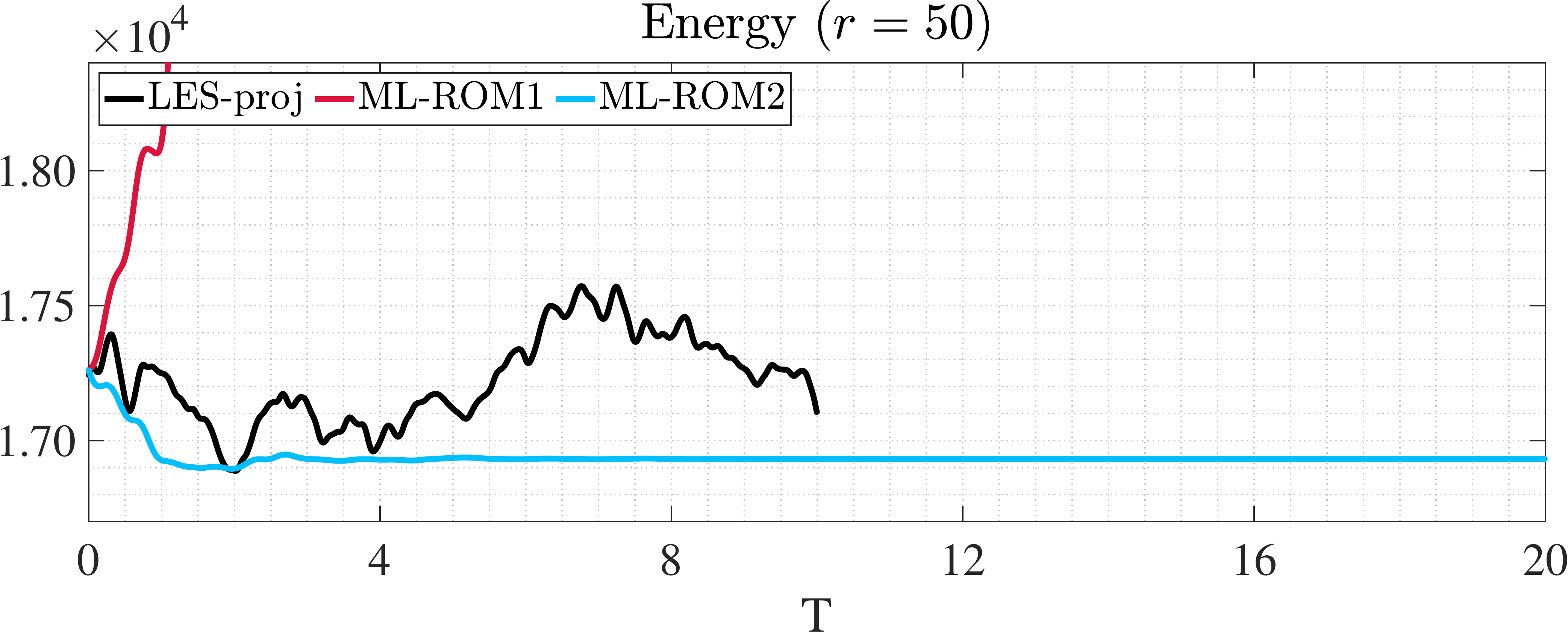}   
    \caption{Time evolution of the kinetic energy for $\alpha=0.2\times 10^{-3}$
    }    
    \label{fig:ke-alpha-5}
\end{figure}

\begin{figure}[H]
\centering
     \begin{subfigure}[b]{0.48\textwidth}
         \centering
    \includegraphics[width=.45\textwidth]{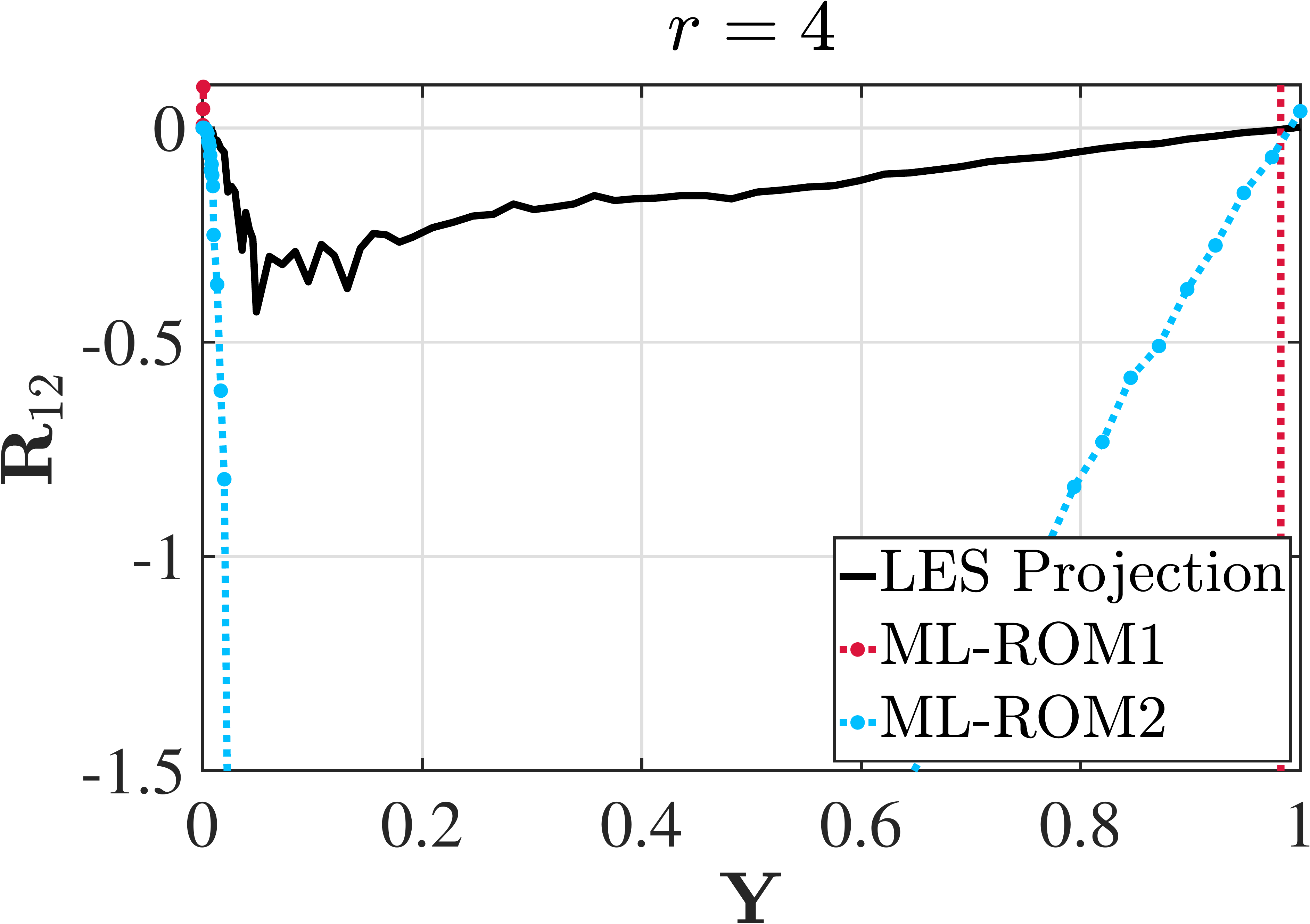}
    \includegraphics[width=.45\textwidth]{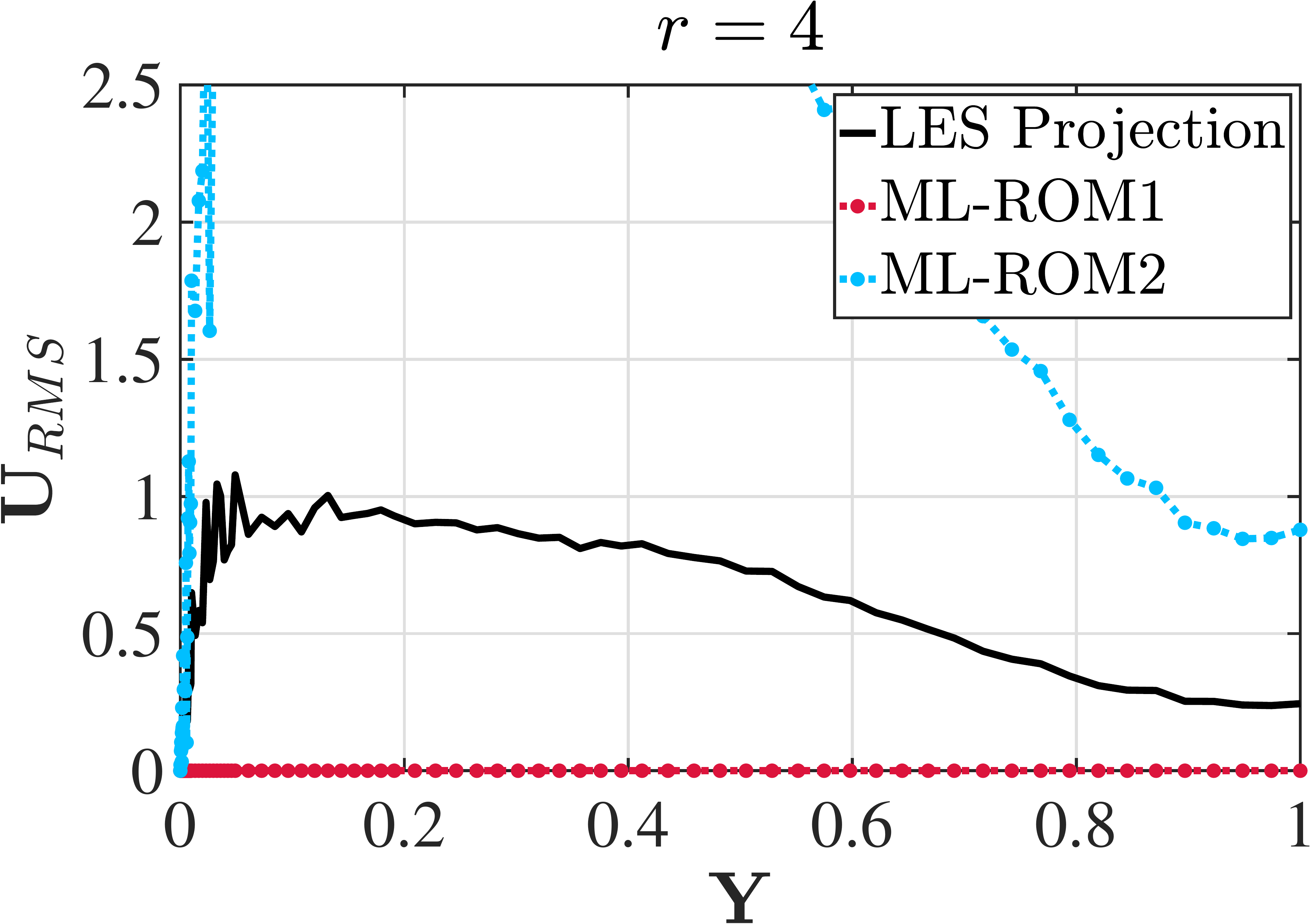}         \caption{$r=4$}
         \label{fig:stat-r-4}
     \end{subfigure}
     \begin{subfigure}[b]{0.48\textwidth}
         \centering
    \includegraphics[width=.45\textwidth]{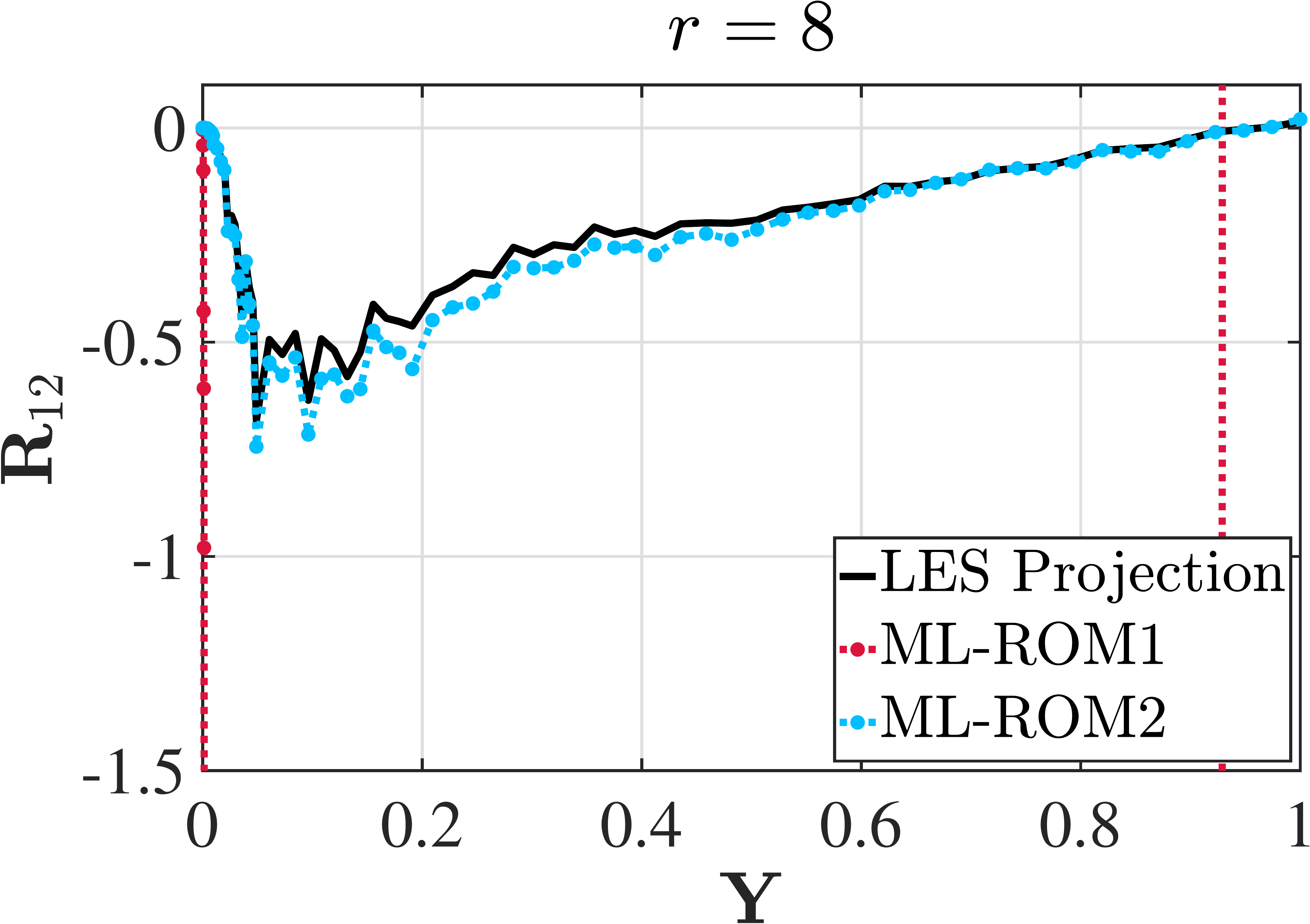}
    \includegraphics[width=.45\textwidth]{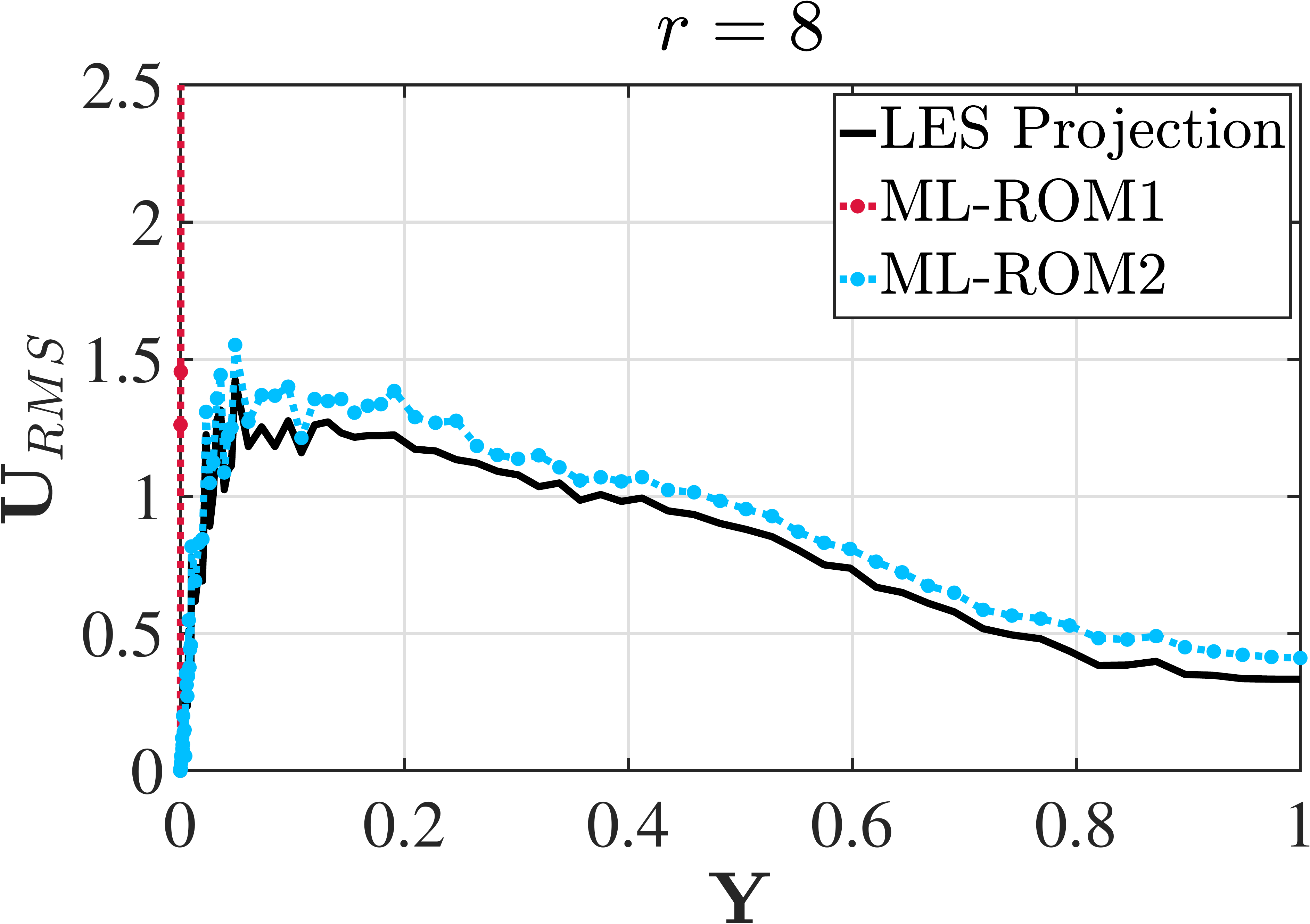}         \caption{$r=8$}
         \label{fig:stat-r-8}
     \end{subfigure}
     \begin{subfigure}[b]{0.48\textwidth}
         \centering
    \includegraphics[width=.45\textwidth]{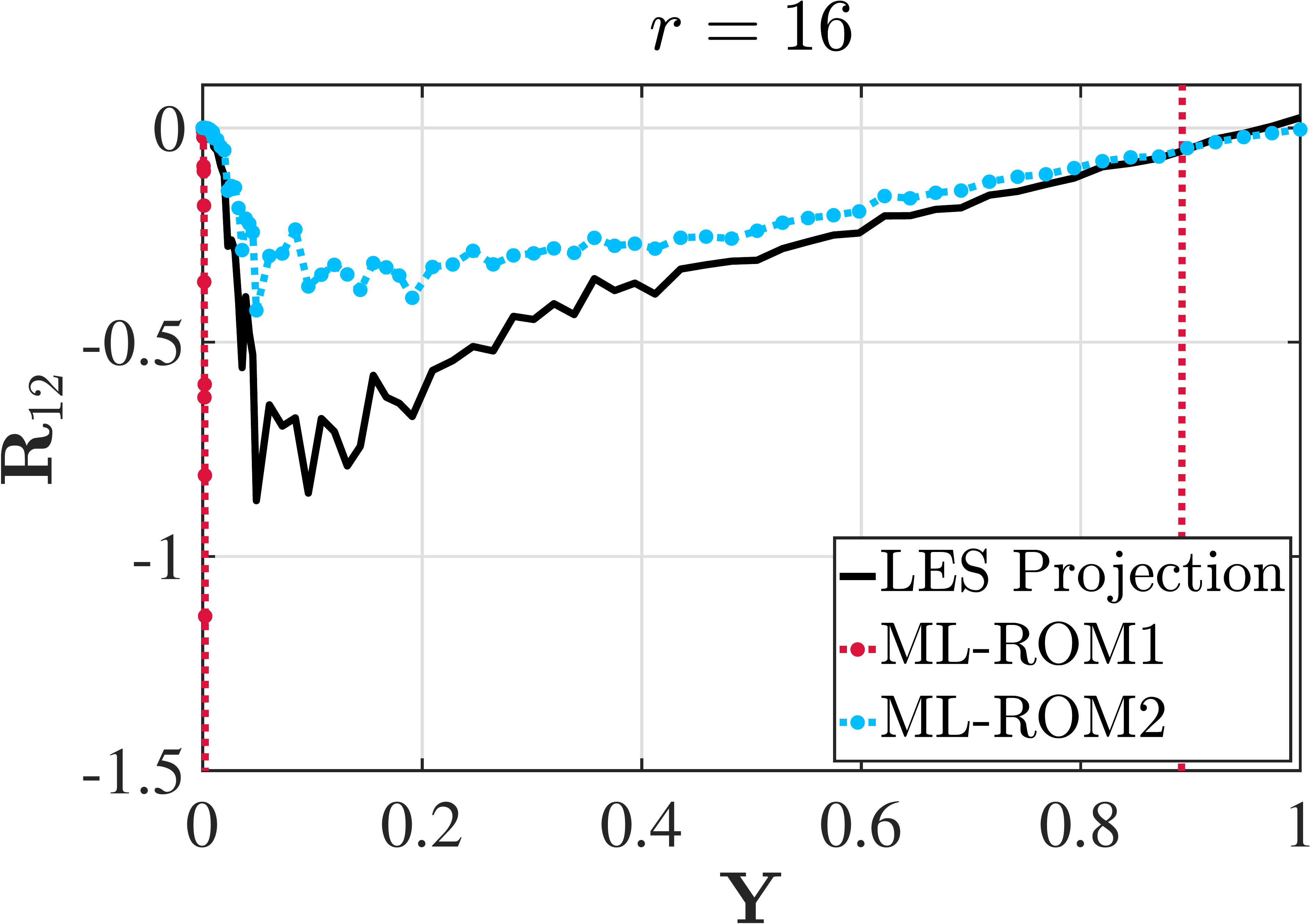}
    \includegraphics[width=.45\textwidth]{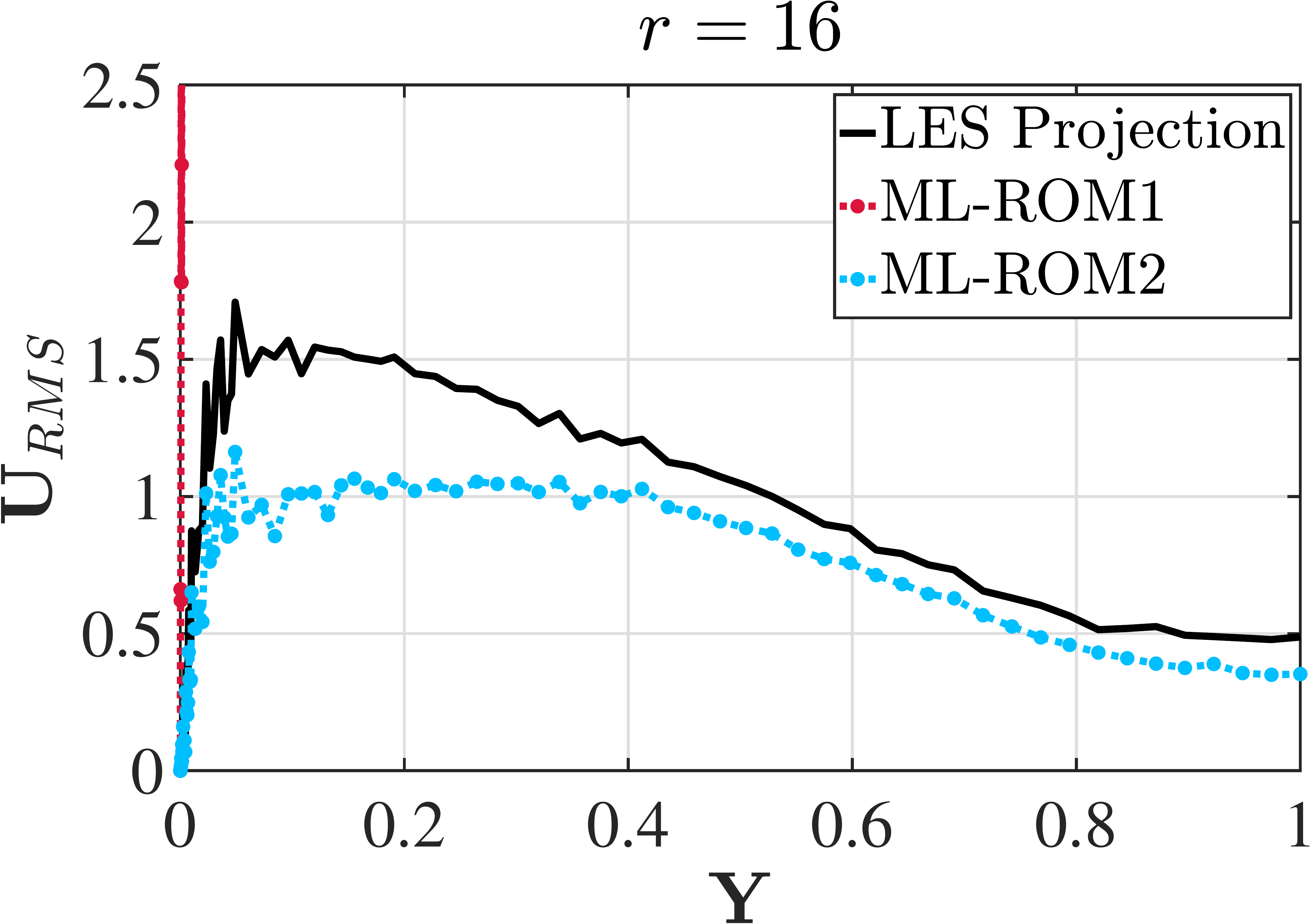}         \caption{$r=16$}
         \label{fig:stat-r-16}
     \end{subfigure}
     \begin{subfigure}[b]{0.48\textwidth}
         \centering
    \includegraphics[width=.45\textwidth]{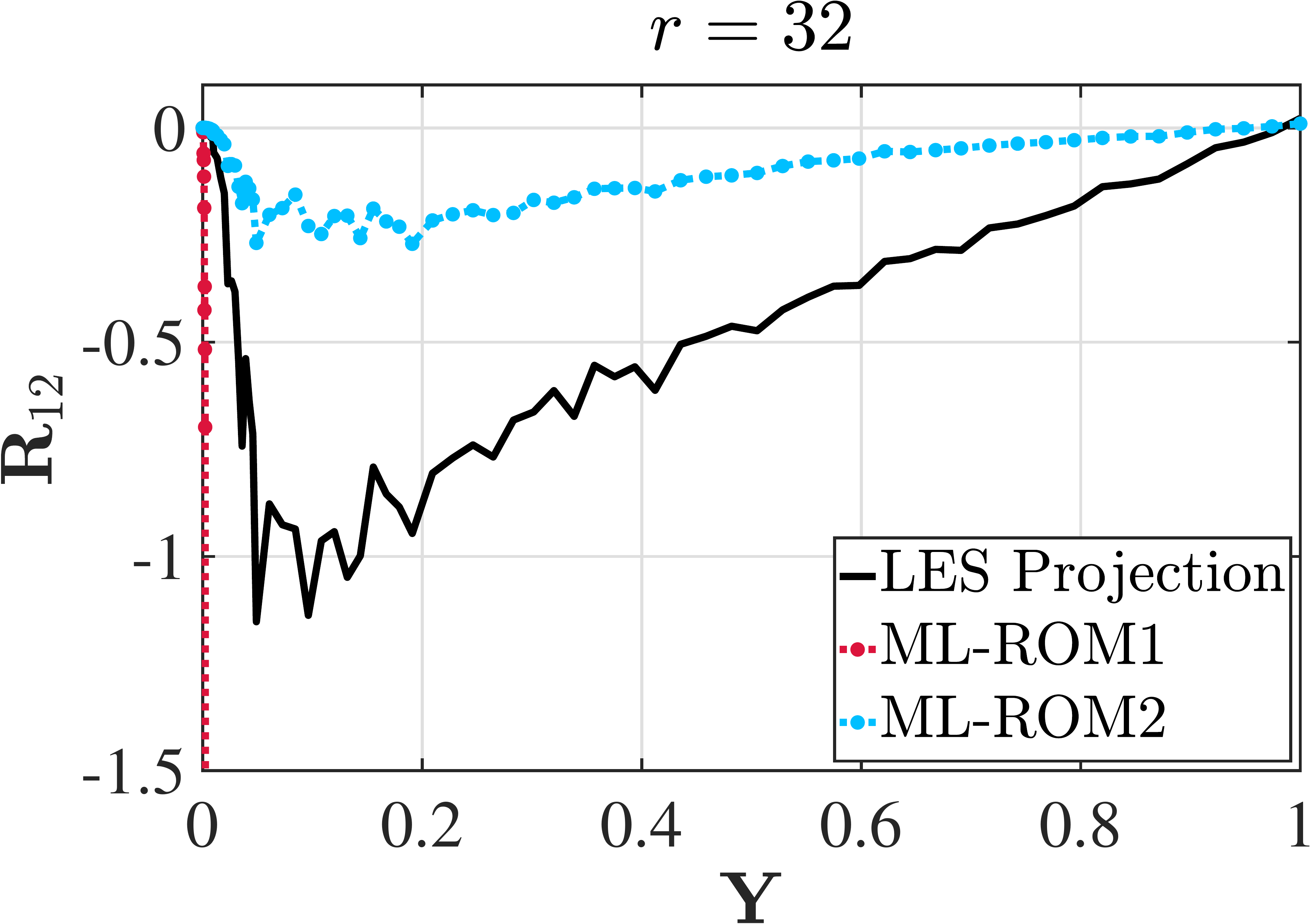}
    \includegraphics[width=.45\textwidth]{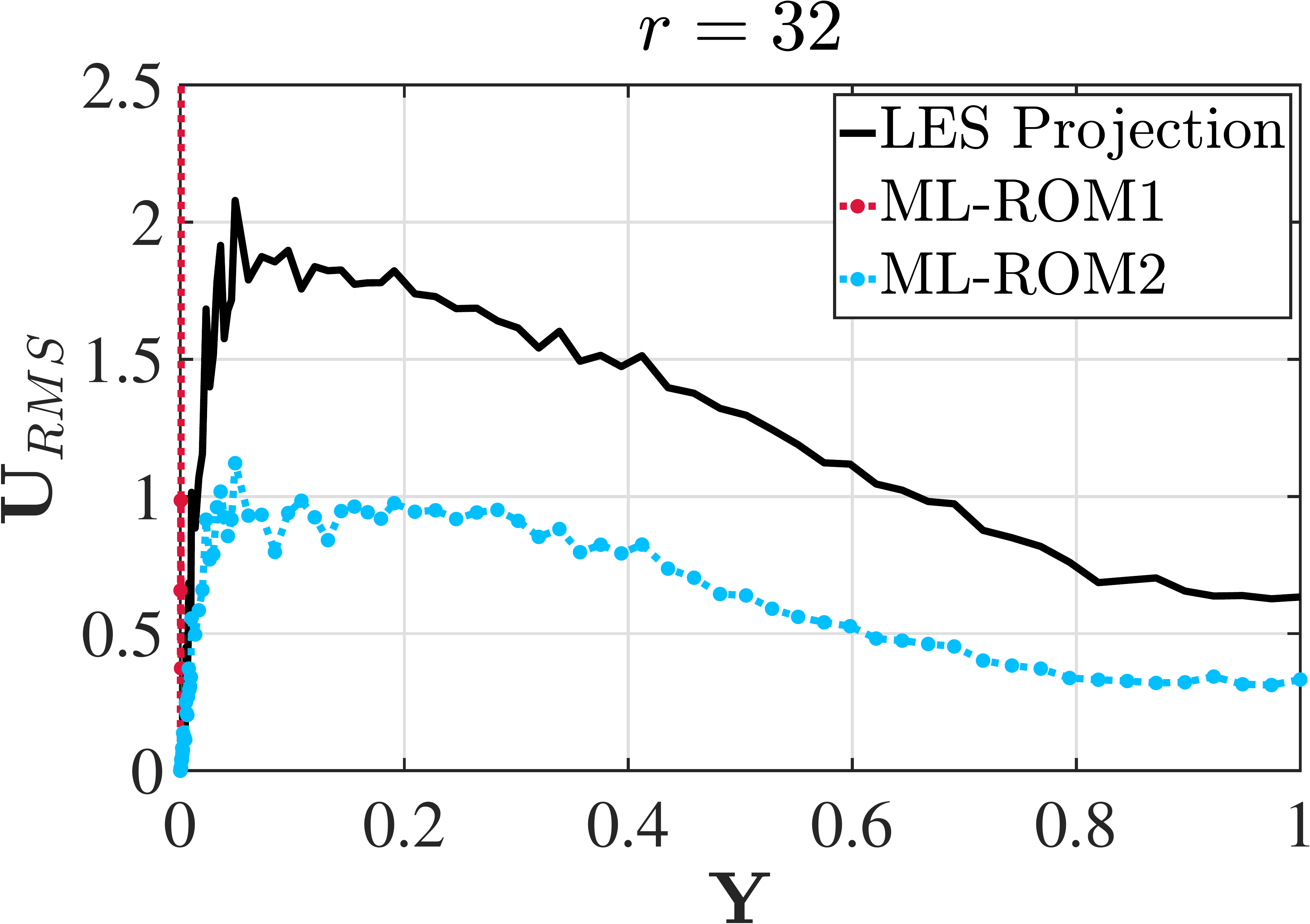}         \caption{$r=32$}
         \label{fig:stat-r-32}
    \end{subfigure}     
     \begin{subfigure}[b]{0.48\textwidth}
         \centering
    \includegraphics[width=.45\textwidth]{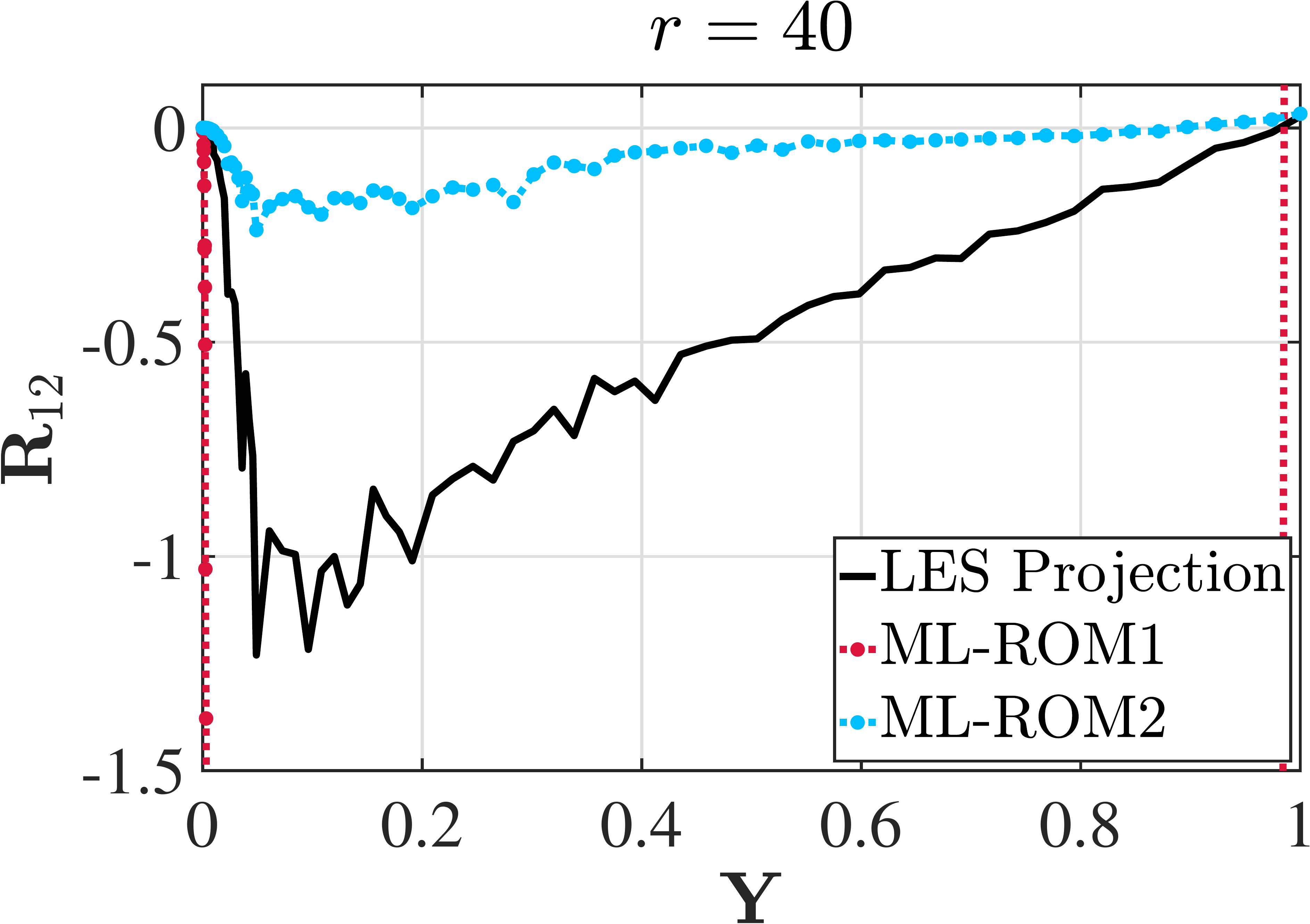}
    \includegraphics[width=.45\textwidth]{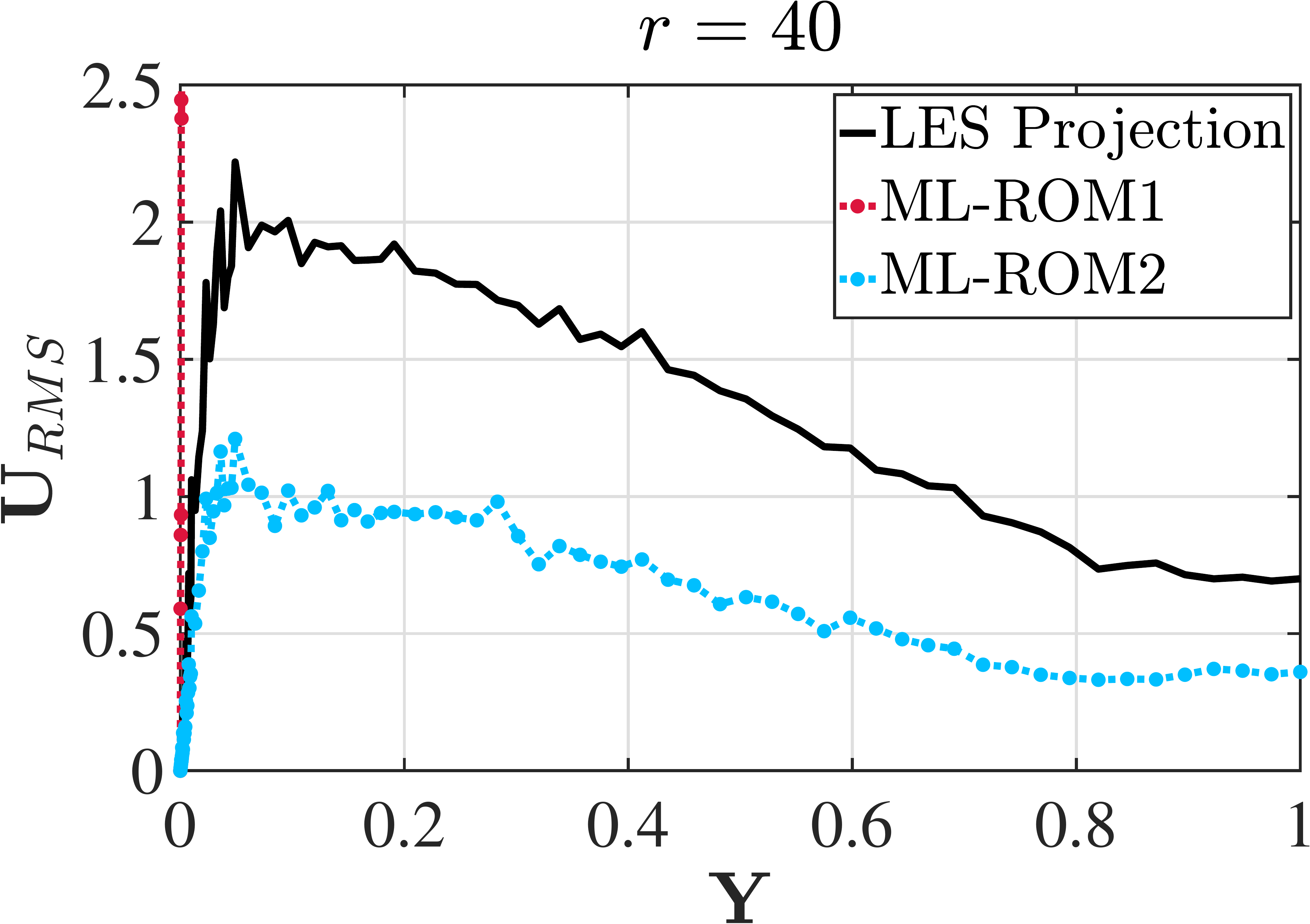}         \caption{$r=40$}
         \label{fig:stat-r-40}
     \end{subfigure} 
     \begin{subfigure}[b]{0.48\textwidth}
         \centering
    \includegraphics[width=.45\textwidth]{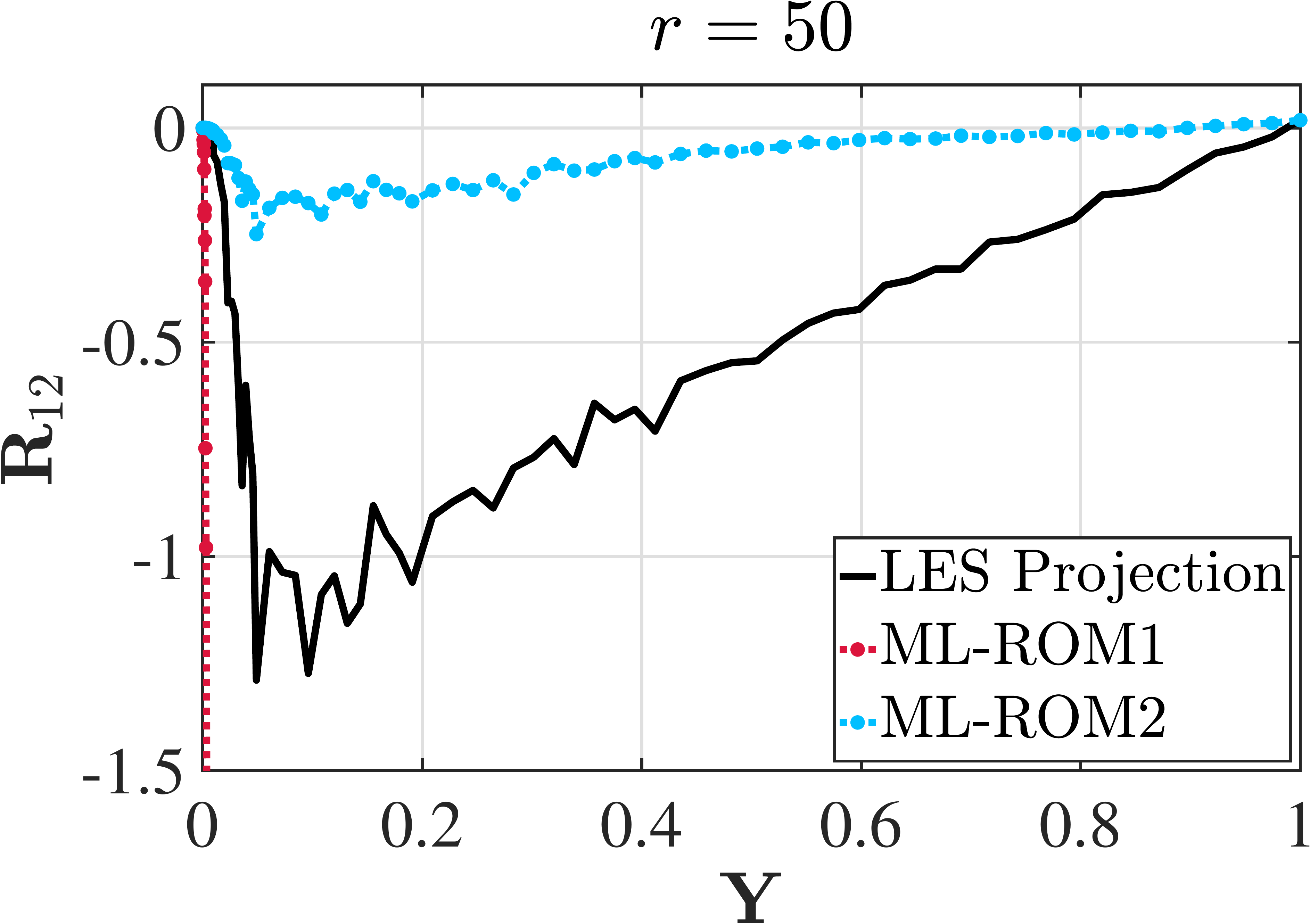}
    \includegraphics[width=.45\textwidth]{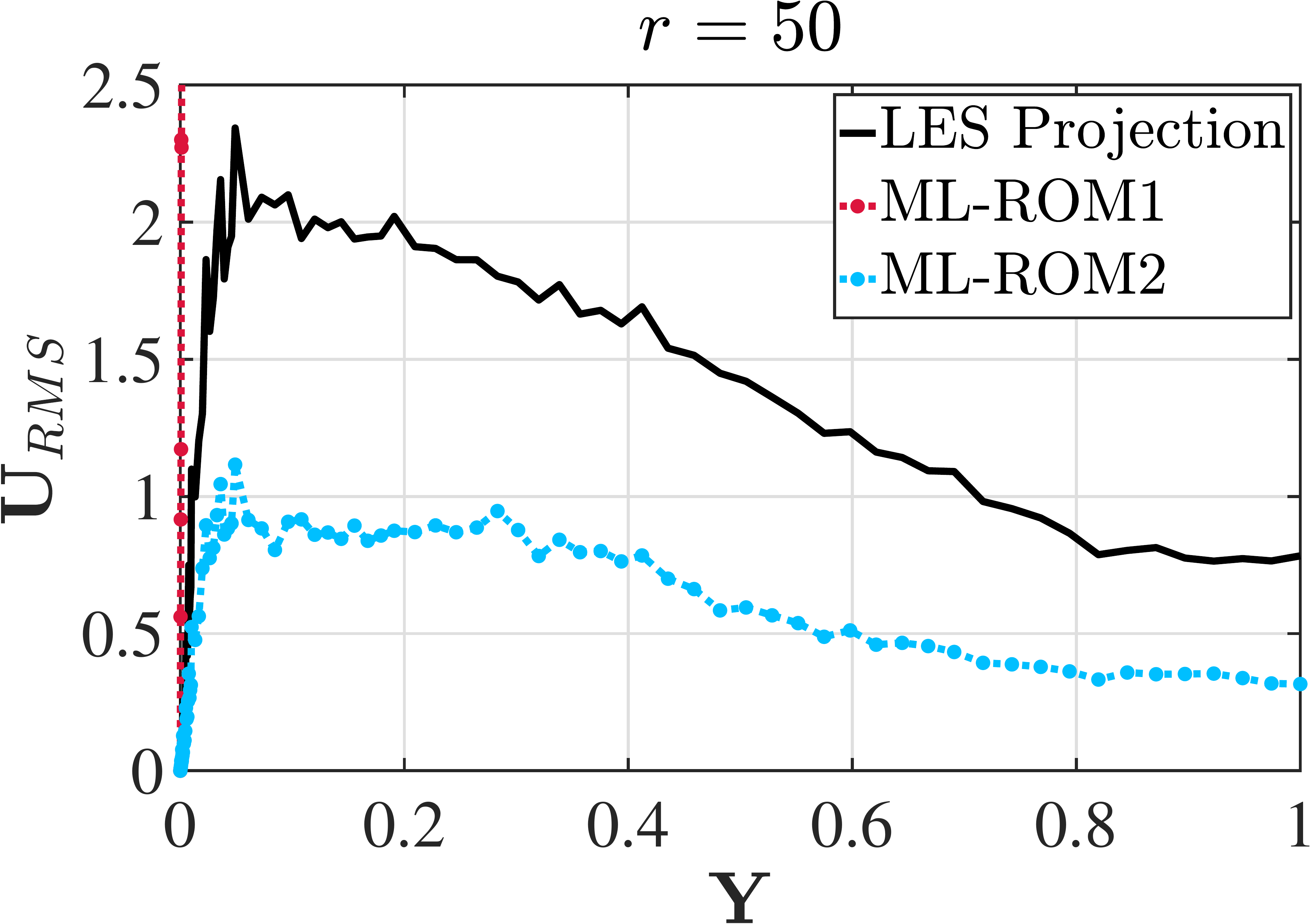}         \caption{$r=50$}
         \label{fig:stat-r-50}
     \end{subfigure} 
     \caption{Second-order statistics for $\alpha=0.2\times 10^{-3}$
    }    
    \label{fig:stat-alpha-5}
\end{figure}

\begin{figure}[H]
\centering
    \includegraphics[width=.45\textwidth]{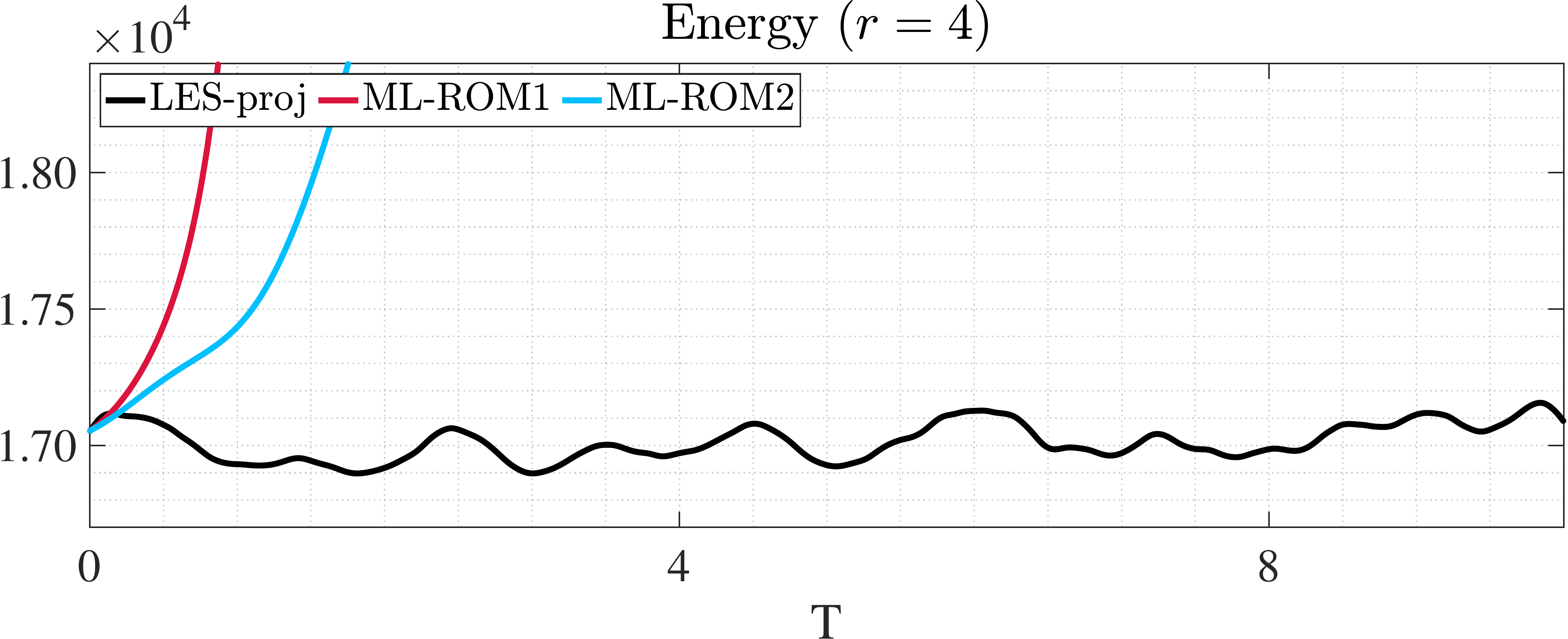}
    \includegraphics[width=.45\textwidth]{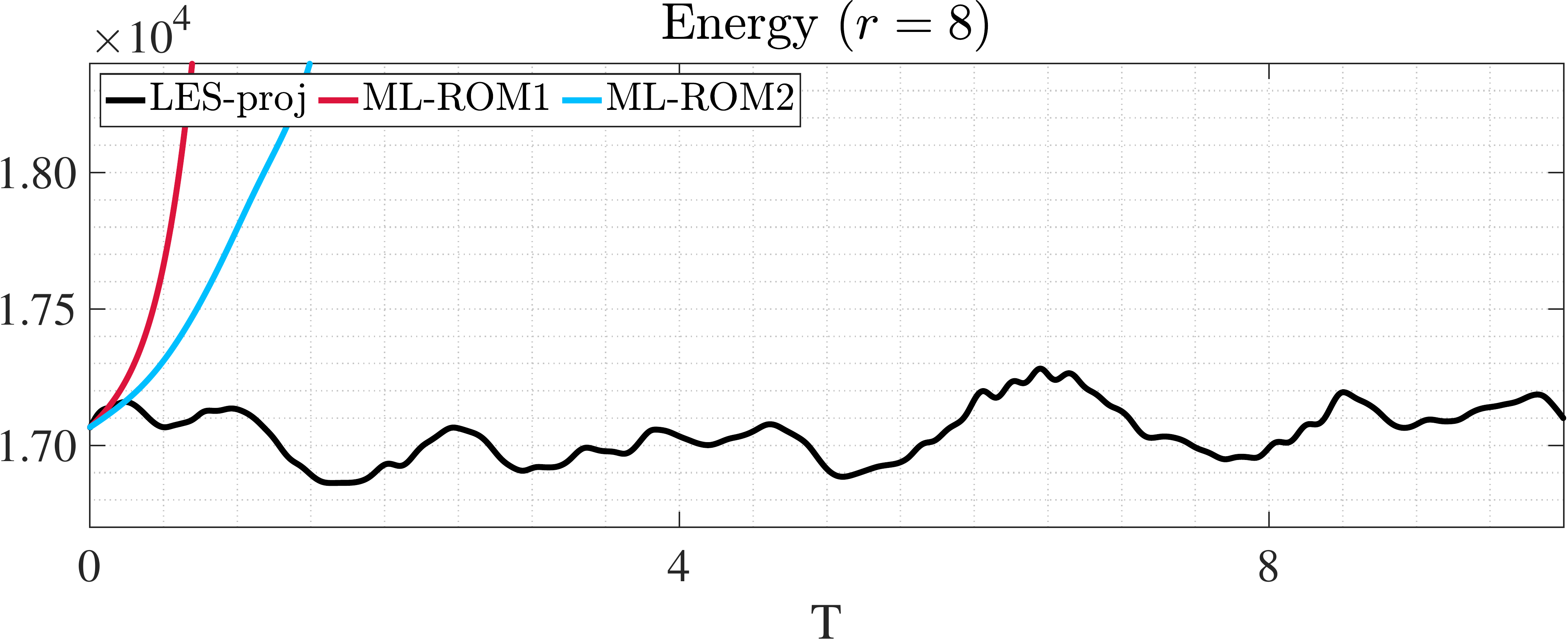}
    \includegraphics[width=.45\textwidth]{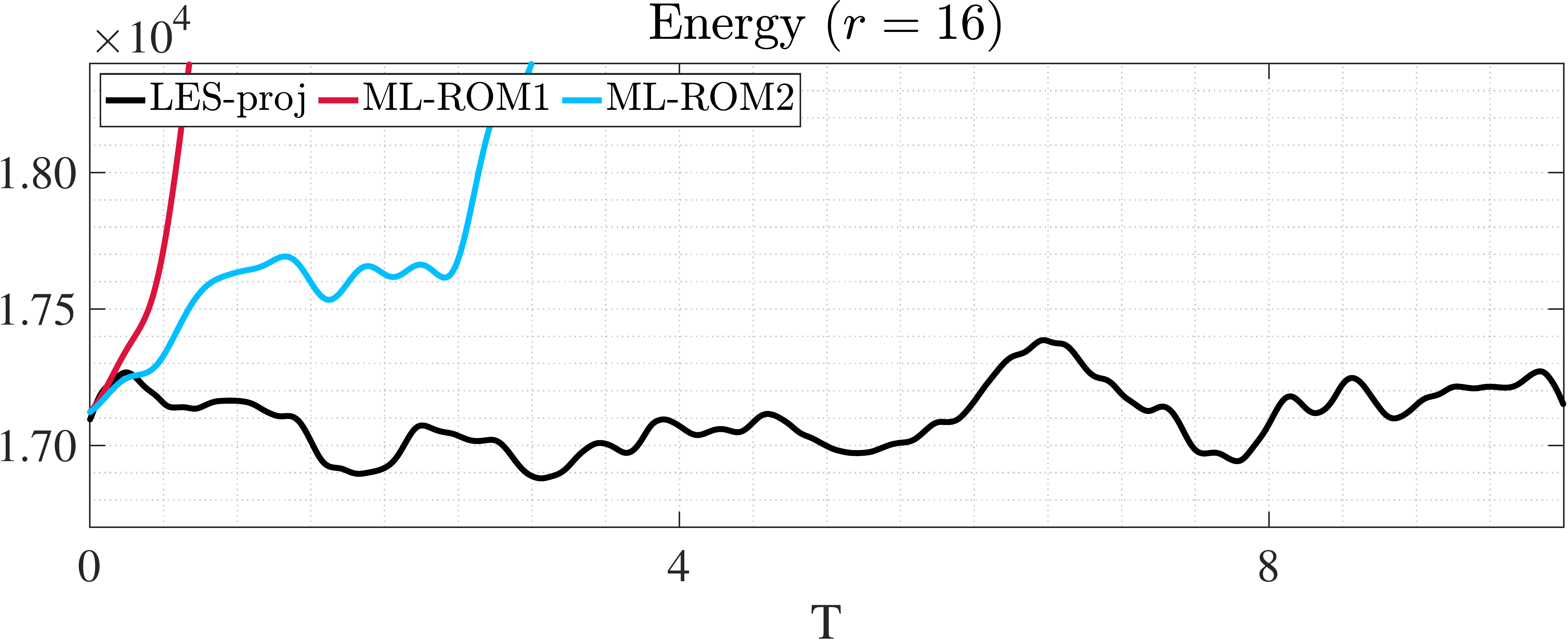}
    \includegraphics[width=.45\textwidth]{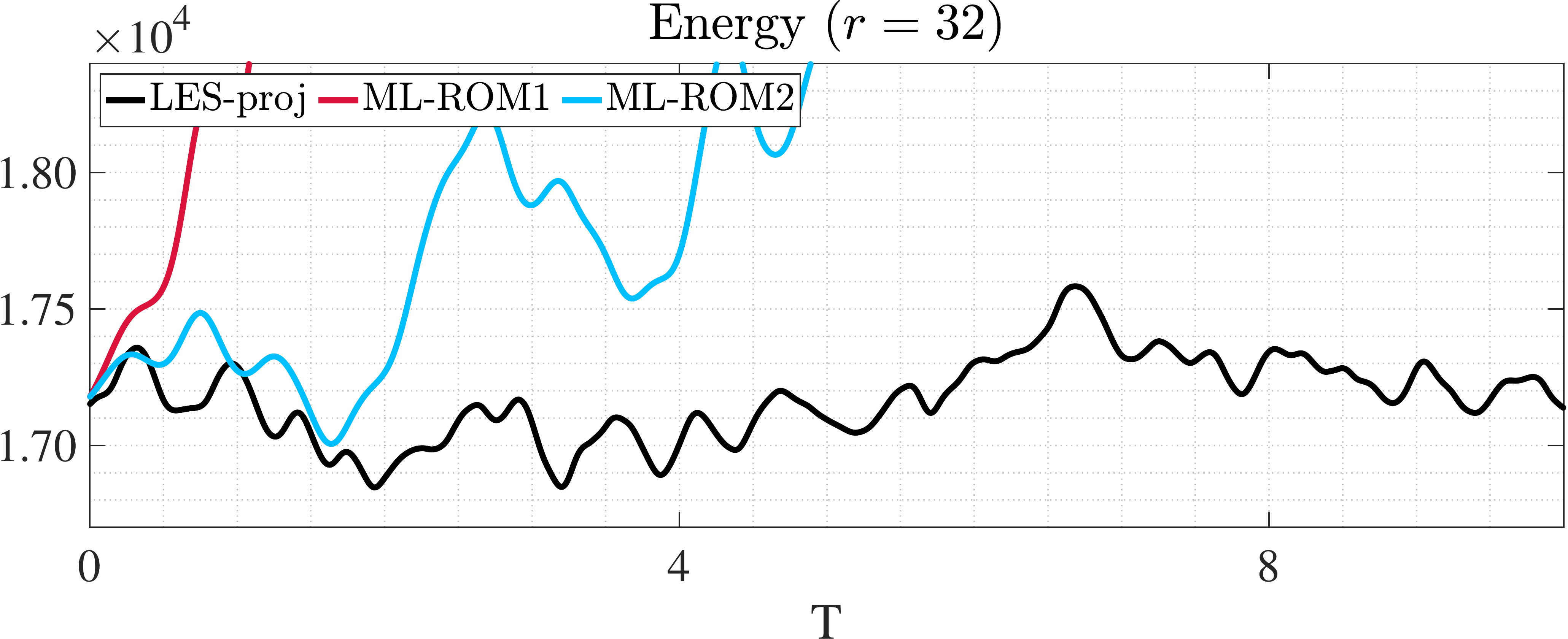}
    \includegraphics[width=.45\textwidth]{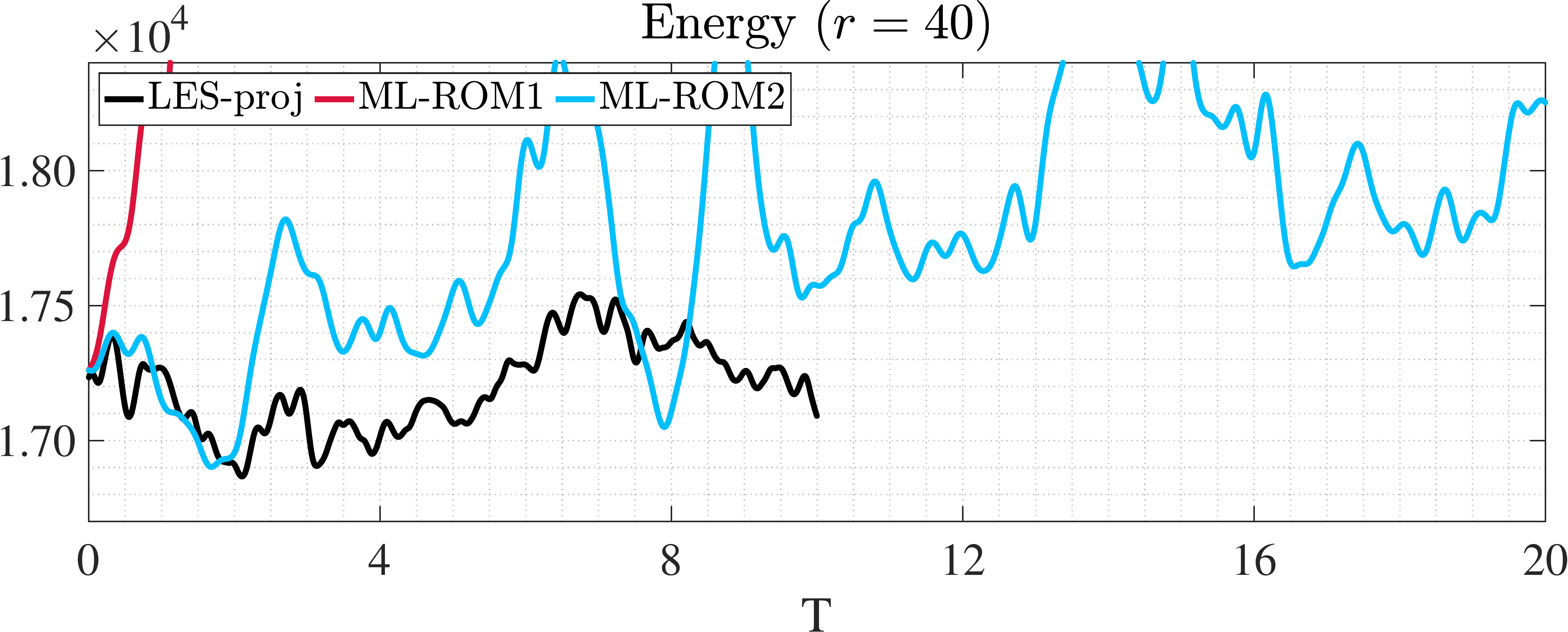}
    \includegraphics[width=.45\textwidth]{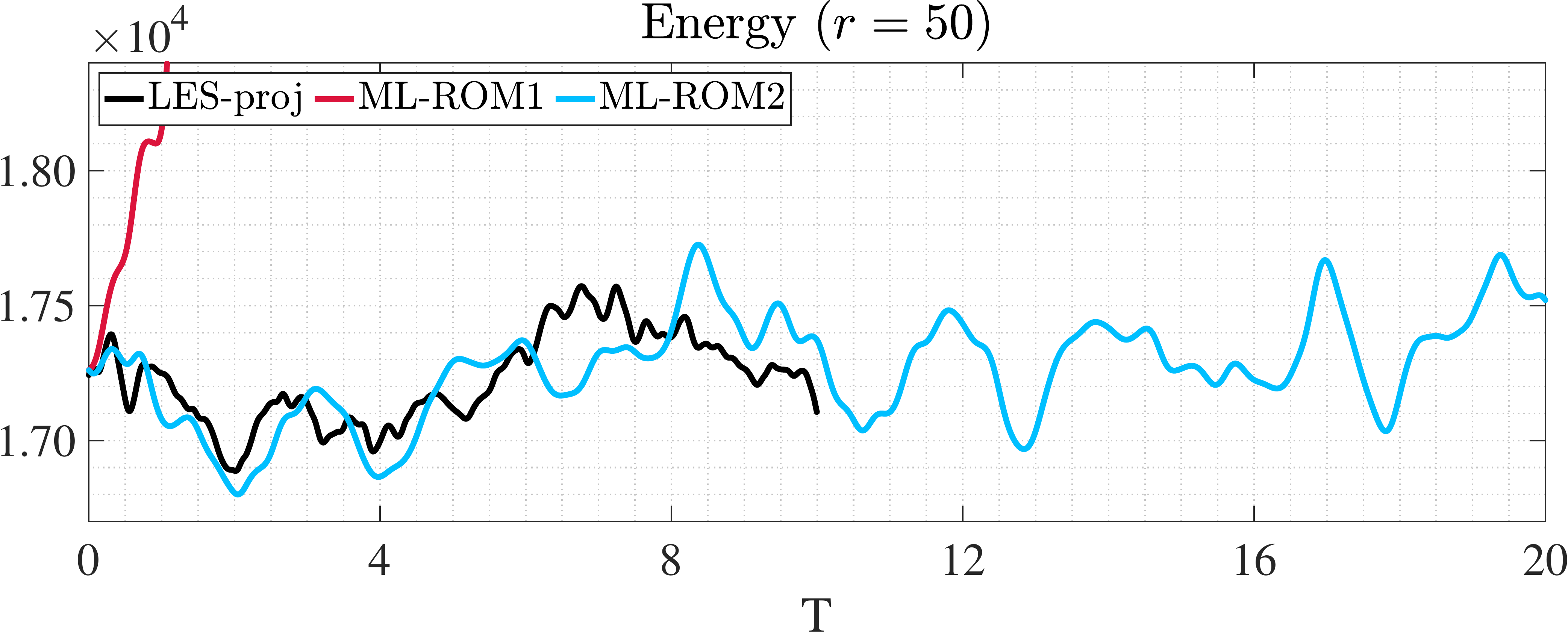}   
    \caption{Time evolution of the kinetic energy for $\alpha=0.1\times 10^{-3}$
    }    
    \label{fig:ke-alpha-7}
\end{figure}

\begin{figure}[H]
\centering
     \begin{subfigure}[b]{0.48\textwidth}
         \centering
    \includegraphics[width=.45\textwidth]{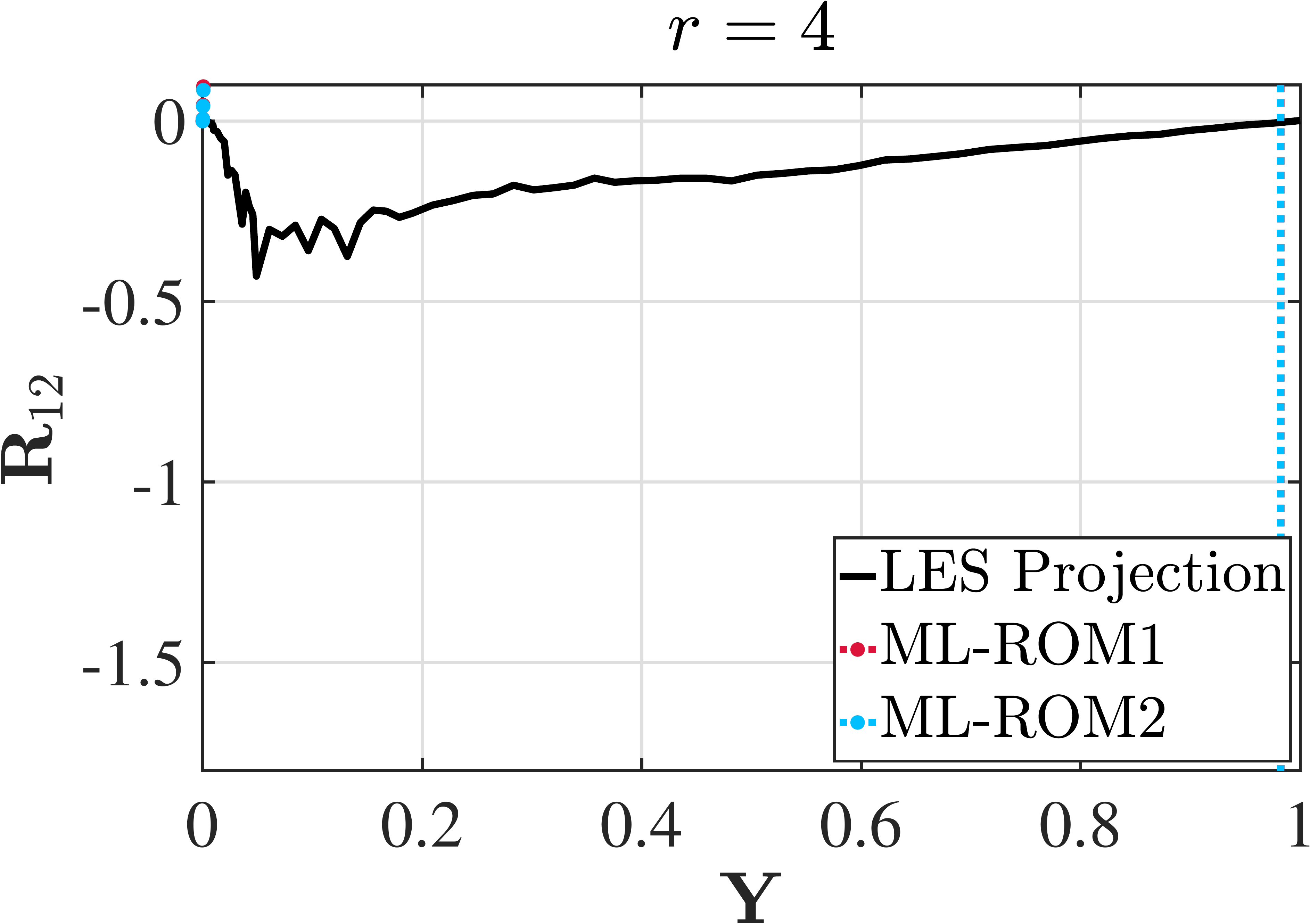}
    \includegraphics[width=.45\textwidth]{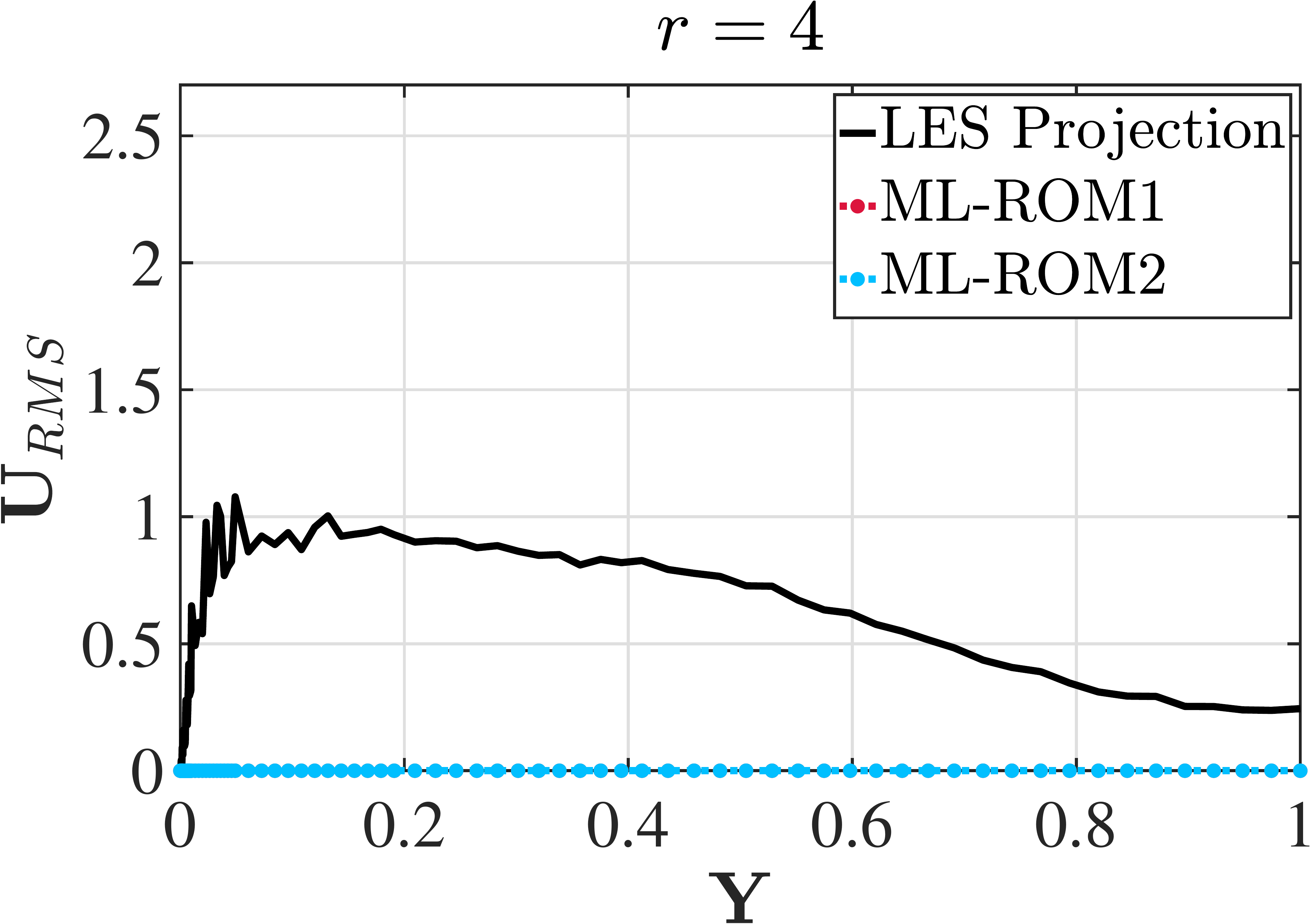}         \caption{$r=4$}
         \label{fig:stat-r-4}
     \end{subfigure}
     \begin{subfigure}[b]{0.48\textwidth}
         \centering
    \includegraphics[width=.45\textwidth]{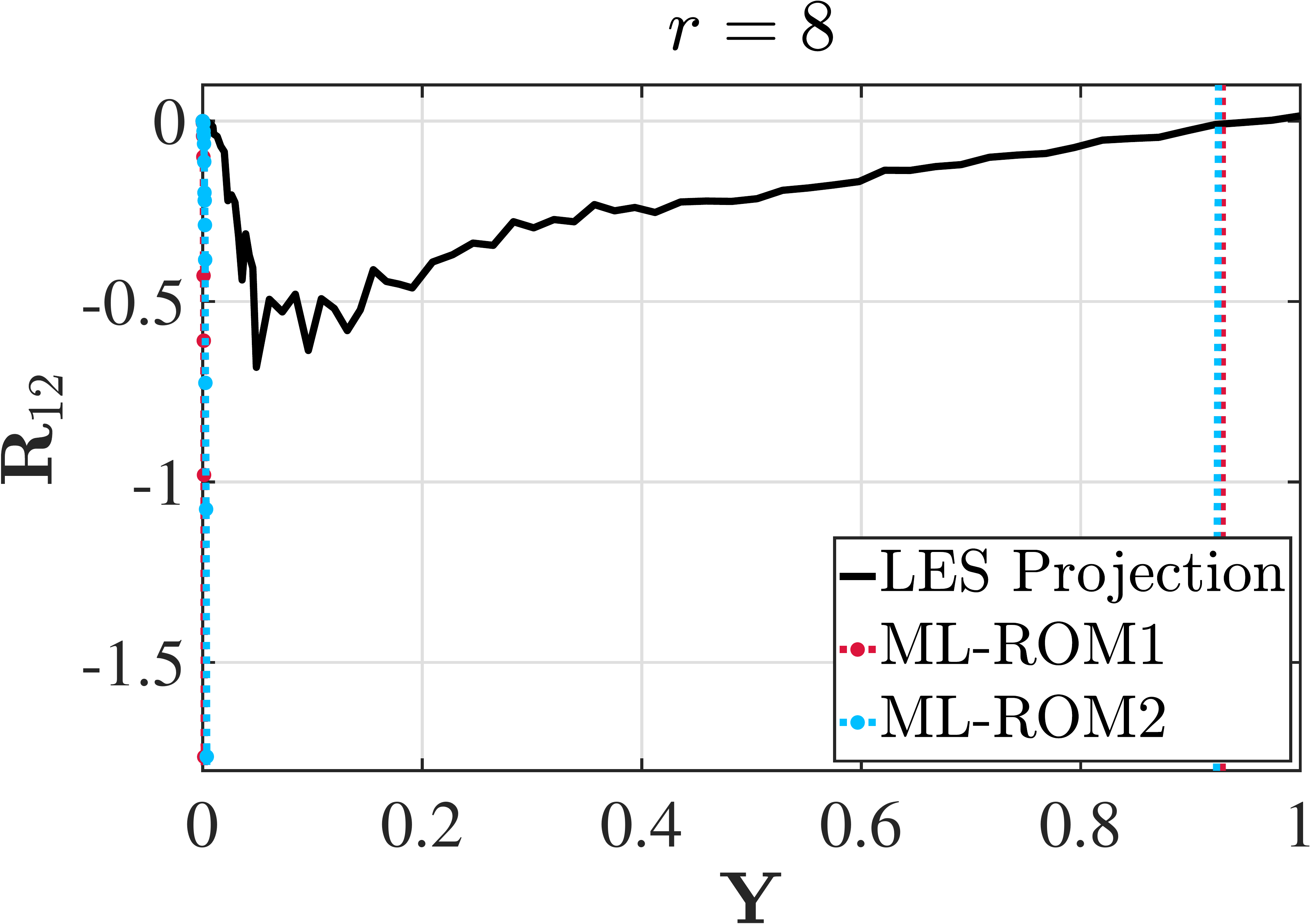}
    \includegraphics[width=.45\textwidth]{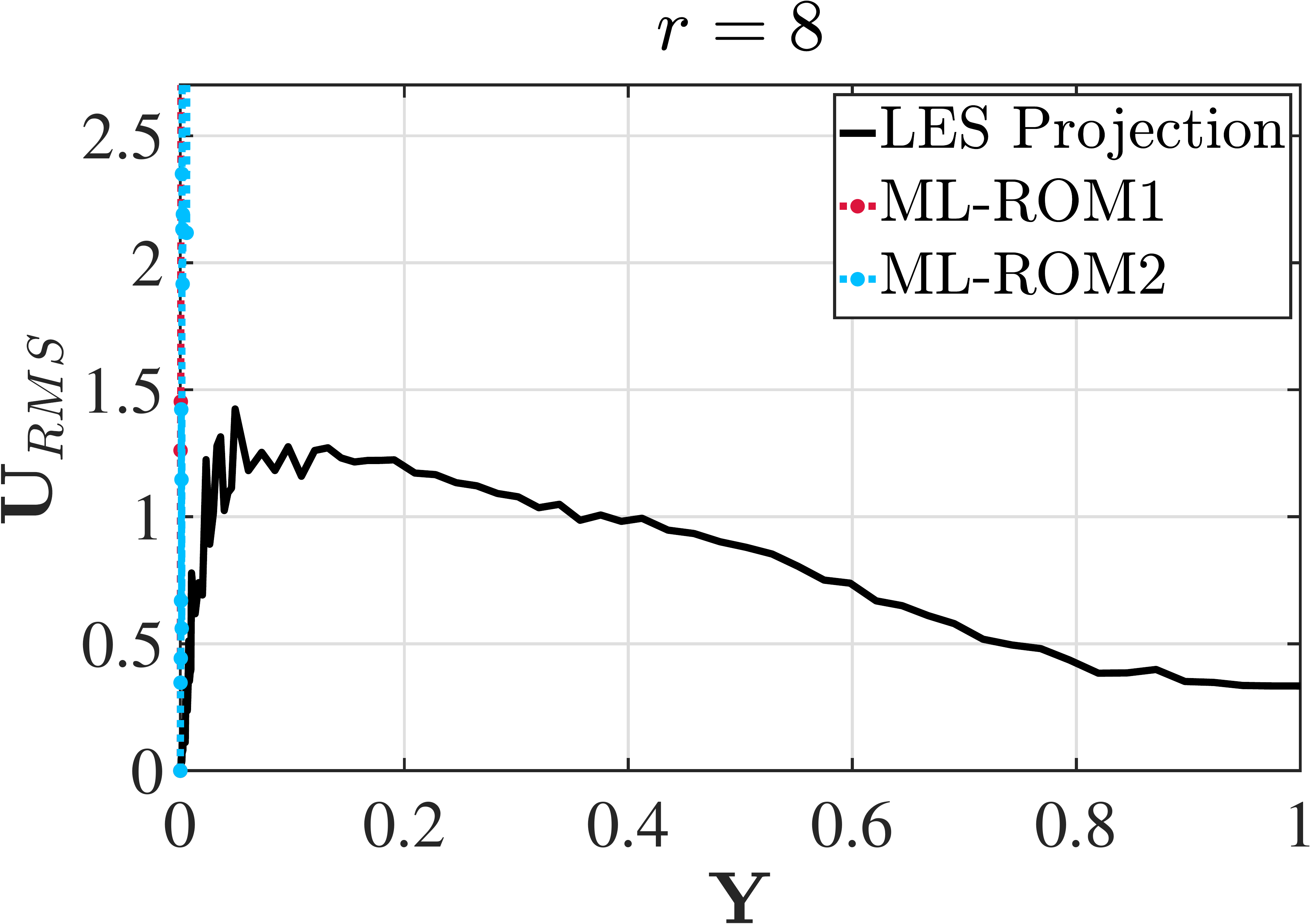}         \caption{$r=8$}
         \label{fig:stat-r-8}
     \end{subfigure}
     \begin{subfigure}[b]{0.48\textwidth}
         \centering
    \includegraphics[width=.45\textwidth]{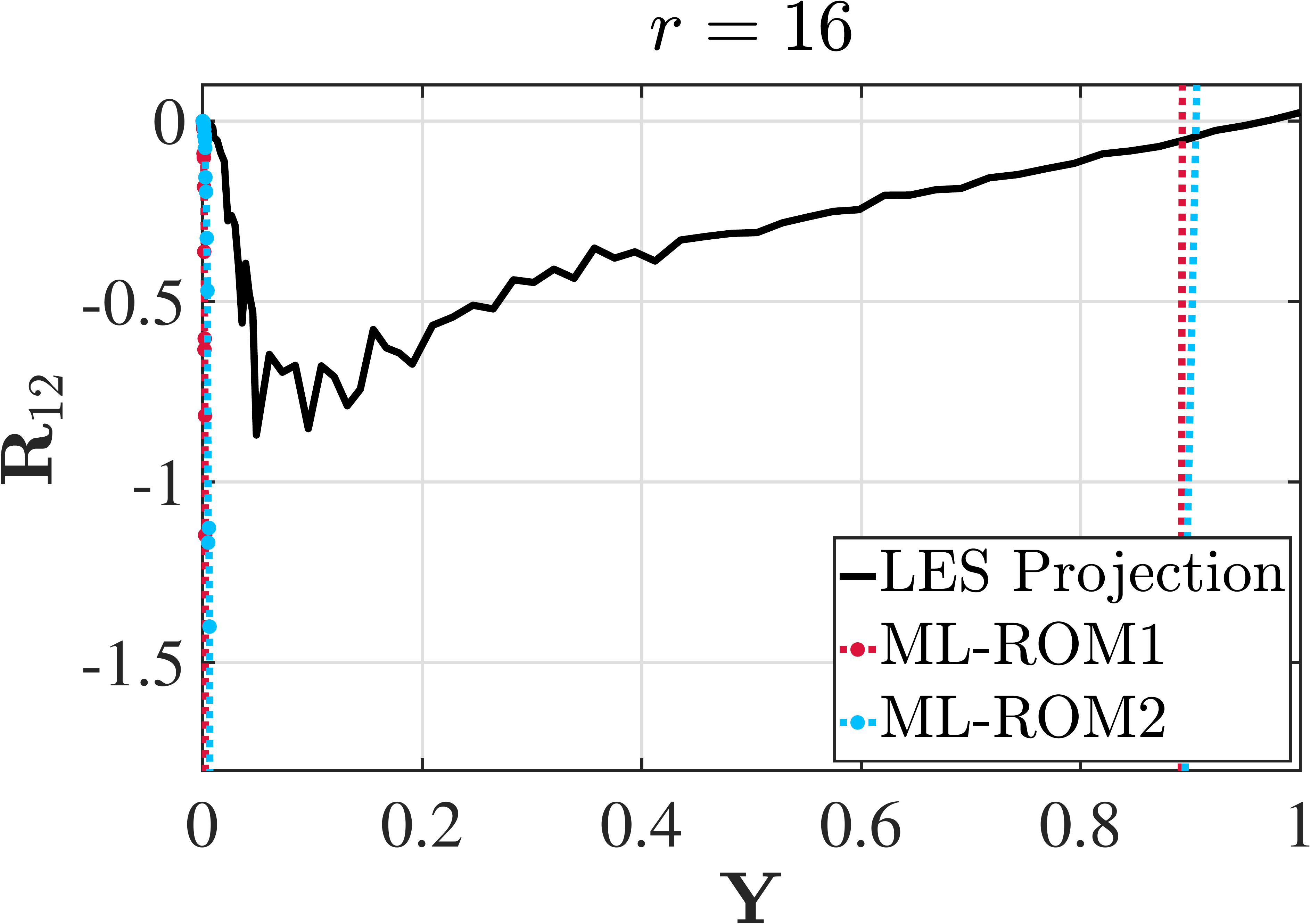}
    \includegraphics[width=.45\textwidth]{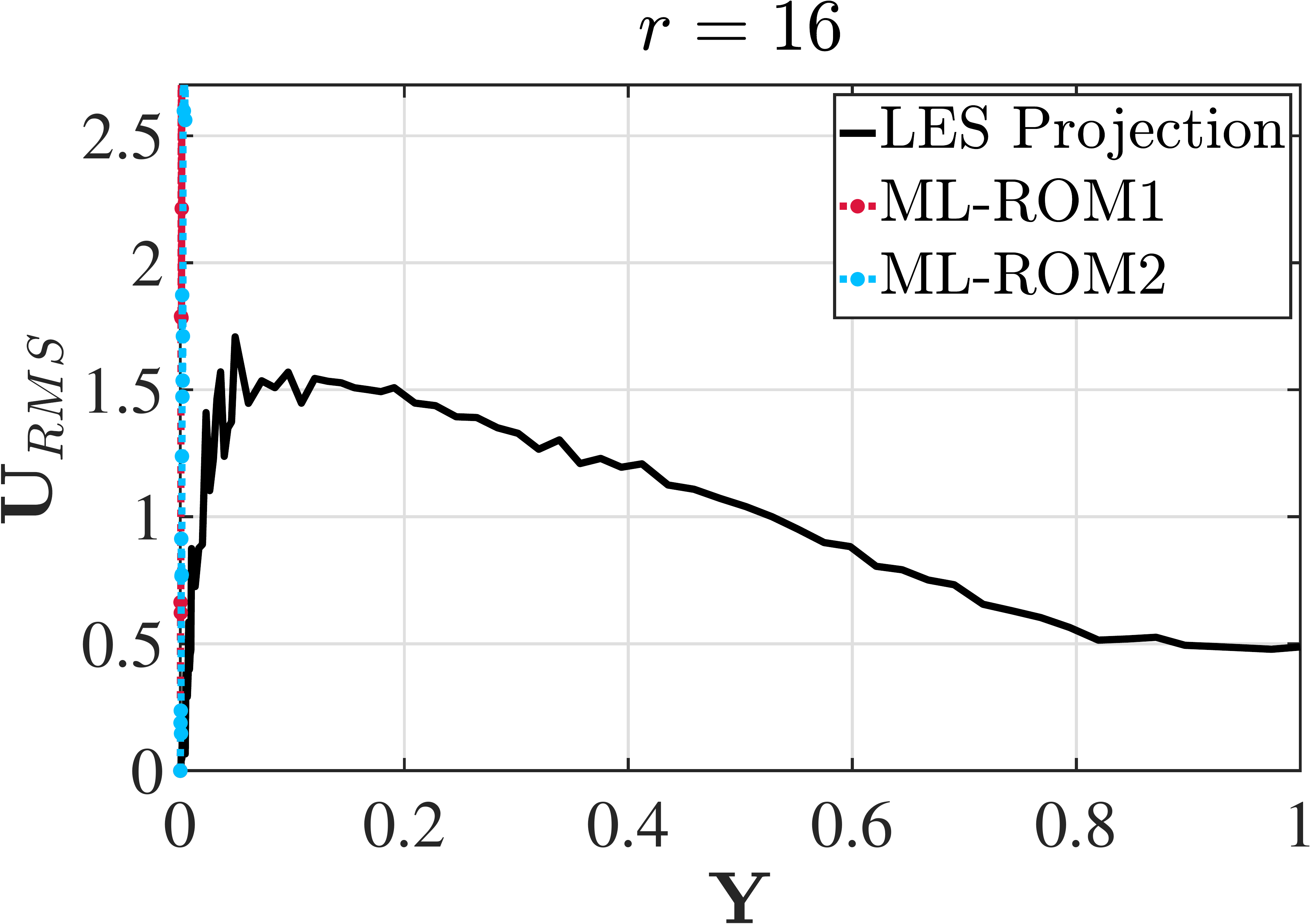}         \caption{$r=16$}
         \label{fig:stat-r-16}
     \end{subfigure}
     \begin{subfigure}[b]{0.48\textwidth}
         \centering
    \includegraphics[width=.45\textwidth]{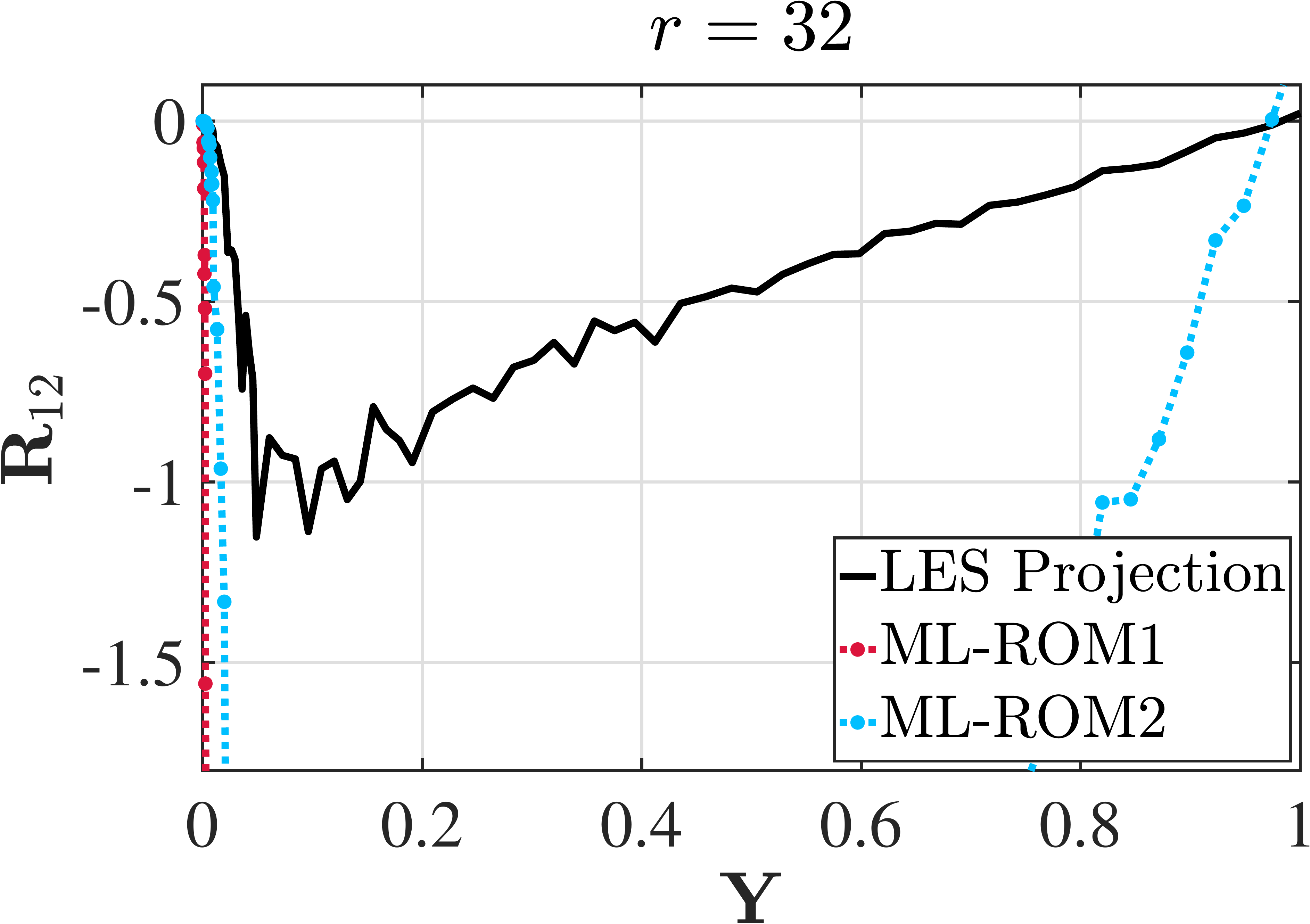}
    \includegraphics[width=.45\textwidth]{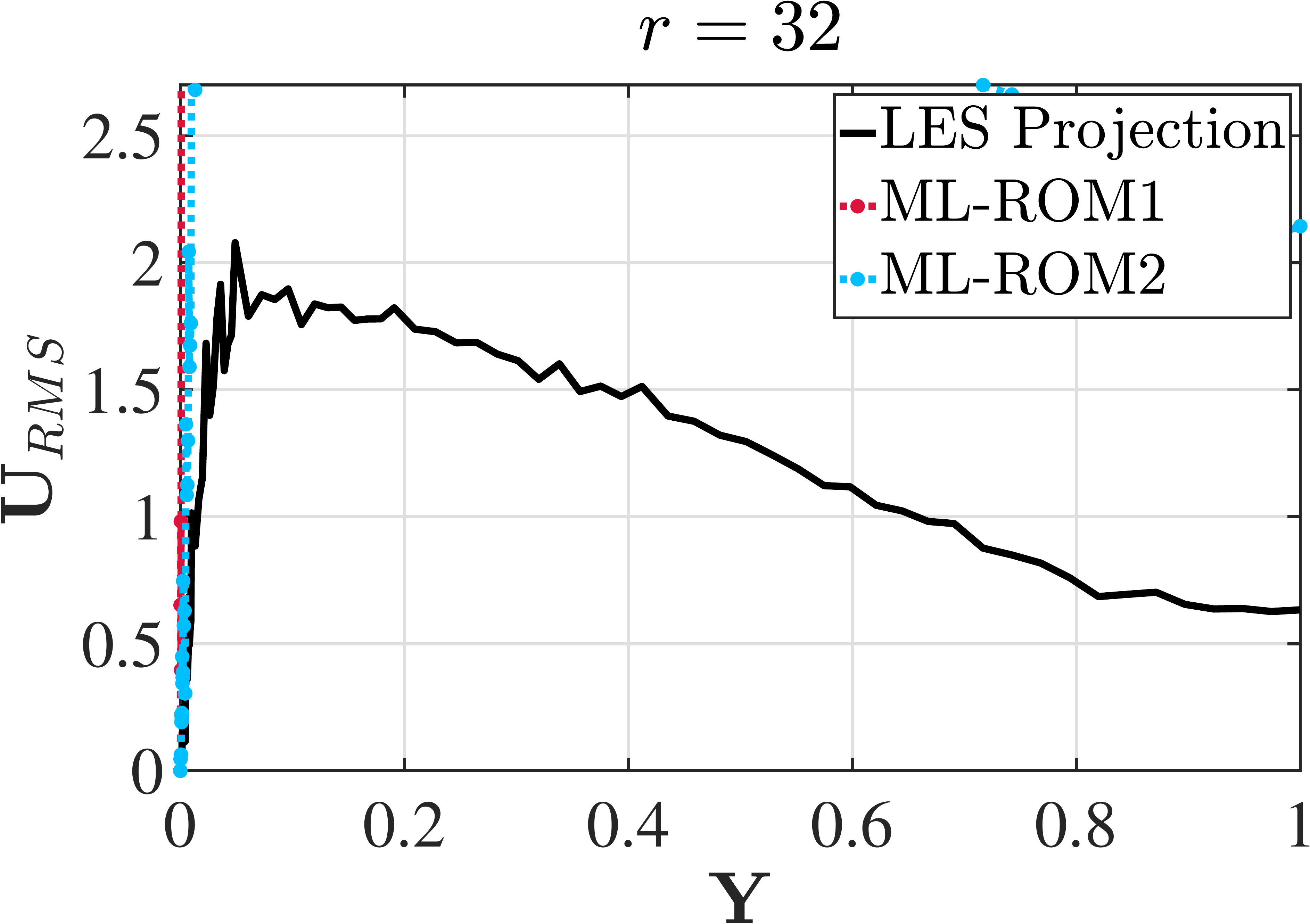}         \caption{$r=32$}
         \label{fig:stat-r-32}
    \end{subfigure}     
     \begin{subfigure}[b]{0.48\textwidth}
         \centering
    \includegraphics[width=.45\textwidth]{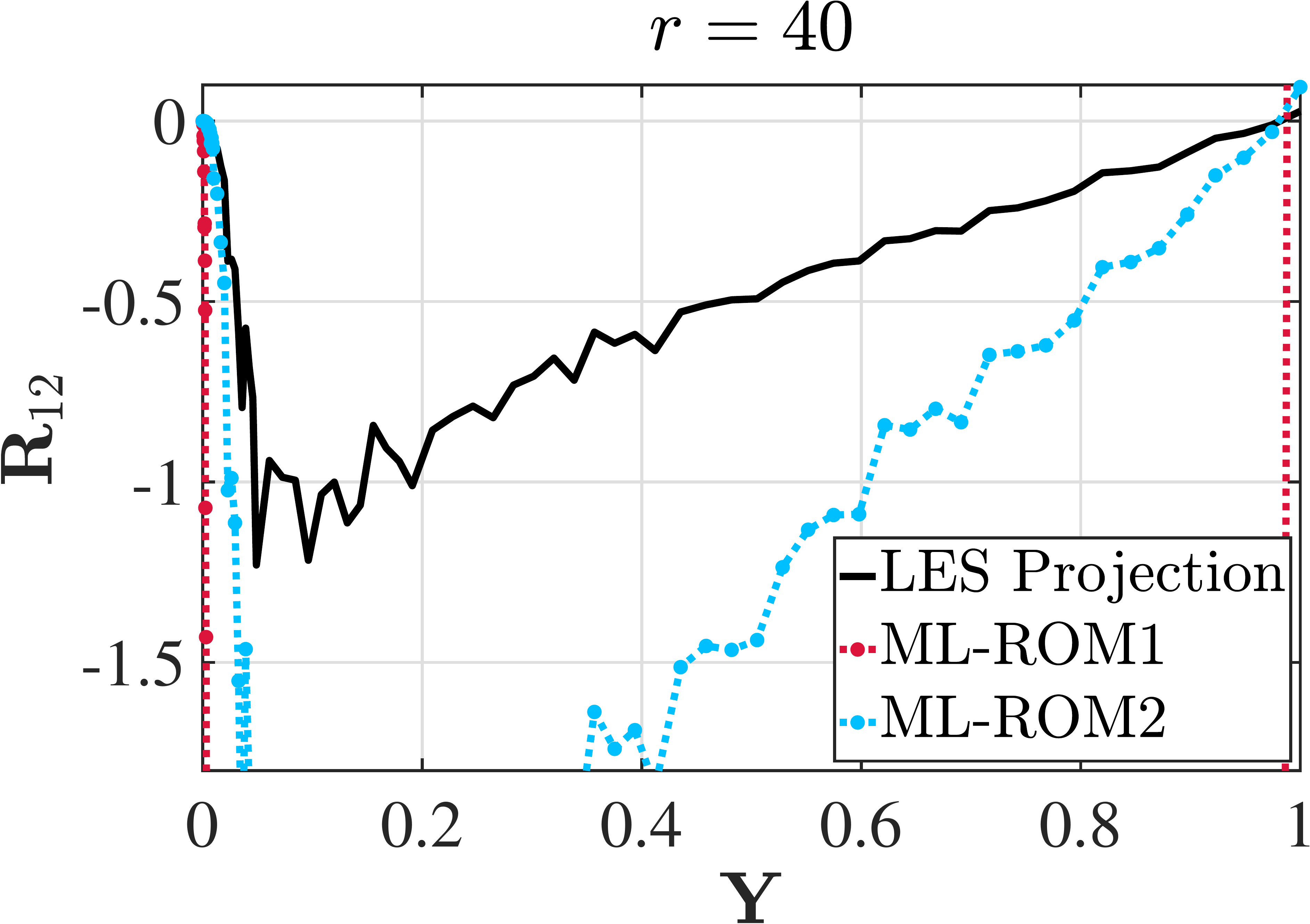}
    \includegraphics[width=.45\textwidth]{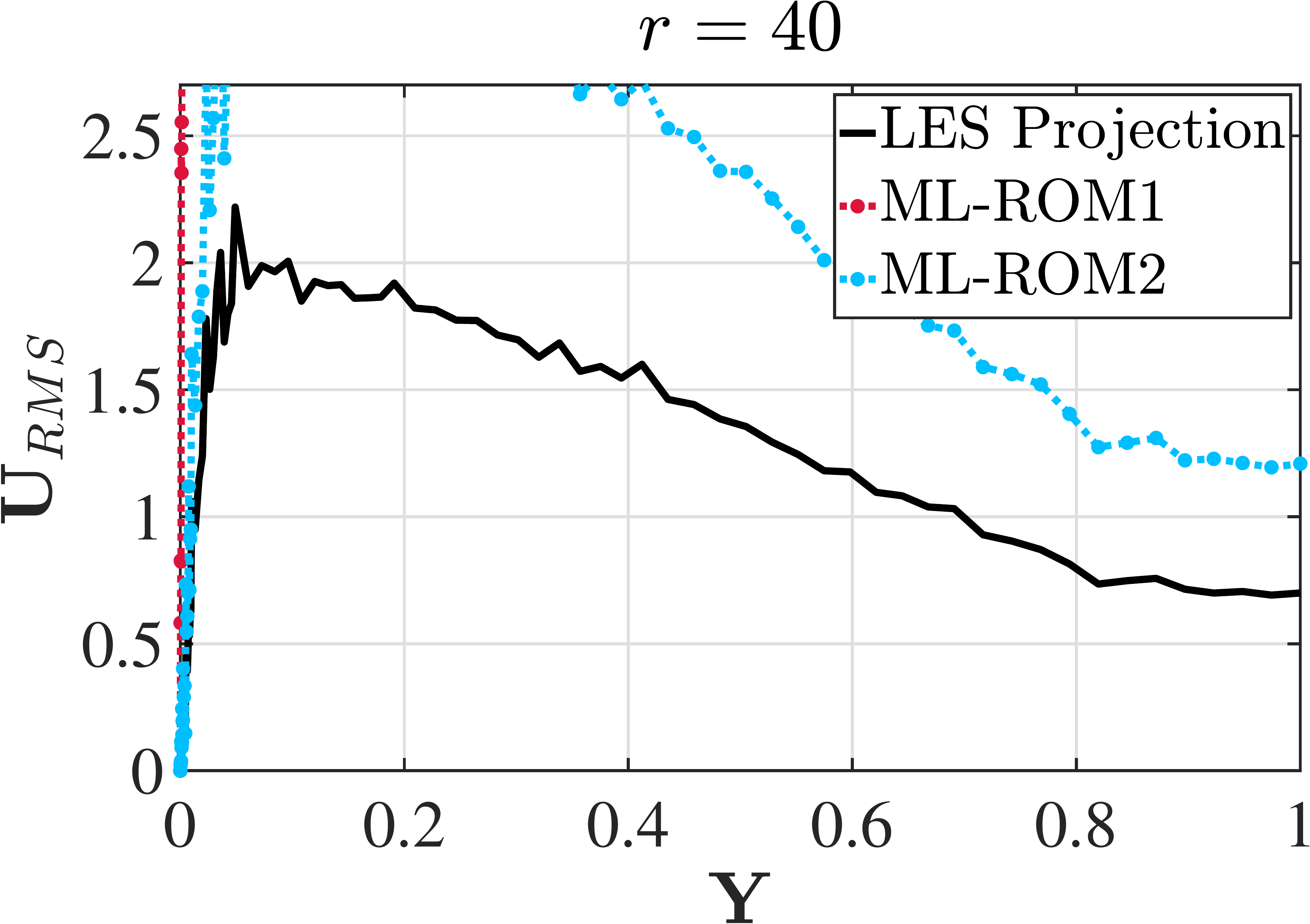}         \caption{$r=40$}
         \label{fig:stat-r-40}
     \end{subfigure} 
     \begin{subfigure}[b]{0.48\textwidth}
         \centering
    \includegraphics[width=.45\textwidth]{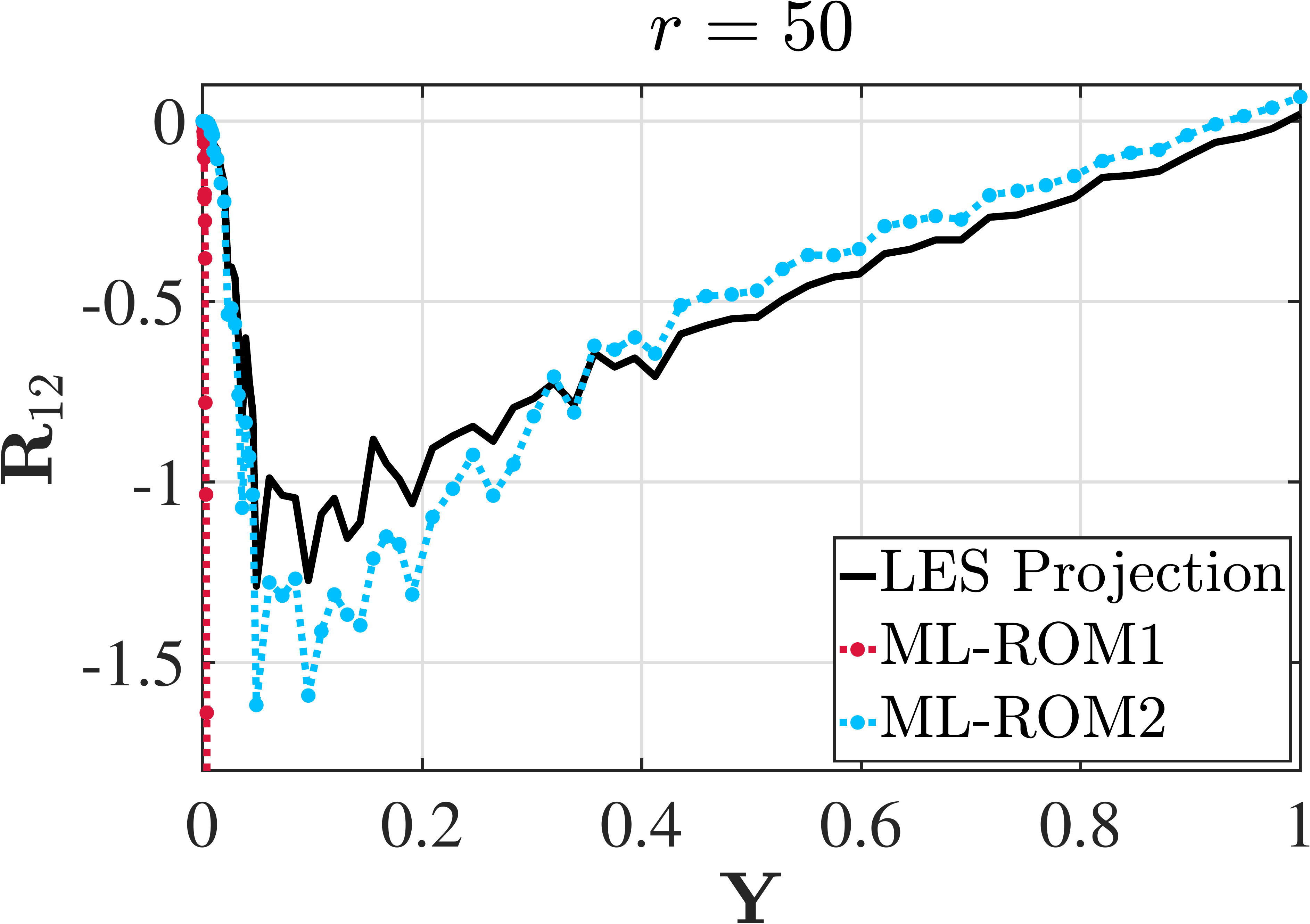}
    \includegraphics[width=.45\textwidth]{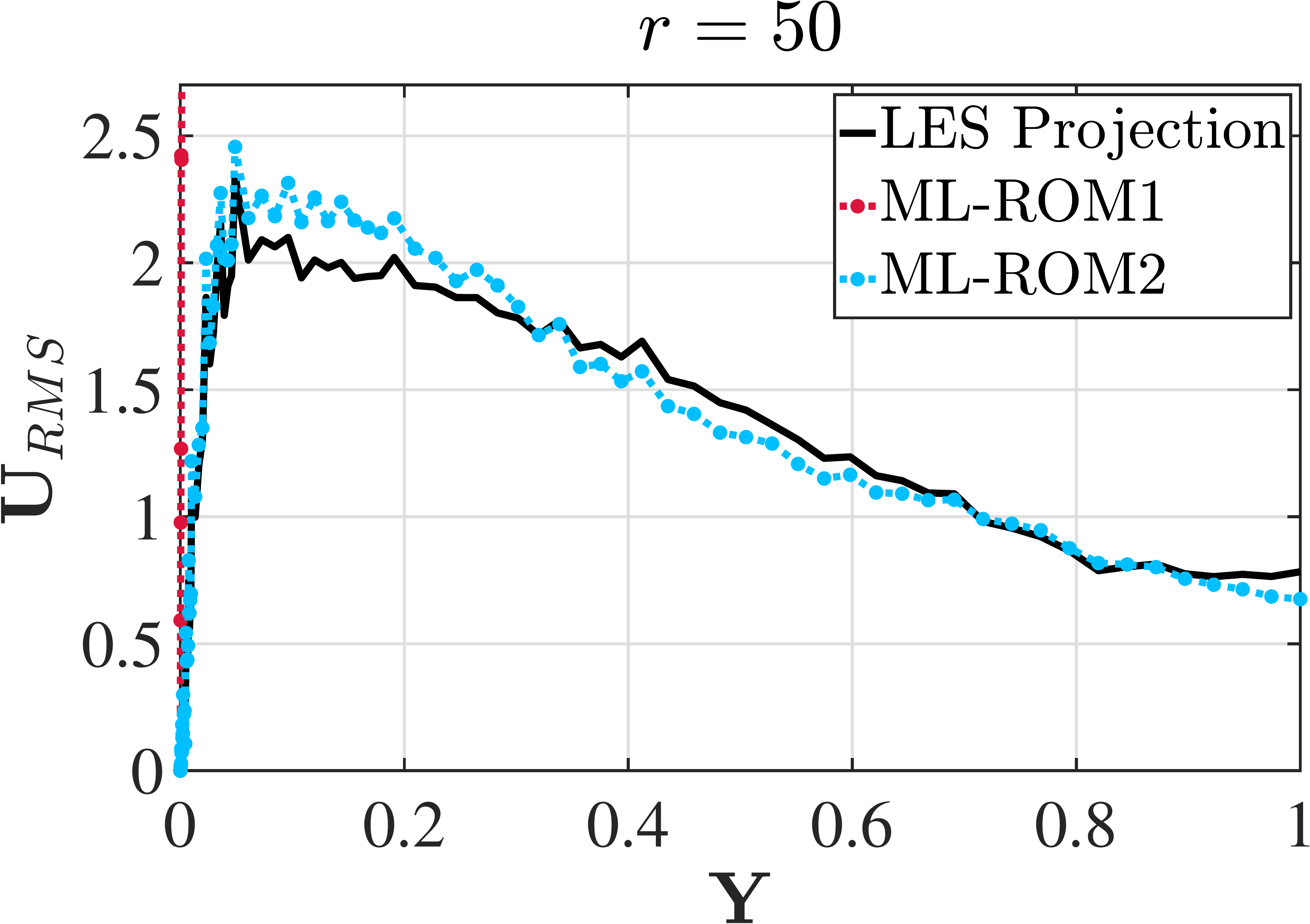}         \caption{$r=50$}
         \label{fig:stat-r-50}
     \end{subfigure} 
     \caption{Second-order statistics for $\alpha=0.1\times 10^{-3}$
    }    
    \label{fig:stat-alpha-7}
\end{figure}

Next, we investigate the role played by the two ROM lengthscales in the ML-ROM~\eqref{eqn:les-rom} in which the ROM closure term is calculated according to~\eqref{eqn:ml-v2}: 
\begin{eqnarray}
	\btau
	= - \bigl( \alpha \, U_{ML} \, L_{ML} \bigr) \, S \, \ba \, .
	\label{eqn:ml-v3}
\end{eqnarray}
We denote ML-ROM1 the ML-ROM in which $L_{ML}=\delta_1$ in~\eqref{eqn:ml-v3} and ML-ROM2 the ML-ROM in which $L_{ML}=\delta_2$ in~\eqref{eqn:ml-v3}.
To ensure a fair comparison of ML-ROM1 and ML-ROM2, we use the same constant $\alpha$ and the same velocity scale $U_{ML}$ (i.e., the time averaged streamwise velocity component) 
in~\eqref{eqn:ml-v3} for both models, and vary only the ROM lengthscale, $L_{ML}=\delta_1$ or $L_{ML}=\delta_2$.
To vary the ROM lengthscale, we vary the $r$ value in the definitions of $\delta_1$ and $\delta_2$.
In Figures~\ref{fig:ke-alpha-1}--\ref{fig:stat-alpha-7}, we plot the time evolution of the kinetic energy and the second-order statistics of the ML-ROM1 and ML-ROM2 for different $r$ values and different $\alpha$ values:
$\alpha = 6 \times 10^{-3}$ (Figures~\ref{fig:ke-alpha-1} and ~\ref{fig:stat-alpha-1}), 
$\alpha = 2 \times 10^{-3}$ (Figure~\ref{fig:ke-alpha-2} and ~\ref{fig:stat-alpha-2}),
$\alpha = 0.2 \times 10^{-3}$
(Figure~\ref{fig:ke-alpha-5} and ~\ref{fig:stat-alpha-5}), and
$\alpha = 0.1 \times 10^{-3}$
(Figure~\ref{fig:ke-alpha-7} and ~\ref{fig:stat-alpha-7}).
We note that the small oscillations in the second-order statistics plots are probably a consequence of considering only the bottom half of the channel instead of averaging over both halves.
As a benchmark for the ROM results, we use the projection of the the FOM results on the ROM basis (denoted as LES-proj in these plots).

Overall, in terms of stability, these plots show that ML-ROM2 yields more stable results than ML-ROM1.
This is clearly shown in Figures~\ref{fig:ke-alpha-1}  and ~\ref{fig:stat-alpha-1}, where ML-ROM2 yields stable results for all $r$ values, whereas ML-ROM2 yields stable results only for the largest values ($r=40$ and $r=50$) and blows up for the other $r$ values.
As we decrease the $\alpha$ value in
Figures~\ref{fig:ke-alpha-2}--\ref{fig:stat-alpha-7},
the ML-ROM1 becomes more unstable.
Indeed, starting with $\alpha = 2 \times 10^{-3}$ (Figures~\ref{fig:ke-alpha-2} and ~\ref{fig:stat-alpha-2}), the ML-ROM1 kinetic energy blows up for all the $r$ values.
This behavior is expected, since decreasing the $\alpha$ value decreases the amount of artificial viscosity in the ML-ROM.
In contrast, the ML-ROM2 is stable for most of the parameter values.
For the largest $\alpha$ values (i.e., $\alpha = 6 \times 10^{-3}$ in Figures~\ref{fig:ke-alpha-1} and ~\ref{fig:stat-alpha-1}, and $\alpha = 2 \times 10^{-3}$ in Figures~\ref{fig:ke-alpha-2} and ~\ref{fig:stat-alpha-2}), ML-ROM2 is stable for all $r$ values.
As we decrease the $\alpha$ value in Figures~\ref{fig:ke-alpha-5}--\ref{fig:stat-alpha-7}, ML-ROM2 becomes 
more unstable and blows up for lower $r$ values: 
For $\alpha=0.2 \times 10^{-3}$ in Figures~\ref{fig:ke-alpha-5} and~\ref{fig:stat-alpha-5}, ML-ROM2 blows up for $r=4, 8$, and $16$;
and for $\alpha=0.1 \times 10^{-3}$ in Figures~\ref{fig:ke-alpha-7} and~\ref{fig:stat-alpha-7}, ML-ROM2 blows up for $r=4, 8, 16$, and $32$.
We emphasize, however, that ML-ROM2 is consistently more stable than ML-ROM1.

Overall, in terms of accuracy, the ML-ROM1 and ML-ROM2 plots in Figures~\ref{fig:ke-alpha-1}--\ref{fig:stat-alpha-7} do not show a clear winner: 
For some $\alpha$ and $r$ values ML-ROM1 is more accurate, for other values ML-ROM2 is more accurate.
Furthermore, it seems that, for both ML-ROM1 and ML-ROM2 and for each $r$ value, one can find the optimal $\alpha$ value that ensures the highest accuracy.
\clearpage

Since the plots in Figures~\ref{fig:ke-alpha-1}--\ref{fig:stat-alpha-7} showed that the ML-ROM1's stability is clearly different from the ML-ROM2's stability, we decided to quantify the stability of the two ML-ROM.
To this end, in Table~\ref{tab:alpha-value-threshold},  
for different $r$ values, we list the threshold $\alpha_0$ value, i.e., the value that ensures that, if $\alpha > \alpha_0$, then the ML-ROM is stable.
These results show that, for each $r$ value, the threshold $\alpha_0$ value is 
more than an order of magnitude lower for ML-ROM2 than for ML-ROM1.
Thus, we conclude that ML-ROM1 is more stable than ML-ROM2. We note that the same conclusion can be drawn from the plots in Figures~\ref{fig:ke-alpha-1}--\ref{fig:stat-alpha-7}.

\begin{table}[H]
    \centering
    \begin{tabular}{c c|c c c c c c c c c c c c }
    \hline\hline
        &$r$ & 4 & 8 & 16 & 32 &40  & 50 
    \\ \hline
        ML-ROM1  &$\alpha_0$ & 10e-3&  9.8e-3& 9.2e-3& 8.5e-3& 6.5e-3& 6.2e-3
    \\ 
        ML-ROM2   &$\alpha_0$
        &2.2e-4 & 2.1e-4 &1.8e-4 & 1.7e-4 &1.2e-4 &1.1e-4
    \\ \hline
    \end{tabular}
    \caption{Threshold $\alpha_0$ values for different $r$ values.}
    \label{tab:alpha-value-threshold}
\end{table}

\section{Conclusions}
    \label{sec:conclusions}

In this paper, we proposed a novel ROM lengthscale definition. 
This new ROM lengthscale, denoted $\delta_2$, is constructed by using energy distribution arguments.
Specifically, we balanced the ROM and FOM energy content with the energy content in the $\delta_2$ and $h$ scales, respectively, where $h$ is the FOM mesh size. 
We emphasize that the novel ROM lengthscale, $\delta_2$, is fundamentally different from the current ROM lengthscales, which are built by using dimensional arguments.

We compared the new ROM lengthscale, $\delta_2$, with a standard dimensional based ROM lengthscale, denoted $\delta_1$.
To this end, we used these two ROM lengthscales to build two  mixing-length ROMs (ML-ROMs) in which all the other parameters were the same.
We investigated the two resulting ML-ROMs in the numerical simulation of the turbulent channel flow at $Re_{\tau} = 395$.
The numerical results showed that the new ROM lengthscale, $\delta_2$, is signficantly different from the standard ROM lengthscale, $\delta_1$.
Furthermore, the ML-ROM based on the new ROM lengthscale was significantly more stable than the ML-ROM based on the standard ROM lengthscale.

This first step in the numerical assessment of the new ROM lengthscale yielded encouraging results.
We plan to further investigate the new ROM lengthscale in the construction of other types of ROMs, e.g., large eddy simulation ROMs~\cite{wang2012proper,xie2017approximate} or regularized ROMs~\cite{wells2017evolve,kaneko2020towards}.
We also plan to leverage the new energy based lengthscale to develop scale-aware ROM strategies that are better suited for flow-specific applications.

\section*{Acknowledgments}

The work of the first and fourth authors was supported by NSF through grant DMS-2012253 and CDS\&E-MSS-1953113.
The third author gratefully acknowledges the U.S. DOE Early Career Research Program support through grant DE-SC0019290 and the NSF support through grant DMS-2012255.
Part of this work was funded under the nuclear energy advanced modeling and simulation program.

\setlength{\baselineskip}{12pt}

\bibliographystyle{plain}
\bibliography{traian}

\end{document}